\documentclass{report}%
\usepackage{graphicx}
\usepackage{amsmath}%
\usepackage{amsfonts}%
\usepackage{amssymb}
\newcommand{\be}{\begin{eqnarray}}
\newcommand{\ee}{\end{eqnarray}}
\begin{document}

\title{A Primer on Instantons in QCD }
\author{Hilmar Forkel\\Institut f\"{u}r Theoretische Physik, Universit\"{a}t Heidelberg, \\Philosophenweg 19, D-69120 Heidelberg, Germany. }
\date{August 2000}
\maketitle
\tableofcontents

\abstract{These are the (twice) extended notes of a set of lectures given at the ``12th
Workshop on Hadronic Interactions'' at the IF/UERJ, Rio de Janeiro (31. 5. -
2. 6. 2000). The lectures aim at introducing essential concepts of instanton
physics, with emphasis on the role of instantons in generating tunneling
amplitudes, vacuum structure, and the induced quark interactions associated
with the axial anomaly. A few examples for the impact of instantons on the
physics of hadrons are also mentioned.}

\chapter{Introduction}

Yang-Mills instantons were discovered a quarter of a century ago \cite{bel75}.
They have furnished the first explicit (and still paradigmatic) example of a
genuinely nonperturbative gauge field configuration and display a wealth of
unprecedented geometrical, topological and quantum effects with fundamental
impact on both ground state and spectrum of nonabelian gauge theories. In
fact, one could argue that the discovery of instantons marked the beginning of
a new era in field theory. In the words of Sidney Coleman \cite{col80}: ``In
the last two years there have been astonishing developments in quantum field
theory. We have obtained control over problems previously believed to be of
insuperable difficulty and we have obtained deep and surprising (at least to
me) insights into the structure of ... quantum chromodynamics.'' Instantons
have remained an active, fascinating and highly diverse research area ever
since, both in mathematics and physics.

The following pages contain the notes of a series of elementary lectures on
instantons in quantum mechanics and QCD which aim at introducing some of the
underlying ideas to a non-expert audience.

\section{Motivation and overview}

QCD instantons are mediating intriguing quantum processes which shape the
ground state of the strong interactions. These localized and inherently
nonperturbative processes can be thought of as an ongoing ``rearrangement'' of
the vacuum. In addition, there exist many other types of instantons in various
areas of physics. This became clear shortly after their discovery when it was
realized that instantons are associated with the semiclassical description of
tunneling processes.

Since tunneling processes abound in quantum theory, so do instantons. We will
take advantage of this situation by explaining the physical role of instantons
in the simplest possible dynamical setting, namely in the one-dimensional
quantum mechanics of tunneling through a potential barrier. Since the
classical aspects of the problem are particularly simple and familiar here, we
can focus almost exclusively on the crucial quantum features and their
transparent treatment. Once the underlying physics is understood, this example
can be more or less directly generalized to the tunneling processes in the QCD
vacuum, where we can then concentrate on the additional complexities brought
in by the field theoretical and topological aspects of classical Yang-Mills
instantons. We will end our brief tour of instanton physics with a short
section on the impact of instantons on hadron phenomenology.

As already mentioned, instantons occur in many guises and we can only give a
small glimpse of this vast field here. (The SLAC high-energy preprint library
lists far beyond 1000 preprints on instantons in the last decade.)\ Moreover,
we will spend a large part of the available time on laying a solid conceptual
foundation and will consequently have less time for applications. To do at
least some justice to the diversity of the subject and to give an idea of its
scope, let us mention a few actively studied\ topics not touched upon in these
lectures: instantons have been found in many field theories, ranging from
scalar field theory via supersymmetric Yang-Mills and gravity to string (or
M-) theory. There are interesting relations and interactions between
instantons and their topological cousins, \ the non-abelian monopoles and
vortices. In several theories, probably including QCD, instantons are
responsible for spontaneous chiral symmetry breaking. The role of instantons
in deep inelastic scattering and other hard QCD processes has been examined,
and also their impact on weak-interaction processes at RHIC,\ LHC and beyond.
In inflationary cosmology and elsewhere relatives of instantons (sometimes
called ``bounces'') describe the ``decay of the false vacuum''.

Moreover, there are fascinating mathematical developments in which instantons
serve as tools. They have been instrumental, e.g., in obtaining profound
results in geometry and topology, including Donaldson theory of four-manifolds
(which led to the discovery of infinitely many new differentiable structures
on $R^{4}$). Instantons play a particularly important role in supersymmetric
gauge, string and brane theories. They saturate, for example, the
nonperturbative sector of the low-energy effective (Seiberg-Witten) action of
$N=2$ supersymmetric Yang-Mills theory.

Theoretical approaches to QCD instantons include well-developed
``instanton-liquid'' vacuum models \cite{sch98}, a variety of hadron models at
least partially based on instanton-induced interactions (see, e.g.
\cite{hor78}), a sum-rule approach based on a generalized operator product
expansion (IOPE) \cite{for00}, and an increasing amount of lattice simulations
\cite{lat}.

There is a vast literature on instantons. In preparing these lectures I have
benefitted particularly from the articles and books by Coleman \cite{col80},
Kleinert \cite{kle94}, Schulman \cite{sch81}, Sakita \cite{sak85}, Polyakov
\cite{pol87}, Vainshtein et al \cite{vai99}, and Zinn-Justin \cite{zin89}.
More advanced and technical material on instantons is collected in
\cite{shi94}. A pedagogical introduction to instantons in supersymmetric field
theories can be found in \cite{shi99}.

The program of these lectures is as follows: first, we will discuss the
semi-classical approximation (SCA)\ in quantum mechanics in the path integral
representation. This will lead us to describe tunneling processes in imaginary
time where we will encounter the simplest examples of instanton solutions. In
the following we will turn to instantons in QCD, their topological properties,
their role in generating the vacuum structure, and their impact on the physics
of hadrons.

\section{WKB reminder}

As we have already mentioned, instantons mediate quantum-mechanical barrier
penetration processes. The semiclassical approximation (SCA) is the method of
choice for the treatment of such processes. It might therefore be useful to
recapitulate some basic aspects of this technique, in the form established by
Wenzel, Kramers and Brioullin (WKB).

The WKB method generates approximate solutions of the Schr\"{o}dinger equation
for wavefunctions with typical wavelength $\lambda$ small in comparison to the
spacial variations of the potential. This situation corresponds to the
semiclassical limit $\hbar\rightarrow0$ where%
\begin{equation}
\lambda=\frac{2\pi\hbar}{p}\rightarrow0.
\end{equation}
Macroscopic structures, for example, behave normally according to classical
mechanics because their wavefunctions vary extremely rapidly compared to the
variations of the underlying potentials or any other length scale in the
problem, even if of microscopic (e.g. atomic) origin. The semiclassical limit
is therefore the analog of the geometrical (ray) limit of wave optics, where
the scatterers are structureless at the scale of the typical wavelengths of
light and where wave phenomena like diffraction disappear.

In order to set up the semiclassical approximation, the WKB method writes the
wave function as%
\begin{equation}
\psi\left(  x\right)  =e^{i\Phi\left(  x\right)  /\hbar}. \label{psiwkb}%
\end{equation}
Since $\Phi\in C$ this ansatz is fully general. In scattering applications
$\Phi$ is often called the ``Eikonal''. Inserting (\ref{psiwkb}) into the
Schr\"{o}dinger equation%
\begin{equation}
\left[  -\frac{\hbar^{2}}{2m}\frac{d^{2}}{dx^{2}}+V\left(  x\right)
-E\right]  \psi\left(  x\right)  =0 \label{schreq}%
\end{equation}
one obtains the WKB ``master equation''%
\begin{equation}
\frac{1}{2m}\Phi^{\prime2}\left(  x\right)  -\frac{i\hbar}{2m}\Phi
^{\prime\prime}\left(  x\right)  =E-V\left(  x\right)  . \label{wkbmeq}%
\end{equation}

This nonlinear differential equation can be solved iteratively. One expands
\begin{equation}
\Phi\left(  x\right)  =\Phi_{0}\left(  x\right)  +\hbar\Phi_{1}\left(
x\right)  +\hbar^{2}\Phi_{2}\left(  x\right)  +...,\label{phiexp}%
\end{equation}
(the series is generally asymptotic) and equates terms of the same order of
$\hbar$ on both sides. For each power of $\hbar$ one thereby obtains a
corresponding WKB equation. The zeroth-order equation has the solution%
\begin{equation}
\Phi_{0}\left(  x\right)  =\pm\int^{x}dx^{\prime}p\left(  x^{\prime}\right)
\text{ \ \ \ \ \ \ \ with \ \ \ \ \ \ \ }p\left(  x\right)  \equiv
\sqrt{2m\left[  E-V\left(  x\right)  \right]  }\label{wkbphase}%
\end{equation}
where $p\left(  x\right)  $ is the momentum of the particle in a constant
potential $U_{x}=const$ whose value equals that of $V$ at $x$. If $V\left(
x\right)  \ $varies slowly compared to $\psi\left(  x\right)  $, $\psi$ indeed
experiences a locally constant $U_{x}$ and the $O\left(  \hbar^{0}\right)  $
solution
\begin{equation}
\psi_{0}\left(  x\right)  =e^{i/\hbar\int^{x}dx^{\prime}\sqrt{2m\left[
E-V\left(  x^{\prime}\right)  \right]  }}\label{psi0wkb}%
\end{equation}
becomes a useful approximation. By inserting it back into the Schr\"{o}dinger
equation (\ref{schreq}),
\begin{equation}
\left[  -\frac{\hbar^{2}}{2m}\frac{d^{2}}{dx^{2}}+V\left(  x\right)
-E\right]  \psi_{0}\left(  x\right)  =-\frac{i\hbar}{2}\frac{V^{\prime}\left(
x\right)  \psi_{0}\left(  x\right)  }{\sqrt{2m\left[  E-V\left(  x\right)
\right]  }},
\end{equation}
we confirm that it is a solution up to the correction term on the right-hand
side, which is indeed of $O\left(  \hbar\right)  $ and proportional to the
variation $V^{\prime}$ of the potential.

The quality of the zeroth-order approximation can be gauged more
systematically by checking whether the neglected term in (\ref{wkbmeq}) is
small, i.e whether
\begin{equation}
\left|  \frac{\frac{i\hbar}{2m}\Phi^{\prime\prime}}{\frac{1}{2m}\Phi^{\prime
2}}\right|  =\left|  \hbar\frac{\Phi^{\prime\prime}}{\Phi^{\prime2}}\right|
=\left|  \frac{d}{dx}\frac{\hbar}{\Phi^{\prime}}\right|  \ll1.
\end{equation}
With $\lambda\left(  x\right)  \equiv2\pi\hbar/p\left(  x\right)  =2\pi
\hbar/\Phi_{0}^{\prime}$ and with $\Phi_{0}^{\prime}\sim\Phi^{\prime}$ (to
leading order) this turns into
\begin{equation}
\frac{1}{2\pi}\left|  \frac{d\lambda}{dx}\right|  =\frac{1}{2\pi}\left|
\frac{\frac{d\lambda}{dx}\cdot\lambda}{\lambda}\right|  \equiv\frac{1}{2\pi
}\left|  \frac{\delta\lambda}{\lambda}\right|  \ll1
\end{equation}
where $\delta\lambda$ is the change of $\lambda$ over the distance of a
wavelength. From the above inequality we learn that the semiclassical
expansion is applicable in spacial regions where the de-Broglie wavelength 1)
is small compared to the typical variations of the potential (therefore highly
excited states behave increasingly classical) and where it 2)\ changes little
over distances of the order of the wavelength. (The latter condition alone is
not sufficient since we have compared two terms in a differential equation,
not in the solution itself.)

The semiclassical treatment of \textit{tunneling} processes, which will be a
recurrent theme throughout these lectures, involves a characteristic
additional step since tunneling occurs in potentials with classically
forbidden regions, i.e. regions $x\in\left[  x_{l},x_{u}\right]  $ where
$E<V\left(  x\right)  $ . The boundaries $x_{l},x_{u}$ are the classical
turning points. Inside the classically forbidden regions, $p\left(  x\right)
$ becomes imaginary and the solution (\ref{psi0wkb}) decays exponentially
as\footnote{Since Gamov's classic treatment of nuclear $\alpha$-decay as a
tunneling process through the Coulomb barrier of the nucleus this amplitude is
also known as the ``Gamov factor''.}%
\begin{equation}
\psi_{0,tunnel}\left(  x\right)  =e^{-1/\hbar\int_{x_{l}}^{x_{u}}dx^{\prime
}\sqrt{2m\left[  V\left(  x^{\prime}\right)  -E\right]  }}.\label{wkbgamov}%
\end{equation}
A comparison of the two solutions (\ref{psi0wkb}) and (\ref{wkbgamov}) shows
that, formally, the meaning of allowed and forbidden regions can be
interchanged by the replacement%
\begin{equation}
t\rightarrow-i\tau\text{ \ \ \ \ \ }\Rightarrow\text{ \ \ \ \ }E\rightarrow
iE,\text{ \ \ \ \ }V\rightarrow iV\label{wkbrepl}%
\end{equation}
which indeed transforms (\ref{psi0wkb}) into the tunneling amplitude
(\ref{wkbgamov}). In other words, we can calculate tunneling amplitudes in SCA
by the standard WKB methods, analytically continued to imaginary time. Below
we will see that this works the same way in the path-integral approach.

Although the WKB procedure is simple and intuitive in principle, it becomes
increasingly involved in more than one dimension (except for special cases
\cite{kap37}) and at higher orders. The $O\left(  \hbar^{n}\right)  $, $n>0$
corrections are calculated by matching the solutions in the classically
allowed and forbidden regions at the classical turning points. This leads to
Bohr-Sommerfeld quantization conditions for $p\left(  x\right)  $ when
integrated over a period of oscillation in the allowed regions. However, the
generalization of this procedure to many-body problems and especially to field
theory is often impractical \cite{ger77}. Fortunately, there is an alternative
approach to the SCA, based on the path integral, which can be directly applied
to field theories. With our later application to QCD in mind, we will now
consider this approach in more detail.

\chapter{Instantons in quantum mechanics}

\section{SCA via path integrals}

One of the major advantages of the path integral representation of quantum
mechanics is that it most transparently embodies the transition to classical
mechanics, i.e. the semiclassical limit. In the following chapter we will
discuss this limit and the associated SCA\ in the simplest dynamical setting
in which instantons play a role, namely in one-dimensional potential problems
with one degree of freedom (a spinless point particle, say).

\subsection{The propagator at real times: path integral and
SCA\label{propreal}}

The key object in quantum mechanics, which contains all the physical
information about the system under consideration (spectrum, wavefunctions
etc.), is the matrix element of the time evolution operator%
\begin{equation}
\left\langle x_{f}\left|  e^{-iHT/\hbar}\right|  x_{i}\right\rangle ,
\label{keyme}%
\end{equation}
i.e. the probability amplitude for the particle to propagate from
\begin{equation}
x_{i}\text{ \ \ \ \ \ \ at \ \ \ \ \ \ }t=-\frac{T}{2}\text{ \ \ \ \ \ \ to
\ \ \ \ \ \ }x_{f}\text{ \ \ \ \ \ \ at \ \ \ \ \ \ }t=\frac{T}{2}.
\end{equation}

(In quantum field theory this matrix element, which is sometimes called the
quantum-mechanical propagator, generalizes to the generating functional.) Its
path integral representation is%

\begin{equation}
\left\langle x_{f}\left|  e^{-iHT/\hbar}\right|  x_{i}\right\rangle
=\mathcal{N}\int D[x]e^{i\frac{S\left[  x\right]  }{\hbar}}=\mathcal{N}\int
D[x]_{\left\{  x\left(  -T/2\right)  =x_{i}|\,x\left(  T/2\right)
=x_{f}\right\}  }e^{\frac{i}{\hbar}\int_{-T/2}^{T/2}dt\mathcal{L}\left(
x,\dot{x}\right)  } \label{pint1}%
\end{equation}
where%
\begin{equation}
S\left[  x\right]  =\int_{-T/2}^{T/2}dt\mathcal{L}\left(  x,\dot{x}\right)
=\int_{-T/2}^{T/2}dt\left\{  \frac{m}{2}\dot{x}^{2}\left(  t\right)  -V\left[
x\left(  t\right)  \right]  \right\}
\end{equation}
is the classical action along a given path. The integration over the paths can
be defined, for example, by discretizing the time coordinate (Trotter formula)
into intervals $\Delta t$ so that $t_{n}=n\Delta t$ and
\begin{equation}
D[x]:=\lim_{N\rightarrow\infty}\left(  \frac{mN}{2\pi i\hbar t}\right)
^{1/2}\prod_{n=1}^{N-1}\left(  \frac{mN}{2\pi i\hbar t}\right)  ^{1/2}dx_{n}%
\end{equation}
where $x_{n}=x\left(  t_{n}\right)  .$ An alternative representation of $D[x]$
in terms of a complete set of functions will be introduced below.

In the semiclassical limit, i.e. for $\hbar\rightarrow0$, the classical action
can become much larger than $\hbar$. As a consequence, the path integral is
dominated by the paths in the vicinity of the stationary point(s)
$x_{cl}\left(  t\right)  $ of the action (if they exist), which satisfy
\begin{equation}
\frac{-\delta}{\delta x\left(  t\right)  }S\left[  x\right]  =m\ddot
{x}+\frac{\partial V}{\partial x}=0 \label{eom}%
\end{equation}
with the boundary conditions%
\begin{equation}
x_{cl}\left(  -\frac{T}{2}\right)  =x_{i},\text{ \ \ \ \ \ \ \ }x_{cl}\left(
\frac{T}{2}\right)  =x_{f}. \label{bc}%
\end{equation}

These classical paths are important not because they themselves give dominant
contributions to the integral (in fact, their contribution vanishes since the
set of classical paths is of measure zero) but rather because the action of
the neighboring paths varies the least around them. Thus, infinitely many
neighboring paths, which lie in the so-called ``coherence region'' with
similar phase factors $\exp\left(  iS/\hbar\right)  $, add coherently. For the
paths outside of the coherence region the phases vary so rapidly that
contributions from neighboring paths interfere destructively and become
irrelevant to the path integral. This exhibits in a transparent and intuitive
way the relevance of the classical paths in quantum mechanics.

\begin{figure}
[ptb]
\begin{center}
\includegraphics[
height=1.9501in,
width=3.1574in
]%
{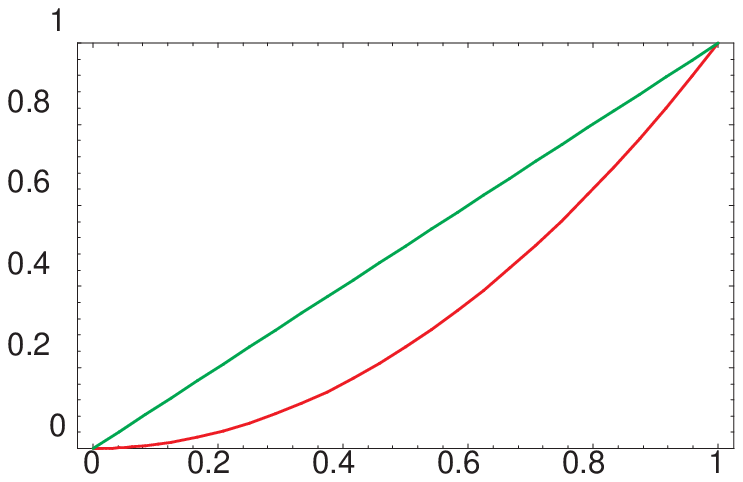}%
\caption{The two paths from $\left(  x,t\right)  =\left(  0,0\right)  $ to
$\left(  1,1\right)  $ whose action is compared.}%
\label{2paths}%
\end{center}
\end{figure}

To get a more quantitative idea of the coherence region, let us approximately
define it as the set of paths whose phases differ by less than $\pi$ from the
phase of the classical path (the ``stationary phase''), which implies
\begin{equation}
\delta S\left[  x\right]  =S\left[  x\right]  -S\left[  x_{cl}\right]  \leq
\pi\hbar.
\end{equation}
Thus, for a macroscopic particle with
\begin{equation}
S\simeq1\text{ erg sec }=1\frac{\text{g cm}^{2}}{\text{sec}^{2}}%
\text{sec}\simeq10^{27}\hbar
\end{equation}
only an extremely small neighborhood of the classical path contributes, since
$\delta\phi=\delta S/\hbar$ gets very sensitive to variations of the path. A
numerical example (borrowed from Shankar \cite{sha94}) makes this more
explicit: compare two alternative paths from $\left(  x,t\right)  =\left(
0,0\right)  $ to $\left(  x,t\right)  =\left(  1\,\text{cm},1\text{\thinspace
s}\right)  $ (see Fig. \ref{2paths}) for a noninteracting particle: the
classical path $x_{cl}\left(  t\right)  =t$ with $S\left[  x_{cl}\right]
=\int_{0}^{T}dt\left(  m/2\right)  v\left[  x_{cl}\right]  ^{2}=m\,$%
cm$^{2}/\left(  2\,\text{sec}^{2}\right)  $ and the alternative path
$x_{alt}\left(  t\right)  =t^{2}$ with $S\left[  x_{alt}\right]  =2m\,$%
cm$^{2}/\left(  3\,\text{sec}^{2}\right)  $.%
For a classical particle with
\begin{equation}
m=1g\text{ \ \ \ \ }\Rightarrow\delta S=S\left[  x_{alt}\right]  -S\left[
x_{cl}\right]  =\frac{m\text{ cm}^{2}}{6\,\text{sec}^{2}}\simeq1.6\times
10^{26}\hbar\text{ \ \ \ \ }\Rightarrow\delta\phi\simeq1.6\times
10^{26}\,\text{rad}\ggg\pi
\end{equation}
the alternative path is extremely incoherent and therefore irrelevant, while
for an electron with%
\begin{equation}
m=10^{-27}g\text{ \ \ \ \ }\Rightarrow\delta S\simeq\frac{1}{6}\hbar\text{
\ \ \ \ }\Rightarrow\delta\phi\simeq\frac{1}{6}\,\text{rad}\ll\pi
\end{equation}
it is well within the coherence region and makes an important contribution to
the path integral!

Obviously, then, the movement of a free electron (even a very fast one) cannot
be described classically. One has to resort to quantum mechanics where the
electron's path is much more uncertain and, in fact, not an observable. There
also exist microscopic situations, on the other hand, where the quantum
fluctuations do not totally wash out the classical results and which can
therefore be treated semiclassically. Typical examples are the scattering off
a slowly varying potential (Eikonal approximation) or the highly excited
electronic orbits in atoms (Rydberg atoms).

\begin{figure}
[ptb]
\begin{center}
\includegraphics[
height=1.9501in,
width=3.1574in
]%
{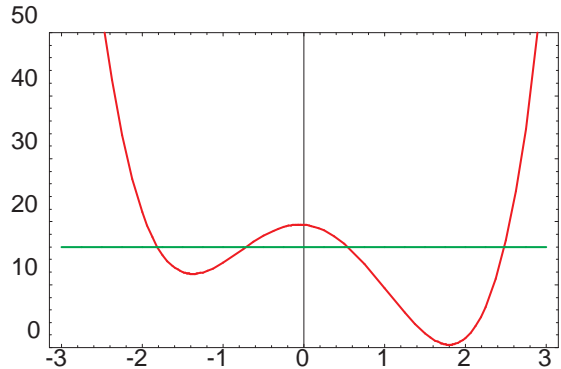}%
\caption{A typical tunneling potential with nondegenerate minima. The total
energy (horizontal line) is smaller than the hump so that a classically
forbidden region exists.}%
\label{tunlpot1}%
\end{center}
\end{figure}

However - and this is a crucial point for our subsequent discussion - the
stationary-phase approximation fails to describe tunneling processes! The
reason is that such processes are characterized by potentials with a
classically forbidden region (or barrier) which cannot be transgressed by
classical particles. In other words, there exist no classical solutions of
(\ref{eom}) with boundary conditions corresponding to barrier penetration in a
potential of the type shown in Fig. \ref{tunlpot1}. As a consequence, the
action $S\left[  x\right]  $ has no extrema with tunneling boundary
conditions, and therefore the path integral (\ref{pint1}) has no stationary
points.%

Does this mean that the SCA becomes unfeasible for tunneling processes?
Fortunately not, of course, as the standard WKB approximation shows. The
problem is particular to the stationary-phase approximation in the path
integral framework, which indeed ceases to exist. The most direct way to
overcome this problem would be to generalize the familiar saddle-point
approximation for ordinary integrals. This method works by deforming the
integration path into the complex plane in such a way that it passes through
the saddle points of the exponent in the integrand. And since there indeed
exist complex solutions of the combined equations (\ref{eom}), (\ref{bc}) in
tunneling potentials (in other words, the complex action still obeys the
Hamilton-Jacobi equation), a generalization of the complex saddle-point method
to integrals over complex paths would seem natural. But unfortunately
integration over complex paths has not been sufficiently developed (although a
combined approach to both diffraction and ray optics in this spirit was
initiated by Keller \cite{kel62}, and McLaughlin \cite{mcl72} has performed a
complex saddle point evaluation of the Fourier-transformed path integral. See
also the recent work of Weingarten \cite{wei02} on complex path integrals.).

However, in many instances - including our tunneling problem - an analytical
continuation of the path integral to imaginary time, as already encountered in
the WKB framework in (\ref{wkbrepl}), serves the same purposes. This is the
way in which we will perform the SCA to tunneling problems in the following sections.

\subsection{Propagator in imaginary time}

As just mentioned, the paths which form the basis of the SCA to tunneling
processes in the path-integral framework can be identified by analytical
continuation to imaginary time. This approach was apparently first used by
Langer \cite{lan67} in his study of bubble formation processes at first-order
phase transitions\footnote{Another early application of similar ideas is due
to W.A. Miller \cite{mil74}. (I thank Takeshi Kodama for drawing my attention
to this reference.)}. In the context of the SCA, tunneling problems in field
theory, and instantons it was introduced by Polyakov et al. \cite{col80,pol77}%
, building on a suggestions by Gribov.

In order to prepare for the use of this approach, let us first see how the
ground-state energy and wavefunction in tunneling situations can be obtained
from the matrix element
\begin{equation}
Z\left(  x_{f},x_{i}\right)  =\left\langle x_{f}\left|  e^{-HT_{E}/\hbar
}\right|  x_{i}\right\rangle \label{imagtprop}%
\end{equation}
which is obtained from (\ref{keyme}) by analytically continuing
\begin{equation}
t\rightarrow-i\tau\text{ \ \ \ \ \ \ }\left(  \Rightarrow T\rightarrow
-iT_{E}\right)  . \label{analytcont}%
\end{equation}
This procedure (together with its counterpart $p_{0}\rightarrow ip_{4}$ in
momentum space) is often called a ''Wick rotation''. (Note, incidentally, that
imaginary-``time'' evolution as generated by $e^{-HT_{E}/\hbar}$ is not
unitary and therefore does not conserve probability. It might also be
worthwhile to recall that the matrix element (\ref{imagtprop}) plays a
fundamental role in statistical mechanics (its trace over state space,
corresponding to the sum over all periodic paths, is the partition function)
although this is not the angle from which we will look at it in the following.)

The matrix element $Z\left(  x_{f},x_{i}\right)  $ has the energy
representation (in the following we drop the subscript $E$ of $T_{E}$)%
\begin{equation}
Z\left(  x_{f},x_{i}\right)  =\sum_{n}e^{-E_{n}T/\hbar}\left\langle
x_{f}|n\right\rangle \left\langle n|x_{i}\right\rangle \label{zspectral}%
\end{equation}
in terms of the spectrum
\begin{equation}
H\left|  n\right\rangle =E_{n}\left|  n\right\rangle ,\text{ \ \ \ \ \ \ }%
1=\sum_{n}\left|  n\right\rangle \left\langle n\right|
\end{equation}
of the (static)\ Hamiltonian $H$. The energies $E_{n}$ are the usual real
eigenvalues, which remain unaffected by the analytical continuation. For large
$T$ the ground state dominates,
\begin{equation}
Z\left(  x_{f},x_{i}\right)  \rightarrow e^{-E_{0}T/\hbar}\left\langle
x_{f}|0\right\rangle \left\langle 0|x_{i}\right\rangle , \label{asympt}%
\end{equation}
and the ground state energy becomes%
\begin{equation}
E_{0}=-\hbar\lim_{T\rightarrow\infty}\frac{1}{T}\ln Z\left(  x_{f}%
,x_{i}\right)  . \label{grsteng}%
\end{equation}
In order to calculate the ground state energy (and wave function) of a
quantum-mechnical system we therefore just need to take the $T\rightarrow
\infty$ limit of the imaginary-time matrix element (\ref{imagtprop}). This
convenient way to obtain ground-state properties is used, e.g., in Euclidean
lattice QCD to calculate hadron masses. In the following section we will show
how to calculate the matrix element $Z\left(  x_{f},x_{i}\right)  $
semiclassically in the path integral framework.

\subsection{Path integral in imaginary time: tunneling in SCA\label{piimagt}}

The quantum mechanical propagator (\ref{imagtprop}) in imaginary time has a
path integral representation which can be obtained from (\ref{pint1}) by
analytical continuation:%
\begin{equation}
Z\left(  x_{f},x_{i}\right)  =\mathcal{N}\int D[x]e^{-\frac{S_{E}\left[
x\right]  }{\hbar}}=\mathcal{N}\int D[x]_{\left\{  x\left(  -T/2\right)
=x_{i}|\,x(T/2)=x_{f}\right\}  }e^{-\frac{1}{\hbar}\int_{-T/2}^{T/2}%
d\tau\mathcal{L}_{E}\left(  x,\dot{x}\right)  }. \label{pint2}%
\end{equation}
(Note, by the way, that this path integral is much better behaved than its
real-time counterpart (and thus can be formalized as an integral over
generalized stochastic Wiener processes) since the oscillating integrand is
replaced by an exponentially decaying one.)

For future applications it is useful to define the ``measure'' of this
integral more explicitly. To this end, we expand $x\left(  \tau\right)  $ into
a complete, orthonormal set of real functions $\tilde{x}_{n}\left(
\tau\right)  $ around a fixed path $\bar{x}\left(  \tau\right)  $ as%

\begin{equation}
x\left(  \tau\right)  =\bar{x}\left(  \tau\right)  +\eta\left(  \tau\right)
\ \ \ \ \ \ \text{where}\ \ \ \ \ \eta\left(  \tau\right)  =\sum_{n=0}%
^{\infty}c_{c}\tilde{x}_{n}\left(  \tau\right)  \ \ \ \ \ \ \ \
\end{equation}
and%
\begin{equation}
\int_{-T/2}^{T/2}d\tau\tilde{x}_{n}\left(  \tau\right)  \tilde{x}_{m}\left(
\tau\right)  =\delta_{mn},\text{ \ \ \ \ \ \ \ \ \ }\sum_{n}\tilde{x}%
_{n}\left(  \tau\right)  \tilde{x}_{n}\left(  \tau^{\prime}\right)
=\delta\left(  \tau-\tau^{\prime}\right)  .\text{ \ \ \ \ \ \ \ \ \ }
\label{orthocompl}%
\end{equation}
Furthermore, $\bar{x}$ is assumed to satisfy the boundary conditions implicit
in the path integral, i.e.
\begin{equation}
\bar{x}\left(  \pm T/2\right)  =x_{f}/x_{i},\text{ \ \ \ \ \ \ \ \ \ \ \ \ \ }%
\tilde{x}_{n}\left(  \pm T/2\right)  =0. \label{bcfluct}%
\end{equation}
(Note that (for now) there are no further requirements on $\bar{x}$.) As a
consequence we have%
\begin{equation}
D\left[  x\right]  =D\left[  \eta\right]  =\prod_{n}\frac{dc_{n}}{\sqrt
{2\pi\hbar}}, \label{meas}%
\end{equation}
where the normalization factor is chosen for later convenience.

Now let us get the explicit form of $\mathcal{L}_{E}\left(  x,\dot{x}\right)
$ (the subscript $E$ stands for ``Euclidean time'', which is used synonymously
with ``imaginary time'' in field theory) in (\ref{pint2}) by analytical
continuation of the integration path as
\begin{equation}
iS=i\int_{-T/2}^{T/2}dt\left(  \frac{m}{2}\dot{x}^{2}-V\left[  x\right]
\right)  \rightarrow i\int_{-\frac{T}{2}e^{-i\pi/2}}^{\frac{T}{2}e^{-i\pi/2}%
}dt\left[  \frac{m}{2}\left(  \frac{dx}{dt}\right)  ^{2}-V\left[  x\right]
\right]  \equiv-S_{E}.
\end{equation}
After substituting $t\rightarrow-i\tau$ according to (\ref{analytcont}) we
obtain%
\begin{equation}
-S_{E}=i\left(  -i\right)  \int_{-T/2}^{T/2}d\tau\left(  -\frac{m}{2}\dot
{x}^{2}-V\left[  x\right]  \right)  \equiv-\int_{-T/2}^{T/2}d\tau
\mathcal{L}_{E}\left[  x\right]  \label{se1}%
\end{equation}
(note that in Eq. (\ref{se1}) and in the following the dot indicates
differentiation with respect to $\tau$) from which we read off%
\begin{equation}
\mathcal{L}_{E}\left[  x\right]  =\frac{m}{2}\dot{x}^{2}+V\left(  x\right)  .
\end{equation}
Thus, besides turning the overall factor of $i$ in the exponent into a minus
sign (as anticipated in (\ref{pint2})), the analytical continuation has the
crucial implication that the potential changes its sign, i.e. that it is
turned upside down!

The consequence of this for the SCA is immediate: now there exist (one or
more) solutions to the imaginary-time equation of motion%
\begin{equation}
\frac{-\delta}{\delta x\left(  \tau\right)  }S_{E}\left[  x\right]  =m\ddot
{x}_{cl}-V^{\prime}\left(  x_{cl}\right)  =0 \label{eomimagt}%
\end{equation}
with tunneling boundary conditions. These solutions correspond to a particle
which starts at $x_{i},$ rolls down the hill and through the local minimum of
$-V\left(  x\right)  $ (which corresponds to the peak of the classically
forbidden barrier of $+V\left(  x\right)  $) and climbs up to $x_{f}$,
according to the boundary conditions
\begin{equation}
x\left(  -\frac{T}{2}\right)  =x_{i},\text{ \ \ \ \ \ \ \ \ }x\left(
\frac{T}{2}\right)  =x_{f}. \label{tunbc}%
\end{equation}

For later reference we note that the solutions of (\ref{eomimagt})\ carry a
conserved quantum number, the ``Euclidean energy''%
\begin{equation}
E_{E}=\frac{m}{2}\dot{x}^{2}-V\left(  x\right)  \label{euclen}%
\end{equation}
which follows immediately from $\dot{E}_{E}=\dot{x}_{cl}\left[  m\ddot{x}%
_{cl}-V^{\prime}\left(  x_{cl}\right)  \right]  $ and the equation of motion
(\ref{eomimagt}).

The essential property of these solutions in our context is that they are
saddle points of the imaginary-time path integral (\ref{pint2}). For
$\hbar\rightarrow0$ the only nonvanishing contributions to the path integral
therefore come from a neighborhood of $x_{cl}$ (since those are least
suppressed by the Boltzmann weight $\exp\left(  -S_{E}/\hbar\right)  $) and
can be calculated in saddle-point approximation.

\subsection{Saddle point approximation}

Now let us perform the saddle-point approximation explicitly. To this end we
write%
\begin{equation}
x\left(  \tau\right)  =x_{cl}\left(  \tau\right)  +\eta\left(  \tau\right)
\end{equation}
and expand the action around the stationary (resp. saddle) point $x_{cl}$
(later we will sum over the contributions from all saddle points) to second
order in $\eta$,%
\begin{align}
S_{E}\left[  x\right]   &  =S_{E}\left[  x_{cl}\right]  +\frac{1}{2}\int
d\tau\int d\tau^{\prime}\eta\left(  \tau\right)  \frac{\delta^{2}S_{E}\left[
x_{cl}\right]  }{\delta x\left(  \tau\right)  \delta x\left(  \tau^{\prime
}\right)  }\eta\left(  \tau^{\prime}\right)  +O\left(  \eta^{3}\right) \\
&  =S_{E}\left[  x_{cl}\right]  +\frac{1}{2}\int_{-T/2}^{T/2}d\tau\eta\left(
\tau\right)  \hat{F}\left(  x_{cl}\right)  \eta\left(  \tau\right)  +O\left(
\eta^{3}\right)  ,
\end{align}
(the first derivative vanishes since $S_{E}$ is minimal at $x_{cl}$) where we
have abbreviated the operator
\begin{equation}
\frac{\delta^{2}S_{E}\left[  x\right]  }{\delta x\left(  \tau\right)  \delta
x\left(  \tau^{\prime}\right)  }=\left[  -m\frac{d^{2}}{d\tau^{2}}%
+\frac{d^{2}V\left(  x_{cl}\right)  }{dx^{2}}\right]  \delta\left(  \tau
-\tau^{\prime}\right)  \equiv\hat{F}\left(  x_{cl}\right)  \delta\left(
\tau-\tau^{\prime}\right)
\end{equation}
which governs the dynamics of the fluctuations around $x_{cl}$.

We now expand the fluctuations $\eta\left(  \tau\right)  $ into the (real)
eigenfunctions of $\hat{F}$,%
\begin{equation}
\eta\left(  \tau\right)  =\sum_{n}c_{n}\tilde{x}_{n}\left(  \tau\right)  ,
\end{equation}
with
\begin{equation}
\hat{F}\left(  x_{cl}\right)  \tilde{x}_{n}\left(  \tau\right)  =\lambda
_{n}\tilde{x}_{n}\left(  \tau\right)  .
\end{equation}
The $\tilde{x}_{n}$ satisfy the boundary conditions (\ref{bcfluct}) and are
normalized according to (\ref{orthocompl}). (Since $\hat{F}$ is real-hermitean
and bounded, it has a complete spectrum.) For the moment we will also assume
that all eigenvalues are positive, $\lambda_{n}>0$. (This is not true in
general. Below we will encounter examples of vanishing eigenvalues. In the
case of so-called ``bounce'' solutions even negative eigenvalues do occur
\cite{col80}.) The action can then be written as
\begin{equation}
S_{E}\left[  x\right]  =S_{E}\left[  x_{cl}\right]  +\frac{1}{2}\sum
_{n}\lambda_{n}c_{n}^{2}+O\left(  \eta^{3}\right)  . \label{actexp}%
\end{equation}

Now we use the definition (\ref{meas}) of the measure to rewrite%
\begin{equation}
Z\left(  x_{f},x_{i}\right)  =\mathcal{N}\int D[x]e^{-\frac{S_{E}\left[
x\right]  }{\hbar}}\simeq\mathcal{N}e^{-\frac{S_{E}\left[  x_{cl}\right]
}{\hbar}}\int D[\eta]e^{-\frac{1}{2\hbar}\int_{-T/2}^{T/2}d\tau\eta\left(
\tau\right)  \hat{F}\left(  x_{cl}\right)  \eta\left(  \tau\right)  }%
\end{equation}
as
\begin{align}
Z\left(  x_{f},x_{i}\right)   &  =\mathcal{N}e^{-\frac{S_{E}\left[
x_{cl}\right]  }{\hbar}}\prod_{n}\int_{-\infty}^{\infty}\frac{dc_{n}}%
{\sqrt{2\pi\hbar}}e^{-\frac{1}{2\hbar}\sum_{n}\lambda_{n}c_{n}^{2}}\\
&  =\mathcal{N}e^{-\frac{S_{E}\left[  x_{cl}\right]  }{\hbar}}\prod_{n}%
\int_{-\infty}^{\infty}dc_{n}\frac{e^{-\frac{1}{2\hbar}\lambda_{n}c_{n}^{2}}%
}{\sqrt{2\pi\hbar}} \label{gaussinte}%
\end{align}
and obtain, after performing the Gaussian integrations (which decouple and can
thus be done independently),%
\begin{align}
Z\left(  x_{f},x_{i}\right)   &  =\mathcal{N}e^{-\frac{S_{E}\left[
x_{cl}\right]  }{\hbar}}\prod_{n}\lambda_{n}^{-1/2}\\
&  \equiv\mathcal{N}e^{-\frac{S_{E}\left[  x_{cl}\right]  }{\hbar}}\left(
\det\hat{F}\left[  x_{cl}\right]  \right)  ^{-1/2} \label{imagampl}%
\end{align}
where a sum $\sum_{x_{cl}}$ is implied if there exists more than one saddle
point.\ The formula (\ref{imagampl}) encapsulates the SCA to $Z\left(
x_{f},x_{i}\right)  $ up to $O\left(  \hbar\right)  $.

Above, we have introduced the determinant of a differential operator as the
product of its eigenvalues%
\begin{equation}
\det\hat{O}=\prod_{n}\lambda_{n},\text{ \ \ \ \ \ \ \ \ \ \ for\ \ \ \ }%
\hat{O}\psi_{n}\left(  x\right)  =\lambda_{n}\psi_{n}\left(  x\right)  ,
\end{equation}
which generalizes the standard definition for quadratic matrices. In
Subsection \ref{det} we will give a more detailed definition of such
determinants (which renders the above (infinite) product finite by adopting a
specific choice for the normalization factor $\mathcal{N}$ ) and show how they
can be calculated explicitly.

Let us summarize what we have accomplished so far. The seemingly artificial
analytical continuation to imaginary times has allowed us to identify those
paths whose neighborhoods give the dominant contributions to the path integral
for a tunneling process in the semiclassical limit, and to evaluate this path
integral to $O\left(  \hbar\right)  $ in the saddle-point approximation. In
more physical terms the situation can be described as follows: for tunneling
problems there exist no minimal-action trajectories (i.e. classical solutions)
with the appropriate boundary conditions in real time. Therefore all
trajectories between those boundary conditions (over which we sum in the
real-time path integral) interfere highly destructively. Still, their net
effect can be approximately gathered in a finite number of regions in function
space, namely those in the neighborhood of the saddle points in imaginary
time. In other words, while the tunneling amplitudes would have to be
recovered at real times from a complex mixture of non-stationary paths (a
forbidding task in practice), they are concentrated around the classical paths
in imaginary time, and are therefore accessible to the saddle-point
approximation. The destructive interference at real times leaves a conspicuous
trace, however, namely the exponential suppression of (\ref{imagampl}) due to
the Gamov factor $\exp\left(  -S_{E}/\hbar\right)  $, which is typical for
tunneling amplitudes.

\section{Double well (hill) potential and instantons\label{dwellsec}}

In the following sections we will apply the machinery developed above to
simple tunneling problems with potentials which resemble as closely as
possible the situation to be encountered later in QCD.

\subsection{Instanton solution\label{qminsoln}}

Let us therefore specialize to tunneling processes between degenerate
potential minima. A simple potential of the appropriate form is%
\begin{equation}
V\left(  x\right)  =\frac{\alpha^{2}m}{2x_{0}^{2}}\left(  x^{2}-x_{0}%
^{2}\right)  ^{2} \label{dwpot}%
\end{equation}%
\begin{figure}
[ptb]
\begin{center}
\includegraphics[
height=1.9501in,
width=3.1574in
]%
{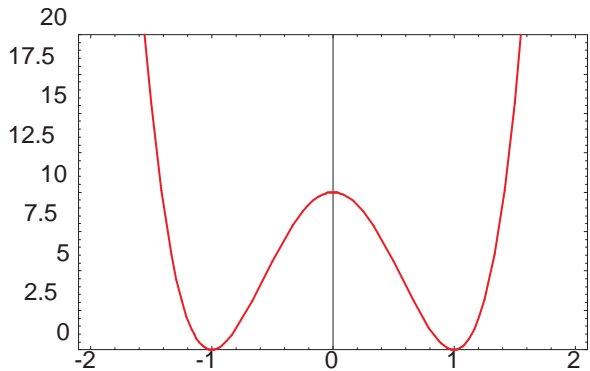}%
\caption{The double-well potential with $\frac{\alpha^{2}m}{2x_{0}^{2}}=10$
and $x_{0}=1$.}%
\label{doubwell}%
\end{center}
\end{figure}
(see Fig. \ref{doubwell}, the coefficient is chosen for later convenience)
which has three saddle points (with finite action, see the comment at the end
of this section). Two of them are trivial
\begin{align}
1)\text{ \ \ \ \ }x_{cl}\left(  \tau\right)   &  =x_{0}\\
2)\text{ \ \ \ \ }x_{cl}\left(  \tau\right)   &  =-x_{0}%
\end{align}
and do not contribute to tunneling (since they cannot satisfy the
corresponding boundary conditions; they do, however, contribute to $Z\left(
x_{0},x_{0}\right)  $ or $Z\left(  -x_{0},-x_{0}\right)  $). The third saddle
point is time-dependent and corresponds to the tunneling solution which
interpolates between both maxima\footnote{Note that, in the exact sense, this
solution exists only for $T\rightarrow\infty$ (which is all we need to extract
the ground-state properties). At large but finite $T$ the corresponding
solution requires a slight change in the boundary conditions to%
\begin{equation}
x_{cl}\left(  \pm T/2\right)  =\mp x_{0}\pm\varepsilon\left(  T\right)  ,
\end{equation}
where $\varepsilon$ is small.} of $-V$,%
\begin{align}
x_{cl}\left(  -T/2\right)   &  =+x_{0},\label{b1}\\
x_{cl}\left(  T/2\right)   &  =-x_{0}, \label{b2}%
\end{align}
or in the opposite direction, i.e. starting at $-x_{0}$ and ending at $x_{0}$.

\begin{figure}
[ptb]
\begin{center}
\includegraphics[
height=1.9501in,
width=3.1574in
]%
{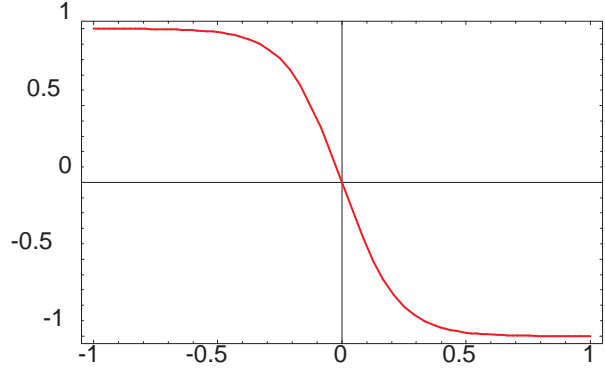}%
\caption{The instanton solution in the double well potential of Fig.
\ref{doubwell}.}%
\label{instfig}%
\end{center}
\end{figure}

The above classification of solutions is common to all potentials which look
qualitatively like (\ref{dwpot}), and all the qualitative results obtained
below will apply to such potentials, too. We have specialized to (\ref{dwpot})
only because for this choice the tunneling solution can be obtained
analytically. The easiest way to get this solution starts from the conserved
Euclidean energy (\ref{euclen}) with $E_{E}=0$, which corresponds to the limit
$T\rightarrow\infty$ we are interested in (since then the particle in the
mechanical analog system starts with (almost) zero velocity at $x_{0}$ and
thus will need (almost) infinite time to reach $-x_{0}$):
\begin{equation}
E_{E}=\frac{m}{2}\dot{x}^{2}-V=0\text{ \ \ \ \ }\Rightarrow\text{
\ \ \ \ }\dot{x}=\mp\sqrt{\frac{2V}{m}} \label{eneq}%
\end{equation}
(as usual, the positive square root is implied) or, after separation of
variables,%
\begin{equation}
\mp\sqrt{\frac{m}{2}}\frac{dx}{\sqrt{V\left(  x\right)  }}=d\tau,
\end{equation}
which can be integrated to become
\begin{equation}
\mp\sqrt{\frac{m}{2}}\int_{x_{cl}\left(  \tau_{0}\right)  }^{x_{cl}\left(
\tau\right)  }\frac{dx}{\sqrt{V\left(  x\right)  }}=\mp\frac{x_{0}}{\alpha
}\int_{x_{cl}\left(  \tau_{0}\right)  }^{x_{cl}\left(  \tau\right)  }%
\frac{dx}{x_{0}^{2}-x^{2}}=\left.  \mp\frac{x_{0}}{\alpha}\frac{1}{x_{0}%
}\arctan h\left(  \frac{x}{x_{0}}\right)  \right|  _{x_{cl}\left(  \tau
_{0}\right)  }^{x_{cl}\left(  \tau\right)  }=\int_{\tau_{0}}^{\tau}d\tau
=\tau-\tau_{0}.
\end{equation}
We now choose $\tau_{0}$, the integration constant, to be the ``center'' of
the solution in imaginary time by requiring $x_{cl}\left(  \tau_{0}\right)
=0$,\ and finally obtain
\begin{equation}
3)\text{ \ \ \ \ }x_{cl}\left(  \tau\right)  \equiv x_{I/\bar{I}}\left(
\tau\right)  =\mp x_{0}\tanh\alpha\left(  \tau-\tau_{0}\right)  . \label{inst}%
\end{equation}
The solution with the minus sign is called the \textbf{instanton} (see Fig.
\ref{instfig}) of the potential (\ref{dwpot}).%
Tunneling from $-x_{0}$ to $x_{0}$ corresponds to the anti-instanton solution
$x_{\bar{I}}\left(  \tau\right)  =-x_{I}\left(  \tau\right)  $. Obviously,
this nomenclature is mere convention. (We repeat that, from the conceptual
point of view, there is nothing special about the potential (\ref{dwpot}).
Similar potentials with degenerate minima will have similar instanton
solutions although those generally cannot be found analytically.)

Note that instantons are by necessity time-dependent since they have to
interpolate between different minima of the potential. The name ``instanton''
(which goes back to 't Hooft) indicates furthermore that the tunneling
transition happens fast, i.e. almost instantaneous\footnote{The instanton
should therefore not be confused with a soliton-like particle. There is an
important general equivalence, however, between instantons and static solitons
in the corresponding (field) theory in one additional space dimension. Our
instanton, for example, is the classical soliton solution (the so-called
``kink'') of the 1+1 dim. $\lambda\phi^{4}$ field theory.}. Indeed, from the
equation of motion (for $E_{E}=0$) for the instanton,%
\begin{equation}
\frac{dx_{I}}{d\tau}=-\sqrt{\frac{2}{m}V\left(  x_{I}\right)  },
\end{equation}
we obtain at large $\tau$ (where $V\left(  x_{I}\right)  $ can be expanded
around $-x_{0}$ with $V\left(  -x_{0}\right)  =V^{\prime}\left(
-x_{0}\right)  =0$ and $V^{\prime\prime}\left(  -x_{0}\right)  =4\alpha^{2}m$)%
\begin{equation}
\frac{dx_{I}}{d\tau}\simeq-2\alpha\left[  x_{I}-\left(  -x_{0}\right)
\right]  .
\end{equation}
Separating variables,
\begin{equation}
\frac{dx_{I}}{x_{I}+x_{0}}=d\ln\left(  x_{I}+x_{0}\right)  =-2\alpha d\tau,
\end{equation}
and integrating reveals that the asymptotic solution for the deviation
\begin{equation}
\Delta x_{I}\left(  \tau\right)  \equiv x_{I}\left(  \tau\right)  +x_{0}%
\end{equation}
of the instanton from its ``vacuum'' value $-x_{0}$ decreases exponentially
for large $\tau_{0}$:
\begin{equation}
\Delta x_{I}\left(  \tau_{0}+\tau\right)  \rightarrow\Delta x_{I}\left(
\tau_{0}\right)  e^{-2\alpha\tau}. \label{dev}%
\end{equation}

Moreover, the characteristic time scale
\begin{equation}
\tilde{\tau}=\frac{1}{2\alpha} \label{deltatau}%
\end{equation}
of the decay of $\Delta x_{I}$ becomes arbitrarily small for large $\alpha$.
In the mechanical analog system, this means that to a particle starting from
$x_{0}$ nothing much happens for a long time since its velocity remains almost
zero. However, when it finally approaches the minimum of $-V$ at $x=0$ it
takes up speed fast, rushes through the minimum, and decelerates equally fast
near to $-x_{0}$, spending all the remaining time to creep up fully and reach
$-x_{0}$ at $T\rightarrow\infty$. The ``abruptness'' of the transition
increases with $\alpha$, the coupling parameter which controlls the height of
the potential barrier.

The (Euclidean) action of the (anti-) instanton solution, which governs the
exponential suppression of the tunneling amplitude, is easily obtained with
the help of the ``first integral'' Eq. (\ref{eneq}),%
\begin{equation}
S_{I}\equiv S_{E}\left[  x_{I}\right]  =m\int_{-T/2}^{T/2}d\tau\dot{x}_{I}%
^{2}=m\int_{x_{0}}^{-x_{0}}dx_{I}\dot{x}_{I}=-\frac{\alpha m}{x_{0}}%
\int_{x_{0}}^{-x_{0}}dx_{I}\left(  x_{0}^{2}-x_{I}^{2}\right)  =\frac{4}%
{3}\alpha mx_{0}^{2}, \label{sinst}%
\end{equation}
and equal to the action of the anti-instanton. Another way of writing the
action,
\begin{equation}
S_{E}\left[  x_{I}\right]  =m\int_{x_{0}}^{-x_{0}}dx_{I}\dot{x}_{I}%
=\int_{-x_{0}}^{x_{0}}dx\sqrt{2mV\left(  x\right)  }=\int_{-x_{0}}^{x_{0}%
}dxp_{E}\left(  x\right)
\end{equation}
with $p_{E}\left(  x\right)  =\sqrt{2mV\left(  x\right)  -E_{E}}$ (recall that
$E_{E}=0$ for the instanton solution) shows that the exponential suppression
factor $\exp\left(  -S_{I}/\hbar\right)  $ is nothing but the Gamov factor
already encountered in Eq. (\ref{wkbgamov}). As expected, the path-integral
approach reproduces the analytically continued WKB result for the tunneling amplitude.

Let us add two comments. First, there exist more saddle points, i.e. solutions
of Eq. (\ref{eomimagt}), than we have discussed above. Those additional
solutions do not correspond to particle motion which both starts and ends at
maxima of $-V$, however. Hence they do not satisfy the tunneling boundary
conditions (\ref{b1},\ref{b2}) although they might involve barrier
penetration. Morevoer, they do not contribute to the SCA for the ground state
since their action become infinite for $T\rightarrow\infty$. Physically, this
is obvious since in such ``runaway'' solutions the particle moves infinitely
far from the center of the potential ($\left|  x\right|  \rightarrow\infty$)
and thereby reaches an infinite velocity at $T\rightarrow\infty$. Finally, we
note that for very low tunneling barriers (i.e. for $\alpha\rightarrow0$) and
correspondingly high tunneling rates semiclassical methods may fail. In such
cases variational approaches can sometimes help \cite{kle94}.

\subsection{Zero mode\label{zeromode}}

The constant solutions 1)\ and 2) above share the full symmetry of the
Hamiltonian $H$ (i.e. of $V$). This is not the case, however, for the
instanton solution since the latter is localized in imaginary time around
$\tau_{0}$ and therefore lacks the continuous time translation invariance of
the $\tau$-independent $H$. As a consequence, the instanton solutions form a
continuous and degenerate one-parameter family whose members are characterized
by their time-center $\tau_{0}$. In the saddle-point approximation to the path
integral we have to sum (i.e. integrate) over the contributions from all
$\tau_{0}$. In the present section we will see how this can be done.

To this end, let us go back to our SCA ``master formula'' (\ref{imagampl})
which was derived under the assumption $\lambda_{n}>0$, i.e. for positive
eigenvalues of the fluctuation operator $\hat{F}$. It is not difficult to see
that this assumption is not valid in the case of the instanton. Since the
instanton ``spontaneously'' breaks the time-translation symmetry of $V$ there
must be a zero mode in the spectrum of $\hat{F}$ (for $T\rightarrow\infty$),
i.e. an eigenfunction
\begin{equation}
\tilde{x}_{0}\left(  \tau\right)  =\sqrt{\frac{m}{S_{I}}}\dot{x}%
_{I}=-\frac{\sqrt{3\alpha}}{2}\frac{1}{\cosh^{2}\left(  \alpha\left(
\tau-\tau_{0}\right)  \right)  } \label{zm}%
\end{equation}
with the eigenvalue $\lambda_{0}=0$ (which we have normalized to one at
$T\rightarrow\infty$, where (\ref{eneq}) applies:%
\begin{equation}
\int_{-T/2}^{T/2}d\tau\left[  \tilde{x}_{0}\left(  \tau\right)  \right]
^{2}=S_{I}^{-1}\int_{-T/2}^{T/2}d\tau\left(  m\dot{x}_{I}^{2}\right)  =1).
\end{equation}
In order to verify that (\ref{zm}) is indeed a zero mode it suffices to take a
time derivative of the equation of motion (\ref{eomimagt}):
\begin{equation}
m\ddot{x}_{I}-V^{\prime}\left(  x_{I}\right)  =0\text{ \ \ }\Rightarrow
\text{\ }\left[  m\partial_{\tau}^{2}-V^{^{\prime\prime}}\left(  x_{I}\right)
\right]  \dot{x}_{I}=-\hat{F}\dot{x}_{I}=0\text{ \ \ }\Rightarrow\text{ }%
\hat{F}\tilde{x}_{0}\left(  \tau\right)  =0.
\end{equation}

The physical origin of this zero mode is quite obvious: it corresponds to an
infinitesimal shift of the instanton center $\tau_{0}$, i.e. an infinitesimal
time translation of the instanton solution. Since the resulting, shifted
instanton is degenerate with the original one, such a fluctuation costs no
action and (\ref{actexp}) implies that the corresponding eigenvalue of
$\hat{F}$, i.e. $\lambda_{0}$, must be zero\footnote{It might be useful to
anticipate already at this stage that zero modes of fluctuatation operators,
especially in quantum field theory, are not always due to continuous
symmetries of the underlying dynamics. We will encounter an intruiguing
example of a topologically induced zero mode below when dealing with
Yang-Mills instantons.}.

The instanton solution (\ref{inst}) is monotonically decreasing, which implies
that the zero mode satisfies $\tilde{x}_{0}\left(  \tau\right)  \leq0$ (cf.
Eq. (\ref{zm})) and therefore has no node. As a consequence, the zero mode is
the (unique)\ eigenfunction of $\hat{F}$ with the lowest eigenvalue. All the
remaining modes have $\lambda_{n}>0$ and for them the Gaussian integrals in
Eq. (\ref{gaussinte}) are well defined. This is not the case, however, for the
$c_{0}$ integration
\begin{equation}
\int_{-\infty}^{\infty}\frac{dc_{0}e^{-\frac{1}{2\hbar}\lambda_{0}c_{0}^{2}}%
}{\sqrt{2\pi\hbar}}=\int_{-\infty}^{\infty}\frac{dc_{0}}{\sqrt{2\pi\hbar}},
\end{equation}
which is not Gaussian at all! And neither should it be, since $\tau_{0}%
$-shifting fluctuations $\eta\left(  \tau\right)  \sim\tilde{x}_{0}\left(
\tau\right)  $ are not damped even when they are large (i.e. for large $c_{0}%
$), due to the $\tau_{0}$-independence of the instanton action. The latter
renders the integrand $c_{0}$-independent and thus the integral divergent.

But this is just the kind of divergence which we should expect anyway from
integrating over the infinite set of saddle points, i.e. over all $\tau_{0}$,
as discussed above. Indeed, it is easy to show that
\begin{equation}
dc_{0}\propto d\tau_{0}%
\end{equation}
by comparing the deviations $dx$ from a given path $x\left(  \tau\right)  $
which are caused by
\begin{align}
\text{small time translations }\tau_{0}  &  \rightarrow\tau_{0}+d\tau
_{0}\text{ \ \ \ }\Rightarrow\text{ \ \ \ }dx=\frac{dx_{I}}{d\tau_{0}}%
d\tau_{0}=-\dot{x}_{I}d\tau_{0},\text{ \ \ \ \ \ \ and}\\
\text{small coefficient shifts }c_{0}  &  \rightarrow c_{0}+dc_{0}\text{
\ \ \ }\Rightarrow\text{ \ \ \ }dx=\frac{dx}{dc_{0}}dc_{0}=\tilde{x}_{0}%
dc_{0}=\sqrt{\frac{m}{S_{I}}}\dot{x}_{I}dc_{0}.
\end{align}
Equating both deviations (and redefining the sign of $dc_{0}$ such that the
integrals over $c_{0}$ and $\tau_{0}$ have the same limits) gives%
\begin{equation}
dc_{0}=\sqrt{\frac{S_{I}}{m}}d\tau_{0}.
\end{equation}

Thus the integration over $c_{0}$ can be done exactly and (in the limit
$T\rightarrow\infty$) just amounts to summing over all one-instanton saddle
points\footnote{Since $c_{0}$ measures how far the instanton is collectively
(i.e. equally at all times) shifted in (imaginary) time, it is an example of a
``collective coordinate''.}. In the presence of a zero mode the expression
(\ref{imagampl}) must therefore be replaced by
\begin{equation}
Z_{I}\left(  -x_{0},x_{0}\right)  =\mathcal{N}e^{-\frac{S_{E}\left[
x_{I}\right]  }{\hbar}}\sqrt{\frac{S_{E}\left[  x_{I}\right]  }{2\pi\hbar m}%
}T\left(  \det\hat{F}\left[  x_{I}\right]  ^{\prime}\right)  ^{-1/2}
\label{zoneinst}%
\end{equation}
where the prime at the determinant indicates that $\lambda_{0}$ is excluded
from the product of eigenvalues. Of course, the factor $T$ in (\ref{zoneinst})
becomes infinite in the limit $T\rightarrow\infty$ which we will take in the
end. This infinity is a consequence of the infinite amount of contributing
saddle points and will be cancelled by other infinities (see e.g. Eq.
(\ref{grsteng})), leaving the observables perfectly finite as it should be.

\section{Fluctuation determinant\label{det}}

Our next task is to make sense of functional determinants as encountered
above, and to calculate%
\begin{equation}
\det\hat{F}\left[  x_{I}\right]  ^{\prime}=\det\left[  -m\frac{d^{2}}%
{d\tau^{2}}+V^{\prime\prime}\left(  x_{cl}\right)  \right]  ^{\prime}
\label{det1}%
\end{equation}
from the leading fluctuations around the instanton (with the zero-mode
contributions removed) explicitly. Together with the zero-mode part, this will
take care of the $O\left(  \hbar\right)  $ contributions to the SCA. Although
the explicit calculation of (\ref{det1}) is a rather technical exercise, we
will go through it in considerable detail. One of the reasons is to develop
intuition for the calculation of determinants in quantum field theory, as they
generally occur during quantization of extended classical solutions (besides
instantons e.g. solitons, monopoles, D-branes etc.). Indeed, all typical
features of the calculation generalize rather directly to QCD. Hence it can
serve as a pedagogical substitute for the actual QCD calculation, which is a
tour de force \cite{tho76} beyond the scope of our introduction to instanton physics.

Despite their uses, the following two subsections lie somewhat outside of our
main focus. Readers not interested in the details of such calculations may
therefore jump ahead to Section \ref{digasec} without compromising their
understanding of the remainder of these lectures.

\subsection{Warm-up: harmonic oscillator}

To prepare for the calculation of the instanton determinant, we first consider
a simpler problem. We look for a nontrivial potential $V$ which nevertheless
leads to a fluctuation operator $\hat{F}$ whose eigenvalues can be obtained
easily and analytically. The most obvious choice is to require $V^{\prime
\prime}$ to be nonzero but $x$-independent, i.e. to work with the potential of
the harmonic oscillator%
\begin{equation}
V_{ho}\left(  x\right)  =\frac{1}{2}m\omega^{2}x^{2}.
\end{equation}
The Euclidean analog potential is an inverted parabola. Of course, there is no
tunneling in this potential. Instead of the tunneling boundary
conditions\ (\ref{tunbc}) we therefore choose
\begin{equation}
x_{cl}\left(  \pm T/2\right)  =0. \label{hobc}%
\end{equation}
The only solution to the equation of motion (\ref{eomimagt}) under the
conditions (\ref{hobc}) is the constant%
\begin{equation}
x_{ho}\left(  \tau\right)  =0, \label{hosoln}%
\end{equation}
corresponding to a particle which stays in ``metastable'' equilibrium at $x=0$
forever. Not surprisingly, its Euclidean action $S_{E,ho}$ vanishes and hence
the exponential suppression factor $\exp\left(  -S_{E}/\hbar\right)  $ of the
tunneling amplitudes is absent, as it should be.

To $O\left(  \hbar\right)  $ the Euclidean time evolution matrix element of
the oscillator becomes%
\begin{equation}
Z_{ho}\left(  0,0\right)  =\left\langle 0\left|  e^{-H_{ho}T/\hbar}\right|
0\right\rangle =\mathcal{N}\left(  \det\hat{F}\left[  x_{ho}\right]  \right)
^{-1/2} \label{zho1}%
\end{equation}
where we have redefined $\hat{F}$ by absorbing a constant factor $\left(  \det
m\right)  ^{-1/2}$ into $\mathcal{N}$:%
\begin{equation}
\hat{F}\left[  x_{ho}\right]  =-\frac{d^{2}}{d\tau^{2}}+\omega^{2}.
\end{equation}
Since the classical solution (\ref{hosoln}) is time-translation invariant,
there is no zero mode and the prime on the determinant can be omitted.

We now apply the formal definition of the determinant of an operator as the
product of its eigenvalues,%
\begin{equation}
\det\hat{F}\left[  x_{ho}\right]  =\prod_{n}\lambda_{ho,n},
\end{equation}
with the $\lambda_{ho,n}$ determined by%

\begin{equation}
\hat{F}\left[  x_{ho}\right]  \tilde{x}_{n}\left(  \tau\right)  =\lambda
_{ho,n}\tilde{x}_{n}\left(  \tau\right)  .
\end{equation}
As intended, the eigenfunctions of $\hat{F}\left[  x_{ho}\right]  $,
\begin{equation}
\tilde{x}_{n}\left(  \tau\right)  =a_{n}\sin\left(  \frac{n\pi\tau}{T}\right)
+b_{n}\cos\left(  \frac{n\pi\tau}{T}\right)  ,
\end{equation}
and the corresponding eigenvalues
\begin{equation}
\lambda_{ho,n}=\left(  \frac{n\pi}{T}\right)  ^{2}+\omega^{2},\text{
\ \ \ \ \ \ \ \ \ \ \ \ }n=1,2,... \label{hoevs}%
\end{equation}
can be read off immediately. The coefficients $a_{n}$, $b_{n}$ are fixed by
the normalization (\ref{orthocompl}) and boundary conditions (\ref{bcfluct}).
In order for $\tilde{x}_{n}$ to satisfy the latter, we have to require
$a_{n}=0$ for $n$ odd, and $b_{n}=0$ for $n$ even.

We then formally have%
\begin{equation}
Z_{ho}\left(  0,0\right)  =\mathcal{N}\left(  \prod_{n=1}^{\infty}\left[
\left(  \frac{n\pi}{T}\right)  ^{2}+\omega^{2}\right]  \right)  ^{-1/2}%
\end{equation}
and since the product is infinite, $\mathcal{N}$ must be infinite, too, in
order to prevent $Z_{ho}$ from vanishing identically. One way of disentangling
the two infinities, or in other words to define the determinant, is to relate
it to the determinant of the fluctuation operator without potential
($\omega=0$): we simply split the above expression as%
\begin{equation}
Z_{ho}\left(  0,0\right)  =Z_{free}\left(  0,0\right)  \left(  \prod
_{n=1}^{\infty}\left[  1+\left(  \frac{\omega T}{n\pi}\right)  ^{2}\right]
\right)  ^{-1/2}, \label{zho2}%
\end{equation}
where the first factor corresponds to the propagator of free motion%
\begin{equation}
Z_{free}\left(  0,0\right)  =\mathcal{N}\left[  \prod_{n=1}^{\infty}\left(
\frac{n\pi}{T}\right)  ^{2}\right]  ^{-1/2}=\left\langle 0\left|  e^{-\left(
\frac{\hat{p}^{2}}{2m}\right)  \frac{T}{\hbar}}\right|  0\right\rangle
=\int_{-\infty}^{\infty}\frac{dp}{2\pi}e^{-\left(  \frac{p^{2}}{2m}\right)
\frac{T}{\hbar}}=\sqrt{\frac{m\hbar}{2\pi T}} \label{zfree}%
\end{equation}
and the forelast equation above is obtained by inserting the momentum
eigenstates $\left\langle x|p\right\rangle =\exp\left(  -ixp/\hbar\right)
/\sqrt{2\pi}$.

Note that (\ref{zfree}) amounts to a calculation of $\mathcal{N}$ from the
requirement that it renders the free propagator finite and equal to the
familiar result from basic quantum mechanics. Note, furthermore, that
$\mathcal{N}$ is a property of the functional integral measure (\ref{meas})
and does not depend on the dynamics specified in $S_{E}\left[  x\right]  $. We
will therefore encounter the same factor below when we return to the instanton problem.

The second, finite factor in (\ref{zho2}) is obtained from the standard
formula
\begin{equation}
\prod_{n=1}^{\infty}\left(  1+\frac{c^{2}}{n^{2}}\right)  =\frac{\sinh\left(
\pi c\right)  }{\pi c}%
\end{equation}
which can be found e.g. in \cite{gra81} or by Mathematica. Putting everything
together, we finally have
\begin{equation}
Z_{ho}\left(  0,0\right)  =\mathcal{N}\left(  \det\hat{F}\left[
x_{ho}\right]  \right)  ^{-1/2}=\sqrt{\frac{m\hbar\omega}{2\pi}}\left[
\sinh\left(  \omega T\right)  \right]  ^{-1/2}.
\end{equation}

Actually, this is the full Euclidean propagator of the harmonic oscillator:
since the corresponding path integral is Gaussian, the SCA becomes exact!
Below, we will be interested in the large-$T$ limit
\begin{equation}
Z_{ho}\left(  0,0\right)  \rightarrow\sqrt{\frac{m\hbar\omega}{2\pi}}\left[
\frac{\exp\left(  \omega T\right)  }{2}\right]  ^{-1/2}=\sqrt{\frac{m\hbar
\omega}{\pi}}e^{-\omega T/2} \label{limitzho}%
\end{equation}
from which we recover, via Eq. (\ref{grsteng}), the familiar ground-state
energy
\begin{equation}
E_{0,ho}=\frac{1}{2}\hbar\omega.
\end{equation}

\subsection{Instanton determinant}

After having calculated two simpler functional determinants (those for the
free particle and the harmonic oscillator), we are now prepared to attack the
original problem. Our task will be to calculate the fluctuation determinant
which appears in the $O\left(  \hbar\right)  $ tunneling amplitude
\begin{equation}
Z_{I}\left(  -x_{0},x_{0}\right)  =Z_{ho}\left(  0,0\right)  e^{-\frac{S_{E}%
\left[  x_{I}\right]  }{\hbar}}\sqrt{\frac{S_{E}\left[  x_{I}\right]  }%
{2\pi\hbar m}}\omega T\left\{  \frac{\det\hat{F}\left[  x_{I}\right]
^{\prime}}{\omega^{-2}\det\hat{F}\left[  x_{ho}\right]  }\right\}
^{-1/2}\label{Z-O(hbar)}%
\end{equation}
around an instanton. Above, we have factored out the root of the
harmonic-oscillator determinant in (\ref{zoneinst}) and used Eq. (\ref{zho1})
for$\ Z_{ho}\left(  0,0\right)  $ to eliminate $\mathcal{N}.$ This
``renormalizes'' $Z_{I}\left(  -x_{0},x_{0}\right)  $ according to the
procedure of the preceding section and leaves us with the above, finite ratio
of determinants. Furthermore, we have factored out the lowest-mode
contribution to the harmonic-oscillator determinant, $\lambda_{ho,0}%
\rightarrow\omega^{2}$ for $T\rightarrow\infty$, to balance the removal of the
zero mode from the determinant of $\hat{F}\left[  x_{I}\right]  $.

In order to calculate $\det\hat{F}\left[  x_{I}\right]  ^{\prime}$, we have to
deal with the spectrum of the operator
\begin{equation}
\hat{F}\left[  x_{I}\right]  =-m\frac{d^{2}}{d\tau^{2}}+V^{\prime\prime
}\left(  x_{I}\right)
\end{equation}
where
\begin{equation}
V^{\prime\prime}\left(  x_{I}\right)  =\frac{2\alpha^{2}m}{x_{0}^{2}}\left(
3x_{I}^{2}-x_{0}^{2}\right)  =2\alpha^{2}m\left[  3\tanh^{2}\alpha\left(
\tau-\tau_{0}\right)  -1\right]  =2\alpha^{2}m\left[  2-\frac{3}{\cosh
^{2}\alpha\left(  \tau-\tau_{0}\right)  }\right]
\end{equation}
($\cosh^{2}x-\sinh^{2}x=1$) describes the interaction of Gaussian fluctuations
with the smooth, step-like potential generated by the instanton. The
eigenvalue equation of $\hat{F}\left[  x_{I}\right]  $ becomes, after
absorbing the factor $\left(  \det m\right)  ^{-1/2}$ as above into the
normalization and specializing to $\tau_{0}=0$,
\begin{equation}
\left[  -\frac{d^{2}}{d\tau^{2}}+4\alpha^{2}-\frac{6\alpha^{2}}{\cosh
^{2}\left(  \alpha\tau\right)  }\right]  \tilde{x}_{\lambda}\left(
\tau\right)  =\lambda\tilde{x}_{\lambda}\left(  \tau\right)  , \label{eveq}%
\end{equation}
subject to the tunneling boundary conditions%
\begin{equation}
\tilde{x}_{\lambda}\left(  \pm T/2\right)  =0 \label{eveqbcs}%
\end{equation}
for $T\rightarrow\infty$ (cf. (\ref{bcfluct})). As usual, the eigenvalue
spectrum will be discrete for the ``bound'' states with $\lambda<4\alpha^{2}$
and (in the limit $T\rightarrow\infty)$ continuous for the scattering states
where $\lambda>4\alpha^{2}$.

Eq. (\ref{eveq}) is of Schr\"{o}dinger type and can be solved analytically in
terms of hypergeometric functions (cf. \S25, Problem 4 in \cite{lan74}, which
makes use of the solution of \S23, Problem 5 in the same book). To obtain the
solutions explicitly, we rewrite (\ref{eveq}) by means of the standard
substitution
\begin{equation}
\xi=\tanh\left(  \alpha\tau\right)  ,\text{ \ \ \ \ \ }\frac{d}{d\tau
}=\frac{\alpha}{\cosh^{2}\left(  \alpha\tau\right)  }\frac{d}{d\xi}%
=\alpha\left(  1-\xi^{2}\right)  \frac{d}{d\xi},\text{ \ \ \ \ \ }\frac{d^{2}%
}{d\tau^{2}}=\alpha^{2}\left(  1-\xi^{2}\right)  \frac{d}{d\xi}\left(
1-\xi^{2}\right)  \frac{d}{d\xi}%
\end{equation}
in the form
\begin{equation}
\left[  \frac{d}{d\xi}\left(  1-\xi^{2}\right)  \frac{d}{d\xi}-\frac{\epsilon
^{2}}{1-\xi^{2}}+6\right]  \tilde{x}_{\lambda}\left(  \xi\right)  =0,
\label{legeq}%
\end{equation}
where
\begin{equation}
\epsilon^{2}=4-\frac{\lambda}{\alpha^{2}}.
\end{equation}
The solutions of (\ref{legeq}) in the interval of interest, $\xi\in\left[
-1,1\right]  $, are associated Legendre functions (see, e.g. \cite{gra81},
chapter 8.7):
\begin{equation}
\tilde{x}_{\lambda}\left(  \xi\right)  =c_{1}P_{2}^{\epsilon}\left(
\xi\right)  +c_{2}Q_{2}^{\epsilon}\left(  \xi\right)  .
\end{equation}

Let us first consider the case $\varepsilon^{2}>0$, corresponding to the
discrete levels. The first of the boundary conditions (\ref{eveqbcs}), i.e.
\begin{equation}
\tilde{x}_{\lambda}\left(  \xi=-1\right)  =0,
\end{equation}
requires $c_{2}=0$. The second one,
\begin{equation}
\tilde{x}_{\lambda}\left(  \xi=1\right)  =0,
\end{equation}
can be nontrivially satisfied only for $\epsilon=1,2$ (for which the
$P_{2}^{\epsilon}\left(  \xi\right)  $ reduce to a polynomials). Thus we have
two discrete levels, with the eigenvalues
\begin{align}
\lambda_{0}  &  =0,\\
\lambda_{1}  &  =3\alpha^{2}.
\end{align}
As anticipated, the lowest level $\lambda_{0}$ is the zero mode which we have
already found in Section \ref{zeromode}. The corresponding (not normalized)
eigenfunction $P_{2}^{2}\left(  \xi\right)  =3\left(  1-\xi^{2}\right)  $
$=3/\cosh^{2}\left(  \alpha\tau\right)  $ is proportional to (\ref{zm}) and,
inserted into (\ref{eveq}), yields indeed the eigenvalue zero. Both discrete
eigenvalues are nonnegative, as expected in a repulsive potential.
Nevertheless, the modified potential corresponding to the eigenvalues
$\epsilon^{2}$ has two ``bound states'', i.e. the discrete levels found above.

The calculation of the contributions from the continuum states with
$\varepsilon^{2}<0$ is more involved but provides an instructive example of
how to obtain a functional determinant in terms of phase shifts. Since the
associated Legendre functions are not very convenient to deal with for
non-integer $\epsilon$, we rewrite $\tilde{x}_{\lambda}\left(  \xi\right)
=\left(  1-\xi^{2}\right)  ^{\epsilon/2}w\left(  \xi\right)  $, which
transforms Eq. (\ref{legeq}) into%
\begin{equation}
\left[  \frac{d}{d\xi}\left(  1-\xi^{2}\right)  \frac{d}{d\xi}\left(
1-\xi^{2}\right)  ^{\epsilon/2}-\epsilon^{2}\left(  1-\xi^{2}\right)
^{\epsilon/2-1}+6\left(  1-\xi^{2}\right)  ^{\epsilon/2}\right]  w\left(
\xi\right)  =0
\end{equation}
or%
\begin{equation}
\left[  \left(  1-\xi^{2}\right)  \frac{d^{2}}{d\xi^{2}}-2\left(
\epsilon+1\right)  \xi\frac{d}{d\xi}-\left(  \epsilon-2\right)  \left(
\epsilon+3\right)  \right]  w\left(  \xi\right)  =0.
\end{equation}
In terms of the new variable $u$ with
\begin{equation}
u=\frac{1-\xi}{2},\text{ \ \ }\xi=1-2u,\text{ \ \ }1-\xi^{2}=4u\left(
1-u\right)  ,\text{ \ \ }\frac{dw}{du}=-2\frac{dw}{d\xi},\text{ \ \ }%
\frac{d^{2}w}{du^{2}}=4\frac{d^{2}w}{d\xi^{2}}%
\end{equation}
this becomes the hypergeometric differential equation%
\begin{equation}
\left[  u\left(  1-u\right)  \frac{d^{2}}{du^{2}}+\left(  \epsilon+1\right)
\left(  1-2u\right)  \frac{d}{du}-\left(  \epsilon-s\right)  \left(
\epsilon+s+1\right)  \right]  w\left(  u\right)  =0
\end{equation}
with $s=2$. Its solutions for positve $\epsilon^{2}$ agree with those given
above. The general solution for $\varepsilon^{2}<0$ reads
\begin{equation}
w\left(  u\right)  =c_{3}\,_{2}F_{1}\left[  \frac{ik}{\alpha}-s,s+\frac{ik}%
{\alpha}+1,1+\frac{ik}{\alpha},u\right]  +c_{4}u^{-ik/\alpha}\,_{2}%
F_{1}\left[  -s,s+1,1-\frac{ik}{\alpha},u\right]  \label{soln1}%
\end{equation}
(see, e.g. chapter 9.1 of \cite{gra81}) where we have defined $\epsilon
=ik/\alpha$. The opposite sign, $k\rightarrow-k$, interchanges incoming and
outgoing solutions (see below). For our case of $s=2$ the solution
(\ref{soln1}) can be expressed in terms of elementary functions,
\begin{align}
w\left(  u\right)   &  =c_{3}\left(  1-u\right)  ^{-ik/\alpha}\left[
k^{2}-3ik\alpha\left(  1-2u\right)  -2\alpha^{2}\left(  1-6\left(  1-u\right)
\right)  \right] \\
&  +c_{4}u^{-ik/\alpha}\left[  1+\frac{6u\alpha\left(  -ik+2\alpha\left(
1-u\right)  \right)  }{\left(  k+i\alpha\right)  \left(  k+2i\alpha\right)
}\right]
\end{align}
or, transforming back to $\xi=1-2u$,
\begin{align}
w\left(  \xi\right)   &  =c_{3}\left(  \frac{1+\xi}{2}\right)  ^{-ik/\alpha
}\left[  k^{2}-3ik\alpha\xi+2\alpha^{2}\left(  2+3\xi\right)  \right] \\
&  +c_{4}\left(  \frac{1-\xi}{2}\right)  ^{-ik/\alpha}\left[  1+\frac{3\alpha
\left(  1-\xi\right)  \left(  -ik+\alpha\left(  1+\xi\right)  \right)
}{\left(  k+i\alpha\right)  \left(  k+2i\alpha\right)  }\right]
\end{align}
and therefore%
\begin{align}
\tilde{x}_{\lambda}\left(  \xi\right)   &  =\left(  1-\xi^{2}\right)
^{ik/\left(  2\alpha\right)  }w\left(  \xi\right) \\
&  =c_{3}\left(  \frac{\sqrt{1+\xi}}{2\sqrt{1-\xi}}\right)  ^{-ik/\alpha
}\left[  k^{2}-3ik\alpha\xi+2\alpha^{2}\left(  2+3\xi\right)  \right] \\
&  +c_{4}\left(  \frac{\sqrt{1-\xi}}{2\sqrt{1+\xi}}\right)  ^{-ik/\alpha
}\left[  1+\frac{3\alpha\left(  1-\xi\right)  \left(  -ik+\alpha\left(
1+\xi\right)  \right)  }{\left(  k+i\alpha\right)  \left(  k+2i\alpha\right)
}\right]  .
\end{align}
To finally restore the original $\tau$-dependence, resubstitute $\xi
=\tanh\left(  \alpha\tau\right)  $ and use%
\begin{equation}
\left(  \frac{\sqrt{1\pm\xi}}{2\sqrt{1\mp\xi}}\right)  ^{-ik/\alpha
}=e^{-ik/\alpha\ln\left(  \frac{\sqrt{1\pm\xi}}{2\sqrt{1\mp\xi}}\right)  }%
\end{equation}
together with%
\begin{equation}
\ln\left(  \frac{\sqrt{1\pm\xi}}{2\sqrt{1\mp\xi}}\right)  =\ln\left(
\frac{1}{2}\sqrt{\frac{\cosh\left(  \alpha\tau\right)  \pm\sinh\left(
\alpha\tau\right)  }{\cosh\left(  \alpha\tau\right)  \mp\sinh\left(
\alpha\tau\right)  }}\right)  =\ln\left(  \frac{e^{\pm\alpha\tau}}{2}\right)
=\pm\alpha\tau-\ln2\text{.}%
\end{equation}
As a result, we can write the full set of solutions with $\varepsilon^{2}<0$
as%
\begin{align}
\tilde{x}_{\lambda}\left(  \tau\right)   &  =\tilde{c}_{3}e^{-ik\tau}\left[
k^{2}-3ik\alpha\tanh\left(  \alpha\tau\right)  +2\alpha^{2}\left(
2+3\tanh\left(  \alpha\tau\right)  \right)  \right] \nonumber\\
&  +\tilde{c}_{4}e^{ik\tau}\left[  1+\frac{3\alpha\left(  1-\tanh\left(
\alpha\tau\right)  \right)  \left(  -ik+\alpha\left(  1+\tanh\left(
\alpha\tau\right)  \right)  \right)  }{\left(  k+i\alpha\right)  \left(
k+2i\alpha\right)  }\right]  . \label{x(tau)}%
\end{align}
(The constants $\tilde{c}_{3}$ and $\tilde{c}_{4}$ are left arbitrary since we
do not need to fix the normalization of the $\tilde{x}_{\lambda}\left(
\tau\right)  $ or to impose the boundary conditions (\ref{eveqbcs}) for our
purposes below.)

In order to calculate the continuum-mode contributions to the determinant, we
have to find and multiply the eigenvalues corresponding to the solutions
(\ref{x(tau)}). Fortunately, there exists an elegant and efficient technique
\cite{vai99} for doing this, which can be applied to functional determinants
in field theory as well. It takes advantage of the fact that (\ref{eveq}) is a
local equation. Thus we can obtain the eigenvalues in the asymptotic region
$\left|  \tau\right|  \rightarrow\infty$, where the potential induced by the
instanton field vanishes (a benefit of expanding around a localized solution)
and (\ref{eveq}) simplifies to%
\begin{equation}
\left[  \frac{d^{2}}{d\tau^{2}}+k^{2}\right]  \tilde{x}_{\lambda}\left(
\tau\right)  =0. \label{asympeq}%
\end{equation}
Here the momentum $k$, introduced above, replaces $\lambda$ as the label of
the continuum modes. Both are related by the ``dispersion relation''
\begin{equation}
k^{2}\equiv\lambda-4\alpha^{2}\text{ \ \ \ \ \ \ }\left(  \geq0\right)  .
\end{equation}
The solutions of Eq. (\ref{asympeq}) are ``plane waves''. Since their
normalization is fixed (elastic scattering), the only effect of the
instanton-induced potential can be a $k$-dependent phase shift. Moreover, the
explicit, asymptotic solution reveals that no reflection occurs in this
potential. Thus we have%
\begin{align}
\tilde{x}_{\lambda}\left(  \tau\right)   &  \propto e^{ik\tau+i\delta_{k}%
}\text{ \ \ \ \ \ \ \ \ for }\tau\rightarrow-\infty,\label{phshdef}\\
\tilde{x}_{\lambda}\left(  \tau\right)   &  \propto e^{ik\tau}\text{
\ \ \ \ \ \ \ \ \ \ \ \ for }\tau\rightarrow+\infty.
\end{align}
(Due to the absence of reflection, the linearly independent solutions
$\propto\exp\left(  -ik\tau\right)  $ do not mix in.) The above discussion
implies that all the required dynamical information about the eigenvalues is
contained in the phase shifts of the solutions (\ref{x(tau)}) in the limit
$\tau\rightarrow-\infty$, where $\tanh\left(  \alpha\tau\right)
\rightarrow-1$ and thus
\begin{equation}
\tilde{x}_{\lambda}\left(  \tau\right)  \rightarrow\tilde{c}_{3}e^{-ik\tau
}\left[  k^{2}+3ik\alpha-2\alpha^{2}\right]  +\tilde{c}_{4}e^{ik\tau}\left[
\frac{\left(  1+ik/\alpha\right)  \left(  2+ik/\alpha\right)  }{\left(
1-ik/\alpha\right)  \left(  2-ik/\alpha\right)  }\right]  . \label{xasympt}%
\end{equation}
By comparing (\ref{phshdef}) and (\ref{xasympt}) we read off the phase shifts%
\begin{equation}
\delta_{k}=-i\ln\left[  \left(  \frac{1+ik/\alpha}{1-ik/\alpha}\right)
\left(  \frac{2+ik/\alpha}{2-ik/\alpha}\right)  \right]  . \label{phshift}%
\end{equation}

How are these phase shifts related to the continuum spectrum the eigenvalues
$k$, and thus to the determinant to be calculated? Since the phase shifts
encode the behavior of the scattering solutions at large $\left|  \tau\right|
$, one would expect the boundary conditions (\ref{eveqbcs}) to play a role.
And indeed, they are all what is needed to establish the relation between the
eigenvalues and $\delta_{k}$. To show this, we start from the general solution%
\begin{equation}
\tilde{x}_{gen,\lambda}\left(  \tau\right)  =A\tilde{x}_{\lambda}\left(
\tau\right)  +B\tilde{x}_{\lambda}\left(  -\tau\right)
\end{equation}
and impose the boundary conditions (\ref{eveqbcs}) to obtain
\begin{equation}
A\tilde{x}_{\lambda}\left(  \frac{T}{2}\right)  +B\tilde{x}_{\lambda}\left(
-\frac{T}{2}\right)  =A\tilde{x}_{\lambda}\left(  -\frac{T}{2}\right)
+B\tilde{x}_{\lambda}\left(  -\frac{T}{2}\right)  =0.
\end{equation}
This implies
\begin{equation}
\frac{\tilde{x}_{\lambda}\left(  -\frac{T}{2}\right)  }{\tilde{x}_{\lambda
}\left(  \frac{T}{2}\right)  }=e^{-ikT-i\delta_{k}}=\pm1
\end{equation}
and has the solutions%
\begin{equation}
k_{n}=\frac{n\pi-\delta_{k}}{T}%
\end{equation}
in terms of the phase shifts $\delta_{k}$. Due to the boundary conditions the
$k_{n}$ are discrete for finite $T$ and become continuous in the limit
$T\rightarrow\infty$\ to be taken at the end.

It remains to calculate
\begin{equation}
\frac{\det\hat{F}\left[  x_{I}\right]  ^{\prime}}{\omega^{-2}\det\hat
{F}\left[  x_{ho}\right]  }=\frac{\lambda_{1}}{\lambda_{ho,2}}\frac{\Pi
_{n=1}\left(  k_{n}^{2}+4\alpha^{2}\right)  }{\Pi_{n=3}\left(  k_{ho,n}%
^{2}+\omega^{2}\right)  }=\frac{3}{4}\frac{\Pi_{n=1}\left(  k_{n}^{2}%
+4\alpha^{2}\right)  }{\Pi_{n=3}\left(  k_{ho,n}^{2}+4\alpha^{2}\right)  },
\label{prods}%
\end{equation}
where we have factored out the contribution of the second levels (for
$T\rightarrow\infty$)
\begin{equation}
\frac{\lambda_{1}}{\lambda_{ho,2}}=\frac{3\alpha^{2}}{\omega^{2}}=\frac{3}{4}%
\end{equation}
to the determinant ratio, using $k_{ho,n}=n\pi/T$ (cf. Eq. (\ref{hoevs})), and
specialized to $\omega=2\alpha$. Since the relevant range of $k$-values in the
eigenvalue products of (\ref{prods}) will turn out not to contain small $k$
(see below), the contribution of the two lowest harmonic-oscillator
eigenvalues can be multiplied to the above denominator with negligible effect
(except for simplifying the ensuing expressions). We are thus led to consider
the ratio%
\begin{equation}
\frac{\Pi_{n=1}\left(  k_{n}^{2}+4\alpha^{2}\right)  }{\Pi_{n=1}\left(
k_{ho,n}^{2}+4\alpha^{2}\right)  }=\exp\sum_{n=1}^{\infty}\ln\left[
\frac{k_{n}^{2}+4\alpha^{2}}{k_{ho,n}^{2}+4\alpha^{2}}\right]  .
\end{equation}
With
\begin{equation}
k_{n}^{2}=\left(  k_{ho,n}-\frac{\delta_{k}}{T}\right)  ^{2}\simeq
k_{ho,n}^{2}-\frac{2\delta_{k}k_{ho,n}}{T}%
\end{equation}
(since $\delta_{k}/T\ll1$ in the $n$-region which contributes non-negligibly
for $T\rightarrow\infty$) we have%
\begin{equation}
\ln\left[  \frac{k_{n}^{2}+4\alpha^{2}}{k_{ho,n}^{2}+4\alpha^{2}}\right]
\simeq\ln\left[  1-\frac{1}{T}\frac{2\delta_{k}k_{ho,n}}{k_{ho,n}^{2}%
+4\alpha^{2}}\right]  \simeq-\frac{1}{T}\frac{2\delta_{k}k_{ho,n}}%
{k_{ho,n}^{2}+4\alpha^{2}}.
\end{equation}
In the continuum limit
\begin{equation}
\sum_{n=1}^{\infty}f\left(  k_{n}\right)  =\frac{1}{\Delta k}\sum
_{n=1}^{\infty}\Delta kf\left(  k_{n}\right)  \rightarrow\frac{T}{\pi}\int
_{0}^{\infty}dkf\left(  k\right)
\end{equation}
($\Delta k=k_{ho,n+1}-k_{ho,n}=\pi/T$) we then obtain
\begin{equation}
\frac{\Pi_{n=1}\left(  k_{n}^{2}+4\alpha^{2}\right)  }{\Pi_{n=1}\left(
k_{ho,n}^{2}+4\alpha^{2}\right)  }=\exp\left(  -\frac{1}{\pi}\int_{0}^{\infty
}dk\frac{2\delta_{k}k}{k^{2}+4\alpha^{2}}\right)  .
\end{equation}

To evaluate the integral in the exponent, we note that
\begin{equation}
\frac{d}{dk}\ln\left[  1+\frac{k^{2}}{4\alpha^{2}}\right]  =\frac{2k}%
{k^{2}+4\alpha^{2}}%
\end{equation}
allows us to rewrite%
\begin{equation}
\int_{0}^{\infty}dk\frac{2k\delta_{k}}{k^{2}+4\alpha^{2}}=-\int_{0}^{\infty
}dk\frac{d\delta_{k}}{dk}\ln\left[  1+\frac{k^{2}}{4\alpha^{2}}\right]
=-\int_{0}^{\infty}d\kappa\frac{d\delta_{\kappa}}{d\kappa}\ln\left[
1+\kappa^{2}\right]
\end{equation}
(the surface term vanishes since $\delta_{k=\infty}=0$) in terms of the
dimensionless variable $\kappa\equiv k/\left(  2\alpha\right)  $. The benefit
of the partial integration is that the $\kappa$-derivative removes the
logarithm in the expression (\ref{phshift}) for the phase shifts from the
integrand,
\begin{equation}
\frac{d\delta_{\kappa}}{d\kappa}=-i\frac{d}{d\kappa}\ln\left[
\frac{1+2i\kappa}{1-2i\kappa}\frac{1+i\kappa}{1-i\kappa}\right]
=\frac{2}{1+\kappa^{2}}+\frac{4}{1+4\kappa^{2}},
\end{equation}
so that the remaining integral can be done analytically:%
\begin{equation}
\int_{0}^{\infty}dk\frac{2k\delta_{k}}{k^{2}+4\alpha^{2}}=-\int_{0}^{\infty
}d\kappa\left(  \frac{2}{1+\kappa^{2}}+\frac{4}{1+4\kappa^{2}}\right)
\ln\left[  1+\kappa^{2}\right]  =\pi\ln9.
\end{equation}
Thus%
\begin{equation}
\frac{\det\hat{F}\left[  x_{I}\right]  ^{\prime}}{\omega^{-2}\det\hat
{F}\left[  x_{ho}\right]  }=\frac{1}{12}%
\end{equation}
and together with (\ref{Z-O(hbar)}) and (\ref{limitzho}) we can now assemble
our final result for $Z_{I}\left(  -x_{0},x_{0}\right)  $ at large $T$,
\begin{equation}
Z_{I}\left(  -x_{0},x_{0}\right)  =\sqrt{\frac{m\hbar\omega}{\pi}}e^{-\omega
T/2}\omega T\text{ }\sqrt{\frac{6S_{E}\left[  x_{I}\right]  }{\pi\hbar m}%
}e^{-\frac{S_{E}\left[  x_{I}\right]  }{\hbar}}. \label{z1instfin}%
\end{equation}

This is the quantum mechanical propagator of the double-well tunneling
problem, to $O\left(  \hbar\right)  $ in the semiclassical approximation
around a single instanton. The exact analogs of both the exponential Gamov
factor and the preexponential factor $\sqrt{S_{I}}$ from the zero mode appear
in the one-instanton sector of QCD.

\section{Dilute instanton gas\label{digasec}}

Up to now, we have concentrated on the saddle points which correspond to
single-instanton solutions. This is not the whole story, however: there are
additional (approximate) saddle points which also contribute\ to the
semiclassical tunneling amplitude for large $T$. We will now analyze those
``multi-instanton solutions'', first in the familiar double-well potential and
subsequently in its periodic extension (which represents the closest
quantum-mechanical analog of the semiclassical Yang-Mills vacuum).

\subsection{Double-well potential}

Since the instanton deviates only in a small time interval $\Delta
\tau=1/\left(  2\alpha\right)  $ appreciably from $x_{0}$ or $-x_{0}$ (cf. Eq.
(\ref{deltatau})), and since the overlap between neighboring instantons and
anti-instantons is exponentially small (cf. (\ref{dev})),
multi-(anti-)-instanton solutions of (\ref{eomimagt}) can be approximately
written as a chain (i.e. an ordered superposition) of $N$ alternating
instantons and antiinstantons, sufficiently far separated in time by the
(average) intervall%
\begin{equation}
\bar{\Delta}_{\tau}=\frac{T}{N}\gg\frac{1}{2\alpha}. \label{dilcond}%
\end{equation}
Those chains correspond to $N$ tunneling processes, back and forth between
both minima of the potential. It is intuitively clear that their importance,
associated with the frequency of their occurence, increases with growing $T$.

The approximate $N$-instanton solutions, composed of single, alternating
instantons and anti-instantons at times $\tau_{0,k}$, can thus be written
as\footnote{Eq. (\ref{multinst2}) is to be understood as a shortcut for the
more precise (and tedious) expression%
\begin{equation}
x_{N}\left(  \tau\right)  =\sum_{k=1}^{\left(  N+1\right)  /2}x_{I}\left(
\tau-\tau_{0,2k-1}\right)  +\sum_{k=1}^{\left(  N-1\right)  /2}x_{\bar{I}}%
\left(  \tau-\tau_{0,2k}\right)
\end{equation}
which takes explicit care of the facts that the first and last transitions in
the chain must be instantons, and that those inbetween consist of an
anti-instanton and $\left(  N-3\right)  /2$ pairs of an anti-instanton
followed by an instanton.}%
\begin{equation}
x_{N}\left(  \tau\right)  =\sum_{k=1}^{N}x_{I,\bar{I}}\left(  \tau-\tau
_{0,k}\right)  , \label{multinst2}%
\end{equation}
where $N$ must be odd in order to satisfy the boundary conditions (\ref{b1})
and (\ref{b2}). They become exact solutions for infinite separations $\left|
\tau_{0,k+1}-\tau_{0,k}\right|  \rightarrow\infty$. The (anti-) instanton
centers are ordered in Euclidean time as
\begin{equation}
-\frac{T}{2}\ll\tau_{0,1}\ll\tau_{0,2}...\ll\tau_{0,N}\ll\frac{T}{2}.
\end{equation}
All multi-instanton solutions have to be included as additional saddle-points
in the SCA. We are now going to derive the corresponding expression for
$Z\left(  -x_{0},x_{0}\right)  $ by using the approximate solutions
(\ref{multinst2}) instead. This simplification goes under the name of
\ ``dilute instanton gas approximation'' (DIGA).

The contribution of the approximate $N$-instanton solution to the path
integral is%
\begin{equation}
Z_{N}\simeq\mathcal{N}\int D[\eta]e^{-S_{E}\left[  x_{N}+\eta\right]  /\hbar}.
\end{equation}
We now write the general fluctuation $\eta\left(  \tau\right)  $ as a sum of
independent, localized fluctuations $\eta_{k}\left(  \tau\right)  $ around the
single (anti-) instantons and $\eta_{0}\left(  \tau\right)  $ around the
approximately constant pieces $x\left(  \tau\right)  =\pm x_{0}$ beween them:%
\begin{equation}
\eta\left(  \tau\right)  =\eta_{0}\left(  \tau\right)  +\sum_{k=1}^{N}\eta
_{k}\left(  \tau\right)  .
\end{equation}
This implies that $\eta_{0}\left(  \tau\right)  $ can be finite all over
$\left[  -T/2,T/2\right]  $ (except at the boundaries)\ while the $\eta
_{k}\left(  \tau\right)  $ are time-localized around the $k$-th (anti-) instanton.

Hence the action approximately decomposes into the sum of actions for single
(well-separated) (anti-) instantons and a ``vacuum'' piece,
\begin{equation}
S_{E}\left[  x_{N}+\eta\right]  \simeq S_{E}\left[  x_{0}+\eta_{0}\right]
+\sum_{k=1}^{N}S_{E}\left[  x_{I}+\eta_{k}\right]
\end{equation}
(recall that $S\left[  x_{cl}=\pm x_{0}\right]  =0$ and $S\left[  x_{0}%
+\eta\right]  =S\left[  -x_{0}+\eta\right]  $, due to the symmetry of the
potential). This formula expresses the (approximate)\ fact that the (anti-)
instantons have too little overlap to interact. Furthermore, the path integral
measure factorizes into integrals over the localized fluctuations around the
instantons and those inbetween\footnote{Due to the limited support of the
localized fluctuations, they do no longer form a complete set on the whole
time interval. Any fluctuation $\tilde{x}\left(  \tau\right)  $ for $\tau
\in\lbrack-\frac{T}{2},\frac{T}{2}]$ must therefore be described piecewise in
bases around every instanton. Since the path integral is the product of
integrals over the coefficients of those base functions, it factorizes
naturally.}. As a consequence, $Z_{N}$ factorizes as
\begin{align}
Z_{N}\left(  -x_{0},x_{0}\right)   &  \simeq\mathcal{N}\int D[\eta
_{0}]e^{-S_{E}\left[  x_{0}+\eta_{0}\right]  /\hbar}\times\prod_{k=1}%
^{N}\mathcal{N}\int D[\eta_{k}]e^{-S_{E}\left[  x_{I}+\eta_{k}\right]  /\hbar
}\\
&  =Z_{0}\left(  x_{0},x_{0}\right)  \left[  Z_{I}\left(  -x_{0},x_{0}\right)
\right]  ^{N}.
\end{align}

We have written the above product in a sloppy manner. Strictly speaking, the
$N$ factors $Z_{I}\left(  -x_{0},x_{0}\right)  $ have different boundary
conditions since the endpoints $\pm x_{0}$ are (almost) reached at different
times in each factor, associated with the time-ordering of the (anti-)
instanton from which they arise. However, due to time translation invariance
this makes no difference for the value of the $Z_{I}\left(  -x_{0}%
,x_{0}\right)  $ except through the zero-mode contribution. Indeed, for the
latter we have to integrate over the temporal position $\tau_{0}$ of the
(anti-) instanton (or, equivalently, over $c_{0}$),
\begin{equation}
Z_{I}=Z_{I}^{^{\prime}}\sqrt{\frac{S_{I}}{2\pi\hbar m}}\int d\tau_{0}%
\equiv\tilde{Z}_{I}\int d\tau_{0},
\end{equation}
where now the range of the center $\tau_{0,k}$ of the $k$-th (anti-) instanton
is restricted by the requirement that it occurs after the $\left(  k-1\right)
$-th, i.e. $\tau_{0,k-1}<\tau_{0,k}<T/2$. The integration over the $\tau
_{0,k}$ thus takes the form\
\begin{equation}
\int_{-T/2}^{T/2}d\tau_{0,1}\int_{\tau_{0,1}}^{T/2}d\tau_{0,2}...\int
_{\tau_{0,N-1}}^{T/2}d\tau_{0,N}=\frac{T^{N}}{N!}%
\end{equation}
which results in%
\begin{equation}
Z_{N}\simeq Z_{0}\frac{\left(  \tilde{Z}_{I}T\right)  ^{N}}{N!}.
\end{equation}
The above expression determines the $N$-instanton contribution explicitly
since we had already calculated $Z_{0}$ in (\ref{limitzho}) (with
$\omega=2\alpha$) and since $\tilde{Z}_{I}$ can be immediately obtained from
(\ref{z1instfin}) and (\ref{sinst}):
\begin{equation}
Z_{0}\left(  \pm x_{0},\pm x_{0}\right)  =\mathcal{N}\left(  \det\left[
-\partial_{\tau}^{2}+\omega^{2}\right]  \right)  ^{-1/2}\rightarrow\left(
\frac{m\hbar\omega}{\pi}\right)  ^{1/2}e^{-\omega T/2}, \label{z0}%
\end{equation}%
\begin{equation}
\tilde{Z}_{I}=2\alpha\sqrt{\frac{6S_{E}\left[  x_{I}\right]  }{\pi\hbar m}%
}e^{-\frac{S_{E}\left[  x_{I}\right]  }{\hbar}}=4\sqrt{\frac{2\alpha^{3}%
x_{0}^{2}}{\pi\hbar}}e^{-\frac{4}{3}\alpha mx_{0}^{2}/\hbar}.
\end{equation}

In order to collect the multi-instanton contributions to $Z\left(
x_{0},-x_{0}\right)  $, we have to sum over all odd$\ N$ (recall that odd
$N$'s are required by the boundary conditions (\ref{b1}), (\ref{b2})) and
obtain
\begin{equation}
Z_{DIGA}\left(  x_{0},-x_{0}\right)  =Z_{0}\sum_{N\text{ \ odd}}\frac{\left(
\tilde{Z}_{I}T\right)  ^{N}}{N!}=\frac{Z_{0}}{2}\left\{  e^{\tilde{Z}_{I}%
T}-e^{-\tilde{Z}_{I}T}\right\}  =Z_{0}\sinh\left(  \tilde{Z}_{I}T\right)  .
\label{digasum}%
\end{equation}
(Note that the second exponential on the RHS above removes the contributions
even in $\tilde{Z}_{I}T$ \ and doubles those odd in $\tilde{Z}_{I}T.$) An
analogous expression for $Z_{DIGA}\left(  -x_{0},-x_{0}\right)  $ is obtained
when only contributions from even numbers of instantons are summed. Both
results can be combined into the expression%
\begin{align}
Z_{DIGA}\left(  \pm x_{0},-x_{0}\right)   &  =\frac{1}{2}\left(
\frac{\hbar\omega}{\pi}\right)  ^{1/2}e^{-\omega T/2}\left\{  e^{\tilde{Z}%
_{I}T}\mp e^{-\tilde{Z}_{I}T}\right\} \\
&  =\frac{1}{2}\left(  \frac{\hbar\omega}{\pi}\right)  ^{1/2}\left\{
e^{-\left(  \omega/2-\tilde{Z}_{I}\right)  T}\mp e^{-\left(  \omega
/2+\tilde{Z}_{I}\right)  T}\right\}  \label{zdiga}%
\end{align}
from which the two lowest energy levels of the system (i.e. those of the
ground state and the first excited state) can be found as%
\begin{align}
E_{0}  &  =-\hbar\lim_{T\rightarrow\infty}\frac{1}{T}\ln Z_{DIGA}\left(  \pm
x_{0},-x_{0}\right)  =-\hbar\lim_{T\rightarrow\infty}\frac{1}{T}\left[
-\left(  \frac{\omega}{2}-\tilde{Z}_{I}\right)  T\right]  =\frac{\hbar\omega
}{2}-\hbar\tilde{Z}_{I},\\
E_{1}  &  =-\hbar\lim_{T\rightarrow\infty}\frac{1}{T}\left[  -\left(
\frac{\omega}{2}+\tilde{Z}_{I}\right)  T\right]  =\frac{\hbar\omega}{2}%
+\hbar\tilde{Z}_{I}.
\end{align}

As expected, the effect of tunneling is to split the degenerate (would-be)
ground state energies $\hbar\omega/2$ of the wave functions $\left|  \pm
x_{0}\right\rangle $ centered in each of the two minima of the
potential\footnote{We note in passing that the energy splitting between the
classical would-be ground states could have been obtained from the
one-instanton approximation alone, i.e. without employing the DIGA. Indeeed,
restricting to small $T$ and expanding
\begin{equation}
\left\langle x_{f}\left|  e^{-HT/\hbar}\right|  x_{i}\right\rangle
=1-\frac{T}{\hbar}\left\langle x_{f}\left|  H\right|  x_{i}\right\rangle
+O\left(  T^{2}\right)
\end{equation}
the propagator is governed by the (time independent) matrix element
\begin{equation}
\left\langle x_{f}\left|  H\right|  x_{i}\right\rangle
\end{equation}
which can be calculated in the one-instanton approximation.}. The
corresponding energy eigenstates are similarly obtained from the prefactors of
the exponential in (\ref{zdiga}) (cf. Eq. (\ref{asympt})). For the ground
(first excited) state we find the symmetric (antisymmetric) linear combination%
\begin{align}
\left|  0\right\rangle  &  =\frac{1}{\sqrt{2}}\left\{  \left|  x_{0}%
\right\rangle +\left|  -x_{0}\right\rangle \right\}  ,\label{grdstate}\\
\left|  1\right\rangle  &  =\frac{1}{\sqrt{2}}\left\{  \left|  x_{0}%
\right\rangle -\left|  -x_{0}\right\rangle \right\}  .
\end{align}
These are the standard WKB results for tunneling amplitudes, with the typical
splitting
\begin{equation}
\Delta E\sim e^{-\frac{S_{E}\left[  x_{I}\right]  }{\hbar}}%
\end{equation}
of the energy levels of the states connected by tunneling, although we have
obtained them in the somewhat less familiar framework of the imaginary-time
path integral. It is remarkable that the SCA works well even for the
lowest-lying states of the system, i.e. for those with the largest de-Broglie
wavelengths $\lambda$, although this $\lambda$ is not small compared to the
size of the potential. Note furthermore that, in contrast to the case without
tunneling and the ground states $\left|  \pm x_{0}\right\rangle $, the above
states are eigenstates of parity (under which $x_{0}\leftrightarrow-x_{0}$).
The new ground state (\ref{grdstate}), in particular, is parity invariant: the
artificially broken parity in the absence of tunneling is restored.

Before leaving this section, we should address a potential concern in summing
over the dilute instanton gas in Eq. (\ref{digasum}). Indeed, for an
increasing number $N$ of (anti-)instantons in the constant interval $T$ the
diluteness condition (\ref{dilcond}) is less and less satisfied, implying that
from some large $N$ onwards the corresponding terms in the sum will violate
this fundamental DIGA requirement. However, Stirling's formula $n!\simeq
\left(  \frac{n}{e}\right)  ^{n}\sqrt{2\pi n}$ for large $n$ shows that the
sum (\ref{digasum}) is dominated by terms with%
\begin{equation}
\frac{\tilde{Z}_{I}T}{N}\sim O\left(  1\right)
\end{equation}
and that contributions from larger $N$ are rapidly suppressed. As a
consequence, only $N$ with%
\begin{equation}
\frac{N}{T}\lesssim Ce^{-\frac{S_{E}\left[  x_{I}\right]  }{\hbar}},
\end{equation}
i.e. with an exponentially small instanton density in the semiclassical limit
(which can be improved at will by increasing $\alpha$ since we had found in
Eq. (\ref{sinst}) that $S_{E}\left[  x_{I}\right]  =\left(  4/3\right)  \alpha
mx_{0}^{2}$), contribute significantly to the sum (\ref{digasum}). This
ensures that $Z_{DIGA}$ is dominated by terms in which the DIGA diluteness
requirement (\ref{dilcond}) is well satisfied\footnote{Unfortunately, it will
turn out below that in QCD, where we do not have an adjustable parameter
analogous to $\alpha$, the situation is much more complex and the
corresponding DIGA does not provide a self-consistent approach to the vacuum
wave functional.}.

Our main motivation for the use of instanton methods in the solution of the
above tunneling problem was that this procedure can be straightforwardly
generalized to gauge theories. In order to develop the analogy with QCD as far
as possible, however, we should still go a step farther in quantum mechanics
and study a periodical potential with degenerate minima. This will be the
subject of the following section.

\subsection{Periodic potential\label{perpot}}

Let us now consider the periodic extension of the double well potential (which
thereby becomes bounded from above). This type of potential most closely
resembles the situation which we will encounter below in the QCD vacuum.

\begin{figure}
[ptb]
\begin{center}
\includegraphics[
height=1.9501in,
width=3.1574in
]%
{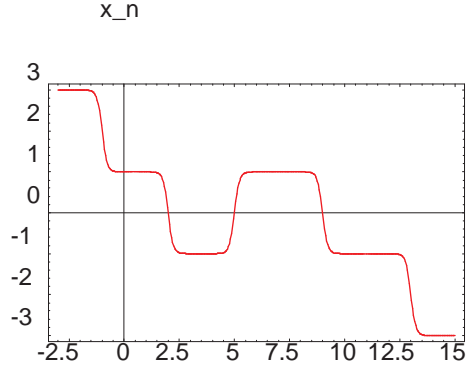}%
\caption{A multi-instanton solution in the periodical potential. The abscissa
denotes the Euclidean time $\tau$ while the ordinate gives the position
variable $x$. The integer values of $x$ correspond to the minima $x_{0,n}$ of
the periodical potential.}%
\label{multinst}%
\end{center}
\end{figure}

In a periodic potential with degenerate minima, instantons and anti-instantons
can arbitrarily follow each other, starting out at the minimum where the
previous one had ended, i.e. connecting adjacent minima $x_{0,n}$ and
$x_{0,n\pm1}$ (see Fig. \ref{multinst}). Generalizing our earlier conventions,
we will refer to an instanton (anti-instanton) as the segment of the solution
which interpolates between neighboring minima to the left (right), i.e. which
decreases (increases) $n$.
Getting from the minimum with index $n_{i}$ to the one with index $n_{f}$,
corresponding to the boundary conditions%
\begin{equation}
x\left(  -\frac{T}{2}\right)  =x_{0,n_{i}},\text{ \ \ \ \ \ \ \ \ \ }x\left(
\frac{T}{2}\right)  =x_{0,n_{f},}%
\end{equation}
therefore requires the number of anti-instantons minus the number of
instantons, $N_{\bar{I}}-N_{I}$, to equal $n_{f}-n_{i}$. As a consequence, the
semiclassical propagator becomes
\begin{equation}
Z_{per}\left(  x_{n_{f}},x_{n_{i}}\right)  \simeq Z_{0}\sum_{N_{I}=0}^{\infty
}\sum_{N_{\bar{I}}=0}^{\infty}\frac{\left(  \tilde{Z}_{I}T\right)
^{N+N_{\bar{I}}}}{N!N_{\bar{I}}!}\delta_{N_{\bar{I}}-N_{I}-\left(  n_{f}%
-n_{i}\right)  }.
\end{equation}

With the representation
\begin{equation}
\delta_{ab}=\int_{0}^{2\pi}\frac{d\theta}{2\pi}e^{i\theta\left(  a-b\right)  }%
\end{equation}
of the Kronecker symbol, $Z_{per}$ can be rewritten as%
\begin{align}
Z_{per}\left(  x_{n_{f}},x_{n_{i}}\right)   &  \simeq Z_{0}\int_{0}^{2\pi
}\frac{d\theta}{2\pi}e^{-i\theta\left(  n_{f}-n_{i}\right)  }\sum_{N_{I}%
=0}^{\infty}\frac{\left(  \tilde{Z}_{I}Te^{-i\theta}\right)  ^{N_{I}}}{N_{I}%
!}\sum_{N_{\bar{I}}=0}^{\infty}\frac{\left(  \tilde{Z}_{I}Te^{i\theta}\right)
^{N_{\bar{I}}}}{N_{\bar{I}}!}\\
&  =Z_{0}\int_{0}^{2\pi}\frac{d\theta}{2\pi}e^{-i\theta\left(  n_{f}%
-n_{i}\right)  }e^{\tilde{Z}_{I}Te^{-i\theta}+\tilde{Z}_{I}Te^{i\theta}}%
\end{align}
and by using again the expression (\ref{z0}) for $Z_{0}$ we arrive at%
\begin{align}
Z_{per}\left(  x_{n_{f}},x_{n_{i}}\right)   &  =\left(  \frac{m\hbar\omega
}{\pi}\right)  ^{1/2}e^{-\omega T/2}\int_{0}^{2\pi}\frac{d\theta}{2\pi
}e^{-i\theta\left(  n_{f}-n_{i}\right)  }e^{2\tilde{Z}_{I}T\cos\theta
}\label{ztheta}\\
&  =\left(  \frac{m\hbar\omega}{\pi}\right)  ^{1/2}\int_{0}^{2\pi
}\frac{d\theta}{2\pi}e^{-i\theta\left(  n_{f}-n_{i}\right)  }e^{-\left(
\omega/2-2\tilde{Z}_{I}\cos\theta\right)  T}.
\end{align}
As before, we can now obtain the low-lying energy levels from the
$T\rightarrow\infty$ limit. The above expression shows that we get, in fact, a
continuous ``band'' of energies parametrized by $\theta$ (since (\ref{ztheta})
is the sum of contributions from the lowest energy levels in (\ref{zspectral}%
))
\begin{align}
E_{0}\left(  \theta\right)   &  =-\hbar\lim_{T\rightarrow\infty}\frac{1}%
{T}\left[  -\left(  \frac{\omega}{2}-2\tilde{Z}_{I}\cos\theta\right)  T\right]
\\
&  =\frac{\hbar\omega}{2}-2\hbar\tilde{Z}_{I}\cos\theta.
\end{align}

Of course, it is well known that the energies in a periodic potential form
continuous ``bands''. Our result is just the analog of, e.g., the lowest-lying
band of electron states in the periodic potential of a metal. Not
surprisingly, then, the corresponding eigenstates are the ``Floquet-Bloch
waves''%
\begin{equation}
\left|  \theta\right\rangle =\frac{1}{\sqrt{2\pi}}\left(  \frac{\hbar\omega
}{\pi}\right)  ^{1/4}\sum_{n}e^{in\theta}\left|  n\right\rangle ,
\end{equation}
where $\left|  n\right\rangle $ is the state localized at the $n$-th minimum
of the periodic potential.

This concludes our discussion of instantons in quantum mechanics. Much more
could be said about them \cite{newqminst}, their cousins (sometimes called
``bounces'') which mediate tuneneling between nondegenerate minima, their
connections to large-order perturbation theory in the real-time theory, etc.
However, we refrain from doing so since we have reached our main objective: to
obtain the semiclassical expansion and the ground state properties in a
potential which mimicks the situation in the QCD vacuum. We will now move on
to discuss ``the real thing'', namely instantons in QCD itself.

\chapter{Instantons in QCD}

What is the relevance of the above tunneling discussion for QCD? As
anticipated, the answer is that several pertinent aspects of the
SCA\footnote{We note in passing that there exists a somewhat complementary way
of looking at the SCA in quantum field theory, namely as an expansion in the
number of Feynman-graph loops (which reflects the non-perturbative nature of
the SCA from a different angle).
\par
\bigskip} generalize to four-dimensional Euclidean (i.e. imaginary-time)
Yang-Mills theory once we have identified the saddle points of the
corresponding functional integral in imaginary time. Our first task will
therefore be to find the minima of the classical Yang-Mills action.
Remarkably, they turn out to be determined by the topology of the gauge group
and to form the minima of a periodic ``potential'' analogous to the one
encountered above.

The corresponding tunneling solutions, the Yang-Mills instantons, are
classical gauge fields with intriguing topological properties. The latter
induce new physical phenomena without analogy in the quantum mechanical
examples. Some of these phenomena, including several ones related to the
light-quark sector of QCD, will be discussed in subsequent sections of this chapter.

\section{Vacuum topology and Yang-Mills instantons}

\subsection{Topology of the Yang-Mills vacuum\label{vactop}}

In order to develop some intuition for semiclassical ground-state properties
of \ QCD, let us start as in the preceding quantum mechanical examples by
searching for the minima of the Euclidean Yang-Mills action. Restricting for
the moment to the gluon sector, the latter reads%
\begin{align}
S\left[  G\right]   &  =\frac{1}{4}\int d^{4}x\left[  G_{\mu\nu}^{a}G_{\mu\nu
}^{a}\right] \label{ymact}\\
&  =\frac{1}{2}\int d^{4}x\left[  E_{i}^{a}E_{i}^{a}+B_{i}^{a}B_{i}%
^{a}\right]  \geq0
\end{align}
where the gluon field strength tensor is%
\begin{equation}
G_{\mu\nu}\left(  x\right)  =\partial_{\mu}G_{\nu}-\partial_{\nu}G_{\mu
}+ig\left[  G_{\mu},G_{\nu}\right]  \equiv G_{\mu\nu}^{a}t^{a},\text{
\ \ \ \ \ \ \ }t^{a}=\frac{\lambda^{a}}{2}, \label{fstrength}%
\end{equation}
and the chromoelectric and -magnetic fields are defined as%
\begin{equation}
E_{i}^{a}=G_{i4}^{a},\,\text{\ \ \ \ \ \ \ \ }B_{i}^{a}=-\frac{1}%
{2}\varepsilon_{ijk}G_{jk}^{a}. \label{ebfields}%
\end{equation}

The corresponding quantum field theory (including the quarks) determines the
structure of the QCD vacuum, i.e. the unique state of lowest energy on which
the Fock space is built. Since QCD is strongly coupled and therefore
non-perturbative at low energies, we expect the vacuum to be populated by
strong fields. (This is in contrast to standard QED, for example, where the
vacuum contains mostly zero-point fluctuations, i.e. weakly interacting
electron-positron pairs and photons which can be handled perturbatively.)

A measure for the strength of the QCD vacuum fields can be obtained from the
trace anomaly, which relates the energy density $\epsilon_{vac}$ of the vacuum
to the phenomenologically known vacuum expectation value of the square of the
gluon field strength tensor, the so-called ``gluon condensate'' (renormalized
at about 1 GeV):%
\begin{equation}
\epsilon_{vac}\simeq-\frac{b_{1}}{128\pi^{2}}\langle0\left|  g^{2}%
G^{2}\right|  0\rangle\simeq-\frac{1}{2}\frac{\text{GeV}}{\text{fm}^{3}}%
\ll\epsilon_{pert}.
\end{equation}
This relation shows that the nonperturbative vacuum fields are indeed
exceptionally strong: they reduce the vacuum energy in a tiny cube of size
10$^{-15}$m by about half a proton mass! As we will see in the remainder of
this section, part of this reduction is due to tunneling processes mediated by instantons.

Strong fields can contain a very large number of quanta, and those quanta can
become coherent and render the corresponding action large compared to
$\hslash$. In other words, such fields behave (semi-) classically since
quantum fluctuations are of $O\left(  \hslash\right)  $ and thus contribute
only relatively small corrections. The above reasoning suggests that insight
into the vacuum fields of QCD may be gained from a semiclassical perspective.
In the following, we will explore this perspective while paying special
attention to robust and generic features which are likely to survive even
stronger quantum fluctuations. The most important such features will turn out
to be ``global'', i.e. topological properties of the vacuum fields which are
invariant under continuous deformations (and therefore in particular under
time evolution).

The first step towards a semiclassical approach to the QCD vacuum
(starting\ at the lowest order of $\hbar$) is to find those classical fields
which minimize the (static) Yang-Mills energy or, equivalently, the Euclidean
action (\ref{ymact}). Such fields are sometimes called ``classical vacua''.
Since the Euclidean action (\ref{ymact}) is non-negative (in contrast to its
counterpart in Minkowski space) its absolute minima will be the gluon fields
with zero action. These fields are the\ ``pure gauges''\
\begin{equation}
G_{\mu}^{\left(  pg\right)  }=\frac{-i}{g}U\partial_{\mu}U^{\dagger}
\label{pureg}%
\end{equation}
where $U\left(  x\right)  \in SU\left(  3\right)  $ is an element of the gauge
group. It is easy to check explicity that the field strength of pure
gauges\ vanishes,%
\begin{equation}
G_{\mu\nu}^{\left(  pg\right)  }=0.
\end{equation}
At first sight, one might think that these fields would be natural candidates
on which to build a semiclassical expansion. A moment's reflection shows,
however, that none of them could generate an acceptable vacuum since they are
neither unique nor gauge invariant.

In fact, it can be easily seen that the $G_{\mu}^{\left(  pg\right)  }$ fall
into an enumerable infinity of topological equivalence classes. In order to
demonstrate this, we start by choosing the temporal gauge
\begin{equation}
G_{0}=0 \label{tempg}%
\end{equation}
so that the residual gauge transformations (which conserve this gauge
condition) are time-independent. Moreover, for our purposes it is useful to
restrict the residual gauge transformations further by letting them approach a
constant (chosen to be unity) at spacial infinity, i.e.%
\begin{equation}
U\left(  \vec{x}\right)  \rightarrow1\text{ \ \ \ \ \ \ \ for \ \ \ \ \ \ }%
\left|  \vec{x}\right|  \rightarrow\infty. \label{bcym}%
\end{equation}
This restriction guarantees that the gauge fields satisfiy definite boundary
conditions at the surface of a large box (which should not affect the local
physics inside).

\begin{figure}
[ptb]
\begin{center}
\includegraphics[
height=2.3454in,
width=5.1361in
]%
{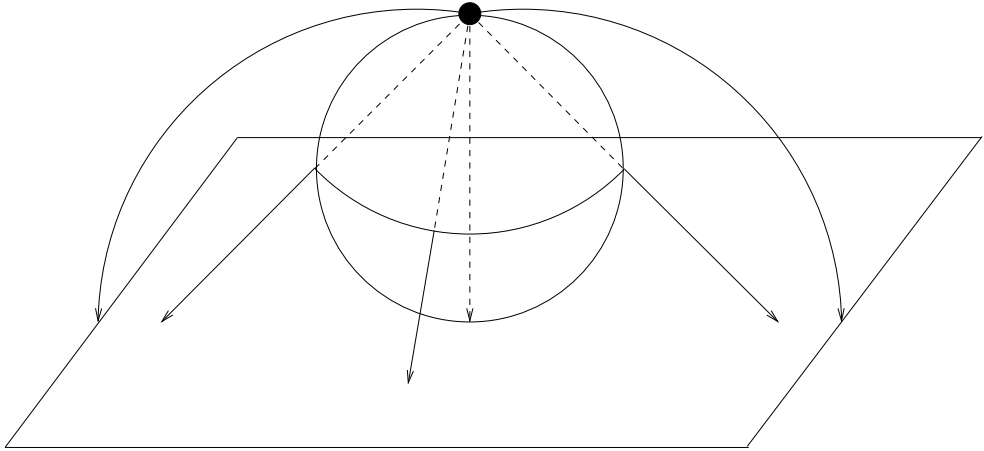}%
\caption{Schematic view of the generalized stereographic projection which
identifies compactified $R^{3}$ with $S^{3}$. (Figure from hep-th/0010225,
courtesy of Falk Bruckmann.)}%
\label{stereo}%
\end{center}
\end{figure}

The effect of the condition (\ref{bcym}) is that, from the point of view of
gauge transformations, the space $R_{space}^{3}$ is effectively compactified
to $S_{space}^{3}$ (i.e. to a 3-dimensional sphere of infinite radius), i.e.
different boundary points at $\left|  \vec{x}\right|  \rightarrow\infty$
cannot be distinguished by $U\left(  \vec{x}\right)  $ and thus can be
identified. This is completely analogous to the compactification
$R^{2}\rightarrow S^{2}$ of the complex plane by Riemann's stereographic
projection, as shown in Fig. \ref{stereo}.
As a consequence, any residual gauge transformation $U\left(  \vec{x}\right)
$ defines a map
\begin{equation}
S_{space}^{3}\rightarrow SU\left(  3\right)  _{color}. \label{map1}%
\end{equation}
For the purpose of the following topological (homotopy) classification, this
map can be furthermore restricted to a $SU\left(  2\right)  $ subgroup of the
gauge group, since according to a powerful theorem by Raoul Bott \cite{bot56}
only this subgroup will be ``topologically active'' in our discussion.
Moreover, the quarternion representation of any $U\in SU\left(  2\right)  $,
\begin{equation}
U=u_{0}+iu_{a}\tau_{a}\,\ \ \ \ \ \ \ \text{with \ \ \ \ \ \ }u_{\alpha}\text{
real and \ \ }u_{0}^{2}+u_{a}u_{a}=1,
\end{equation}
(in terms of the Pauli matrices $\tau_{a}$) shows that $SU\left(  2\right)  $
can be mapped onto the 3-sphere $S_{group}^{3}$ (which is its group manifold).
Thus the map (\ref{map1}) is topologically (in the sense of homotopy theory)
equivalent to
\begin{equation}
S_{space}^{3}\rightarrow S_{group}^{3} \label{s3map}%
\end{equation}
i.e. any residual gauge transformation maps the spacial 3-sphere into the
group 3-sphere. In general, two such mappings $U\left(  \vec{x}\right)  $
cannot be continuously deformed into each other, which implies that they fall
into different homotopy classes. The same holds for spheres of any dimension
$n$, i.e. for any map $S^{n}\rightarrow S^{n}$.

\begin{figure}
[ptb]
\begin{center}
\includegraphics[
height=3.8943in,
width=2.7614in
]%
{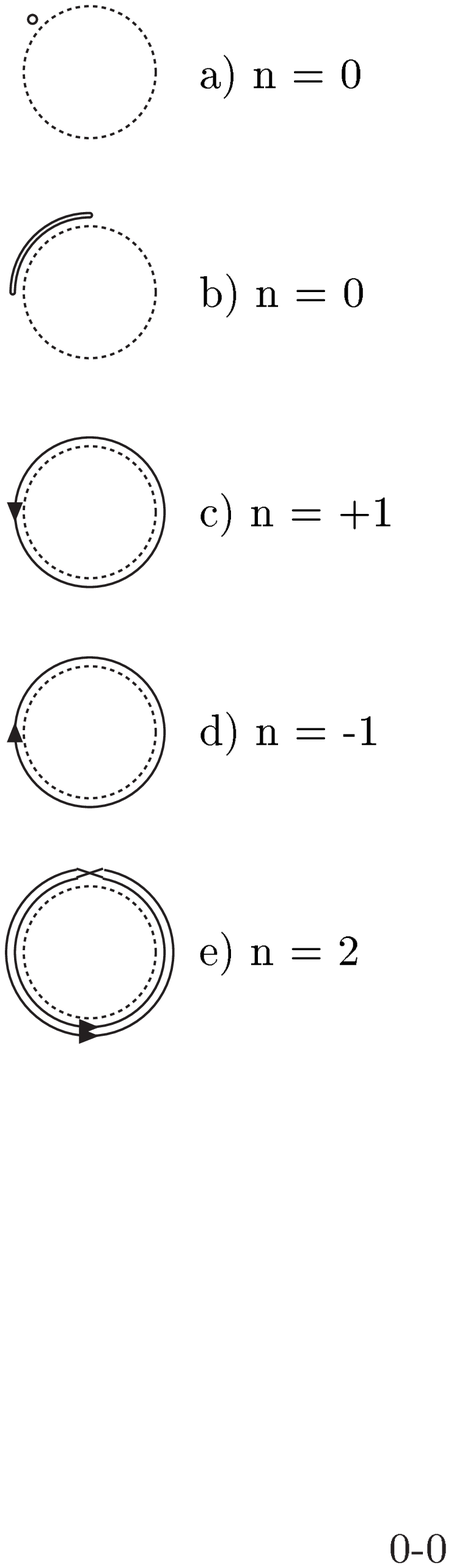}%
\caption{Some characteristic examples of the mapping $S^{1}\rightarrow S^{1},$
$\exp\left(  i\chi\right)  \mapsto\exp\left(  i\phi\left(  \chi\right)
\right)  $. The inner, dotted circle is the range of values of $\phi$, while
the outer, solid ``circles`` symbolize the domain of $\chi$. }%
\label{windingn}%
\end{center}
\end{figure}

To get a visual understanding of this topological classification, it is thus
sufficient to consider just the simplest case%
\begin{equation}
S_{sp}^{1}\rightarrow S_{gr}^{1} \label{circlemap}%
\end{equation}
of maps between two circles. Such mappings can equivalently be thought of as
phase fields $U\left(  \chi\right)  =e^{i\phi\left(  \chi\right)  }\in
U\left(  1\right)  $ defined on a circle with coordinate angle $\chi\in\left[
0,2\pi\right]  $. In Fig. \ref{windingn} various maps of this type are
graphically represented by drawing the domain space circle (which can be
imagined as a rubber band) in contact with the target space (the dotted circle
which represents the phase of the field) such that the mapping occurs between
those points on the circles which touch each other. Note that we consider only
continuous maps which implies that every point of the target circle must
somewhere touch the domain circle.%
The examples drawn in Fig. \ref{windingn} illustrate the topological
properties of such maps. We start with Fig. \ref{windingn}a where the whole
domain space shrinks to a point (which we draw as a tiny circle with radius
$r\rightarrow0$ implied) and touches the target space at one and the same
point. Clearly this graph represents the constant map: all points of ``space''
are mapped into one and the same point in the field or target space.

Let us now consider Fig. \ref{windingn}b where the mapping (field) becomes
space-dependent. We define that fields which can be continuously deformed into
each other (i.e. such that the all points of the domain space stay in contact
with points on the domain circle) belong to the same ``homotopy class''. Now
the map of Fig. \ref{windingn}b can clearly be continuously deformed into the
identity map and thus belongs to the same homotopy class. This is in distinct
contrast to the remaining maps of Fig. \ref{windingn}. In Fig. \ref{windingn}c
the domain space wraps once counterclockwise around the target circle and
therefore cannot be continuously shrunk to a point. Therefore, the
\ corresponding map lies in a disjoint homotopy class with which we will
associate a ``winding number'' $n=+1$. The mapping of Fig. \ref{windingn}d
wraps clockwise\footnote{Note that the direction in which a point on the curve
proceeds if the curve parameter increases does not change under continuous
deformations.} around the target circle and belongs to the homotopy class with
winding number $n=-1$. (The trivial class containing the identity map
obviously has $n=0$.)

The next graph, Fig. \ref{windingn}e, shows a map with winding number $n=+2$
where the target space wraps twice counterclockwise around the domain space.
Now it should be clear how this classification proceeds to higher winding
numbers. We have thus made plausible that the maps (\ref{circlemap}) fall into
an enumerable infinity of disjoint homotopy classes characterized by an
integer winding number $n\in Z$. If one additionally defines a composition law
for such maps by concatenation (under which the winding numbers of the maps to
be composed simply add) the homotopy classes become elements of a group, the
so-called homotopy group
\begin{equation}
\pi_{1}\left(  S^{1}\right)  =Z.
\end{equation}
Here the subscript of $\pi$ indicates the dimension of the domain sphere (in
the present case equal to one), and its argument denotes the target space
(here $S^{1}$). Now, according to what we have said before this result
generalizes to the homotopy groups for mappings between spheres of dimension
$d$, $\pi_{d}\left(  S^{d}\right)  =Z$, which includes the maps (\ref{s3map})
which interest us in the context of the QCD vacuum. We thus conclude that%
\begin{equation}
\pi_{3}\left(  S^{3}\right)  =Z.
\end{equation}

Let us summarize what we have learned so far: the pure-gauge fields
(\ref{pureg}), constructed from gauge transformations $U^{\left(  n\right)
}\left(  x\right)  $ which satisfy the boundary conditions (\ref{bcym}),
minimize the Euclidean QCD action and fall into disjoint homotopy classes
characterized by an integer winding number $n$ which derives from the
topological properties of the $U^{\left(  n\right)  }\left(  x\right)  $. Thus
the topology of the gauge group on the compactified space $S^{3}$ induces the
semiclassical vacuum structure with its periodic ``potential'' (action).

\begin{figure}
[ptb]
\begin{center}
\includegraphics[
height=2.2606in,
width=5.2926in
]%
{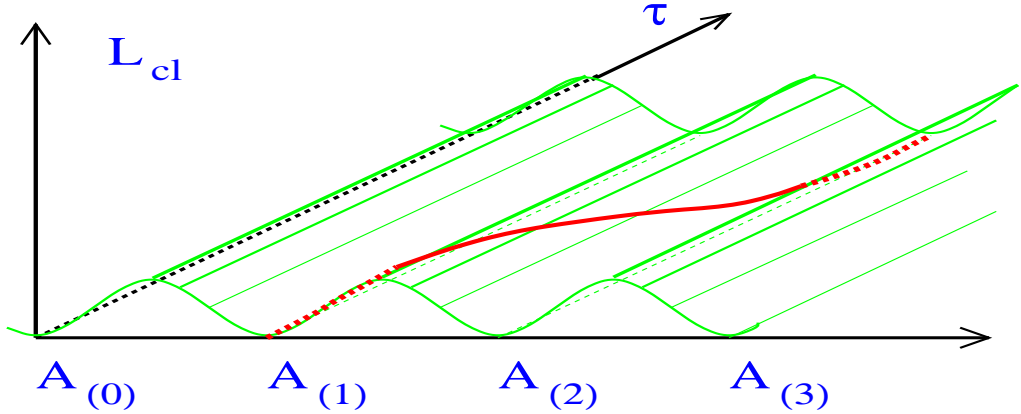}%
\caption{The classical, Euclidean QCD Lagrangian for different classes of
gauge field configurations. The degenerate absolute minima correspond to pure
gauges $A_{\left(  n\right)  ,\mu}\left(  \equiv G_{\mu}^{\left(  n\right)
}\right)  $ with winding number $n$. An instanton trajectory, interpolating
between $A_{\left(  1\right)  }$ and $A_{\left(  2\right)  }$, is also shown.
}%
\label{tunnel}%
\end{center}
\end{figure}

Note that the above arguments do \textit{not} prevent us from transforming a
pure-gauge field of class $n$, i.e.
\begin{equation}
G_{\mu}^{\left(  n\right)  }=\frac{-i}{g}U^{\left(  n\right)  }\partial_{\mu
}U^{\left(  n\right)  \dagger},
\end{equation}
by continuous deformation into one of class $m\neq n$, despite the fact that
we have just shown that we cannot continuously deform $U^{\left(  n\right)  }$
into $U^{\left(  m\right)  }$. Indeed, the latter implies only that we cannot
go from $G_{\mu}^{\left(  n\right)  }$ to $G_{\mu}^{\left(  m\right)  }$
without leaving pure gauge, i.e. we cannot keep the field always in the form
(\ref{pureg}). This also means that we cannot stay in the sector of fields
with zero action (recall that the pure gauges are the only gluon fields which
minimize the Euclidean action). In other words, while continuously deforming
$G_{\mu}^{\left(  n\right)  }$ into $G_{\mu}^{\left(  m\right)  }$ with $n\neq
m$ we necessarily encounter field configurations with non-minimal action,
$S_{E}>0$: each topological class corresponds to an absolute action minimum,
and those minima are separated by a finite-action barrier (the so-called
sphaleron barrier). This situation is depicted in Fig. \ref{tunnel}.%

\subsection{Yang-Mills instanton solution\label{instsoln}}

The reader has probably already noted the analogy between the degenerate
classical ground state found in Section \ref{vactop} and the periodic
potential discussed in Section \ref{perpot}. In both cases, the Euclidean
action has an enumerable infinity of degenerate minima. And as in the quantum
mechanical examples, the degeneracy between the classical ``would-be'' or
``candidate'' vacua of QCD is lifted by tunneling of gluons through the finite
action barriers.

Thus we know already, at least in principle, how to obtain the semiclassical
approximation to the tunneling amplitude between the classical minima $G_{\mu
}^{\left(  n\right)  }$: it is generated by a saddle-point approximation to
the Euclidean functional integral around stationary (and approximately
stationary) field configurations, i.e. the solutions of the Euclidean
Yang-Mills equation%
\begin{equation}
\frac{\delta S}{\delta G_{\nu}}=\partial_{\mu}G_{\mu\nu}+g\left[  G_{\mu
},G_{\mu\nu}\right]  \equiv D_{\mu}G_{\mu\nu}=0 \label{ymeq}%
\end{equation}
with the boundary conditions%
\begin{align}
G_{\mu}\left(  \vec{x},T=-\infty\right)   &  =G_{\mu}^{\left(  n\right)
}\left(  \vec{x}\right)  ,\label{bc3}\\
G_{\mu}\left(  \vec{x},T=+\infty\right)   &  =G_{\mu}^{\left(  m\right)
}\left(  \vec{x}\right)  . \label{bc4}%
\end{align}
According to the above boundary conditions, the solutions start out at
$T\rightarrow-\infty$ as pure-gauge fields constructed from a gauge
transformation with winding number $n,$ and they end up in a pure gauge
associated with winding number $m$ at $T\rightarrow+\infty$. Moreover, they do
by ``paying'' the minimal amount of action possible. These solutions are the
QCD-instantons, and they can be found analytically \cite{bel75,shi94}. For
$m=n+1$, the instanton (in a non-singular gauge) has the explicit form%
\begin{equation}
G_{\mu}^{\left(  I\right)  }\left(  x\right)  =\frac{2}{g}\frac{\eta_{a\mu\nu
}\left(  x_{\nu}-z_{\nu}\right)  t^{a}}{\left(  x-z\right)  ^{2}+\rho^{2}}.
\label{yminst}%
\end{equation}
An instanton with $n=1$, $m=2$ is drawn in Fig. \ref{tunnel}.

A couple of remarkable features can be read off directly from the instanton
solution (\ref{yminst}):\ first, its ``spin'' (i.e. the Lorentz vector index)
is coupled to the color orientation by the 't Hooft symbol $\eta_{a\mu\nu}$.
Second, its nonperturbative character (as expected in the context of
tunneling) reveals itself in the diverging weak-coupling limit. (The $1/g$
behavior is common to all solutions of the classical Yang-Mills equation
(\ref{ymeq}), as can be seen by rescaling the gluon field.) Finally, the
solution makes explicit that the tunneling process is localized in a region of
size $\rho$ in space and time, around its center $z$.

By using the definition
\begin{equation}
\eta_{a\mu\nu}=\delta_{a\mu}\delta_{4\nu}-\delta_{a\nu}\delta_{4\mu
}+\varepsilon_{a\mu\nu} \label{thsymb}%
\end{equation}
of the 't Hooft symbol, it is easy to verify that the instanton (\ref{yminst})
indeed satisfies the boundary conditions (\ref{bc3}), (\ref{bc4}) with $n=0$.
Thus, QCD instantons describe localized, flash-like rearrangements of the
vacuum which mediate tunneling processes between topologically distinct
pure-gauge sectors.

In analogy with the quantum-mechanical example of the periodic potential,
those tunneling processes lift the degeneracy between the pure-gauge
``would-be'' vacua $\left|  n\right\rangle $. Instead, superpositions of all
those states, the ``theta vacua''%
\begin{equation}
\left|  0\right\rangle _{\theta}=\mathcal{N}\sum_{n}e^{i\theta n}\left|
n\right\rangle , \label{thetvac}%
\end{equation}
become the gauge-invariant eigenstates. Although part of our above discussion
(including the tunneling interpretation) depended on the gauge choice
(\ref{tempg}), the ensuing vacuum structure is therefore gauge-independent,
too. We will further elaborate on the $\theta$ vacuum structure in Section
\ref{theta}. Here we only note that the analogy with the Bloch waves of the
quantum-mechanical periodic potential is limited. While the symmetry
generators in the periodic potential (i.e. the generators of periodic
translations) are physical, they just generate gauge transformations of
nontrivial topology in the Yang-Mills case. Moreover, the analog of the Bloch
momentum, the angle $\theta$, classifies different theories and cannot be
changed inside a given ''$\theta$-world''. At present, the value of $\theta$
can only be determined phenomenologically. This task is made easier by the
fact that finite values of $\theta$ induce CP-violating amplitudes (see
Section \ref{thangleclus}).

Evidence from instanton-liquid vacuum models suggests that approximate saddle
points, namely superpositions of instantons and anti-instantons
\cite{dia84,sch98}, dominate the SCA to the generating functional of QCD. Such
superpositions are approximate solutions of the Yang-Mills equation if the
typical separation between the (anti-) instantons is much larger than their
average size, a condition which seems to hold in the QCD vacuum (see below).
In contrast to the quantum mechanical examples of Sections \ref{dwellsec} and
\ref{digasec}, however, the classical interactions between Yang-Mills
instantons are of dipole type (at large separation) and therefore of much
longer range than the exponentially suppressed overlaps between quantum
mechanical instantons (cf. Eq. (\ref{dev})). As a consequence, the dilute
instanton gas approximation fails in QCD. The strong correlations among
QCD-instantons generate an ensemble which probably resembles more a liquid
than a gas.

The size distribution $n\left(  \rho\right)  $ of instantons in the vacuum is
currently under active investigation, mainly on the lattice \cite{lat}. The
obtained results are not yet fully consistent with each other but conform more
or less to the older results of instanton vacuum models \cite{sch98}. (Some
calculations have led to a larger density which, however, is difficult to
measure reliably on the lattice.) For the lowest moments of the distribution,
the standard values are%
\begin{align}
\bar{n} &  =\int d\rho n\left(  \rho\right)  \sim1\text{ fm}^{-4}%
,\label{nbar}\\
\bar{\rho} &  =\frac{1}{\bar{n}}\int d\rho\rho n\left(  \rho\right)
\sim\frac{1}{3}\text{ fm.}\label{rhobar}%
\end{align}

Thus, in the QCD vacuum tunneling happens on average in about 5\% of
spacetime, and it happens very rapidly: the action barrier is penetrated
almost instantaneously ($\tau_{tunnel}\sim0.3\times10^{-15}$m $\sim10^{-24}%
$s), therefore the name ``instanton''\textbf{. }

\section{Including quarks}

Some of the most striking effects of QCD instantons are associated with the
existence of light quarks, and not even a terse introduction to instanton
physics like the present one would be complete without mentioning at least a
few prominent examples. In the following two sections we will therefore try to
give a flavor of the most important quark-related instanton effects.

\subsection{The chiral anomaly\label{axanomsec}}

Behind many essential instanton effects in the light-quark sector lurks the
axial anomaly, i.e. the fact that the (flavor-singlet) axial quark current,
which is conserved in the classical theory, ceases to be so at the quantum
level. In the following we will derive this anomaly and some of its
implications\footnote{A transparent introduction to QCD anomalies (and
low-energy theorems) can be found in \cite{shi91}.}. Also from a more general
point of view, the discussion of anomalies fits well into our context since it
exhibits a particular strength of the semiclassical approximation: all known
anomalies are fully determined by one-loop Feynman diagrams, and as such
appear exclusively at $O\left(  \hbar\right)  $ of the SCA.

To set the stage for the following discussion, let us switch on the fermionic
part of the QCD\ action,%
\begin{equation}
S_{q}\left[  q,\bar{q},G\right]  =\int d^{4}x\bar{q}\left[  i\gamma_{\mu
}\left(  \partial_{\mu}+igG_{\mu}\right)  \right]  q,
\end{equation}
which couples the previously considered gluon fields to light quarks (we omit
the small quark mass term). The Noether procedure and the Dirac equation imply
that this action has a classically conserved axial current, i.e.%
\begin{equation}
\partial_{\mu}\left(  \bar{q}i\gamma_{\mu}\gamma_{5}q\right)  =0.
\end{equation}
(Including the factor $i$ is a convenient convention in Euclidean spacetime.)
Now let us compute the quantum expectation value of this current in the
background of a fixed, classical gauge field\footnote{Note that this is the
first step towards a full quantum calculation, where we would afterwards
integrate over all classical gauge fields. In order to derive the anomaly,
however, it is sufficient to quantize just the fermion sector.} $G_{\mu},$
i.e.%
\begin{equation}
J_{5,\mu}\left(  x|G\right)  =Z^{-1}\left[  G\right]  \int\mathcal{D}%
q\mathcal{D}\bar{q}\left(  \bar{q}i\gamma_{\mu}\gamma_{5}q\right)
e^{-S_{q}\left[  q,\bar{q},G\right]  } \label{cur1}%
\end{equation}
where%
\begin{equation}
Z\left[  G\right]  =\int\mathcal{D}q\mathcal{D}\bar{q}e^{-S_{q}\left[
q,\bar{q},G\right]  }.
\end{equation}

In order to perform the standard Grassmann integration over the quark fields
in (\ref{cur1}), we introduce Grassmann-valued sources $\eta,\bar{\eta}$ and
write%
\begin{equation}
J_{5,\mu}\left(  x|G\right)  =\left.  Z^{-1}\left[  G\right]  \left(
\frac{-\delta}{\delta\eta}i\gamma_{\mu}\gamma_{5}\frac{\delta}{\delta\bar
{\eta}}\right)  \int\mathcal{D}q\mathcal{D}\bar{q}e^{-\int d^{4}x\left[
\bar{q}\mathcal{D}q+\bar{\eta}q+\bar{q}\eta\right]  }\right|  _{\eta,\bar
{\eta}=0} \label{jfi}%
\end{equation}
where we have defined the Euclidean Dirac operator%
\begin{equation}
\mathcal{D}=i\gamma_{\mu}\left[  \partial_{\mu}+igG_{\mu}\left(  x\right)
\right]  .
\end{equation}
With the help of its inverse, the Dirac propagator $S\left(  x,y|G\right)  $
in the background field $G_{\mu}$, defined by
\begin{equation}
\left[  i\gamma_{\mu}\left(  \partial_{x,\mu}+igG_{\mu}\left(  x\right)
\right)  \right]  S\left(  x,y|G\right)  =\delta^{4}\left(  x-y\right)  ,
\end{equation}
we can rewrite the exponent in the integrand of (\ref{jfi}) as
\begin{equation}
\bar{q}\mathcal{D}q+\bar{\eta}q+\bar{q}\eta=\left(  \bar{q}+\bar{\eta
}S\right)  \mathcal{D}\left(  q+S\eta\right)  -\bar{\eta}S\eta
\end{equation}
and, after redefining $q\rightarrow q^{\prime}=q+S\eta$ (this is a constant
shift of the functional variable $q$ which leaves the measure invariant), we
obtain%
\begin{equation}
\int\mathcal{D}q\mathcal{D}\bar{q}e^{-\int d^{4}x\left[  \bar{q}%
\mathcal{D}q+\bar{\eta}q+\bar{q}\eta\right]  }=Z\left[  G\right]  e^{\int
d^{4}xd^{4}y\left[  \bar{\eta}\left(  x\right)  S\left(  x,y|G\right)
\eta\left(  y\right)  \right]  }.
\end{equation}
Inserting this identity into Eq. (\ref{jfi}), we find
\begin{equation}
J_{5,\mu}\left(  x|G\right)  =-N_{f}tr_{\gamma,c}\left\{  i\gamma_{\mu}%
\gamma_{5}S\left(  x,x|G\right)  \right\}  \label{cur2}%
\end{equation}
($N_{f}$ is the number of quark flavors; the trace is over Dirac and color
indices). This expression is still formal: the singular coincidence limit of
the propagator has to be regularized, and we will do so below. In terms of
Feynman graphs, Eq. (\ref{cur2}) states that the expectation value of the
axial current can be calculated from a closed quark loop in the background of
a (classical) gluon field $G$, with one insertion of the axial vertex
$\gamma_{\mu}\gamma_{5}$.

For the further evaluation of (\ref{cur2}) we employ the spectral
representation of $S\left(  x,y|G\right)  $ in terms of the normalized
eigenfunctions $\psi_{n}$ of the Dirac operator $\mathcal{D}$,
\begin{equation}
\mathcal{D}\psi_{n}\left(  x\right)  =\lambda_{n}\psi_{n}\left(  x\right)  ,
\label{diraceq}%
\end{equation}
($\mathcal{D}$ is hermitean ($\gamma_{\mu}^{+}=\gamma_{\mu}$ and $G_{\mu}%
^{+}=G_{\mu}$) so that its spectrum is real) which reads
\begin{equation}
S\left(  x,y|G\right)  =\sum_{n}\frac{\psi_{n}\left(  x\right)  \bar{\psi}%
_{n}\left(  y\right)  }{\lambda_{n}}. \label{specprop}%
\end{equation}
(For the moment we do not worry about vanishing eigenvalues, although they
will play a prominent role later on\footnote{Alternatively, one could
regularize the divergence they cause in (\ref{specprop}) by introducing a
small quark mass.}.) In order to regularize this infinite sum we introduce a
cutoff function
\begin{equation}
f_{\varepsilon}\left(  \lambda_{n}^{2}\right)  =e^{-\varepsilon\lambda_{n}%
^{2}}%
\end{equation}
which suppresses the contributions from large $\left|  \lambda_{n}\right|  $
and goes to unity in the limit $\varepsilon\rightarrow0$ which we are going to
take in the end. Inserted into (\ref{cur2}) this yields%
\begin{equation}
J_{5,\mu}\left(  x|G\right)  =-N_{f}\sum_{n}\frac{\bar{\psi}_{n}\left(
x\right)  i\gamma_{\mu}\gamma_{5}\psi_{n}\left(  x\right)  }{\lambda_{n}%
}e^{-\varepsilon\lambda_{n}^{2}}.
\end{equation}

Let us now obtain the divergence of this current by noting that (\ref{diraceq}%
) implies%
\begin{equation}
\partial_{\mu}\left(  \bar{\psi}_{n}i\gamma_{\mu}\gamma_{5}\psi_{n}\right)
=2\lambda_{n}\left(  \bar{\psi}_{n}\gamma_{5}\psi_{n}\right)  ,
\end{equation}
and leads to%
\begin{equation}
\partial_{\mu}J_{5,\mu}\left(  x|G\right)  =-2N_{f}\sum_{n}\bar{\psi}%
_{n}\gamma_{5}e^{-\varepsilon\lambda_{n}^{2}}\psi_{n}=-2N_{f}\sum_{n}\bar
{\psi}_{n}\gamma_{5}e^{-\varepsilon\mathcal{D}^{2}}\psi_{n}=2N_{f}%
tr_{\gamma,c}\left\langle x\left|  \gamma_{5}e^{-\varepsilon\mathcal{D}^{2}%
}\right|  x\right\rangle \label{specsum}%
\end{equation}
where%
\begin{equation}
\mathcal{D}^{2}=-\left(  \partial_{\mu}+igG_{\mu}\right)  ^{2}-\frac{g}%
{2}\sigma_{\mu\nu}G_{\mu\nu} \label{dopsq}%
\end{equation}
($\sigma_{\mu\nu}=\left(  i/2\right)  \left[  \gamma_{\mu},\gamma_{\nu
}\right]  $).

We can now expand the exponential\footnote{Note that the $G$-dependence of
(\ref{specsum}) originates solely from the exponential since the implicit
$G$-dependence of the $\psi_{n}$ cancels in the sum. (This is because the
matrix element $\left\langle x\left|  \gamma_{5}e^{-\varepsilon\mathcal{D}%
^{2}}\right|  x\right\rangle $ is independent of the basis of functions in
which one may decide to evaluate it.)} in powers of the classical background
field $G$. In Feynman-diagram language, this corresponds to expanding the
quark loop in the number of $G$ insertions. Each additional $G$ insertion
implies an additional quark propagator in the loop and therefore makes the
loop integral converge faster at high momenta. Thus there can be at most a few
UV-divergent terms, associated with the lowest orders of $G$. These terms
require regularization, and it is at this point where the classical axial
symmetry gets broken (if we insist on keeping the vector current conserved)
and the anomaly emerges.

\begin{figure}
[ptb]
\begin{center}
\includegraphics[
height=1.0075in,
width=1.7703in
]%
{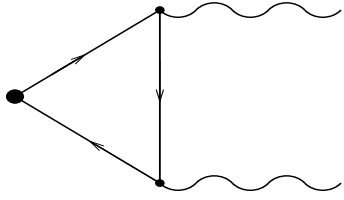}%
\caption{The triangle graph correponding to the UV-divergent quark loop which
generates the chiral anomaly. The bigger blob denotes the insertion of the
axial-vector vertex $\gamma_{\mu}\gamma_{5}$, the smaller one the insertion of
the vector vertex $\gamma_{\mu}$ to which the gluons are attached. }%
\label{triangle}%
\end{center}
\end{figure}

The above reasoning is reflected in our calculation by the fact that only the
leading nonvanishing term in the $G$ expansion of (\ref{specsum}) will survive
the limit $\varepsilon\rightarrow0$ which we are going to take in the end
(since higher-order terms in the expansion of the exponential (\ref{specsum})
also introduce higher orders of $\varepsilon$). We therefore just have to
identify and calculate this single contribution. Since all the terms from the
first part of (\ref{dopsq}) vanish due to $tr_{\gamma}\left(  \gamma
_{5}\right)  =0,$ and since those from the first-order contribution of the
second term vanish due to $tr_{\gamma}\left(  \gamma_{5}\sigma_{\mu\nu
}\right)  =0$, the first nonvanishing contribution originates from the
second-order piece of the $\sigma_{\mu\nu}$-term, i.e.
\begin{equation}
tr_{\gamma,c}\left\langle x\left|  \gamma_{5}e^{-\varepsilon\mathcal{D}^{2}%
}\right|  x\right\rangle =\left\langle x\left|  e^{\varepsilon\partial^{2}%
}\right|  x\right\rangle \left[  \frac{g^{2}\varepsilon^{2}}{8}tr_{\gamma
}\left(  \gamma_{5}\sigma_{\mu\nu}\sigma_{\rho\sigma}\right)  tr_{c}\left(
G_{\mu\nu}G_{\rho\sigma}\right)  +O\left(  \varepsilon^{3}\right)  \right]  ,
\end{equation}
which corresponds to a quark loop with two (vector) interactions with the
background gluon field (cf. Fig. \ref{triangle}).
With
\begin{equation}
tr_{\gamma}\left(  \gamma_{5}\sigma_{\mu\nu}\sigma_{\rho\sigma}\right)
=4\varepsilon_{\mu\nu\rho\sigma},\text{ \ \ \ \ \ \ \ \ \ }\tilde{G}_{\mu\nu
}=\frac{1}{2}\varepsilon_{\mu\nu\rho\sigma}G_{\rho\sigma} \label{dualfs}%
\end{equation}
and
\begin{equation}
\left\langle x\left|  e^{\varepsilon\partial^{2}}\right|  x\right\rangle =\int
d^{4}p\left\langle x|p\right\rangle e^{\varepsilon\partial^{2}}\left\langle
p|x\right\rangle =\int\frac{d^{4}p}{\left(  2\pi\right)  ^{4}}e^{-\varepsilon
p^{2}}=\frac{1}{16\pi^{2}\varepsilon^{2}}%
\end{equation}
(note that this term contains the UV singularity for $\varepsilon\rightarrow
0$), where we have used $\left\langle x|p\right\rangle =\exp\left(  -ip_{\mu
}x_{\mu}\right)  /\left(  2\pi\right)  ^{2}$, we finally obtain (in the limit
$\varepsilon\rightarrow0$)%
\begin{equation}
\partial_{\mu}J_{5,\mu}\left(  x|G\right)  =\frac{g^{2}N_{f}}{8\pi^{2}}%
tr_{c}\left(  G_{\mu\nu}\tilde{G}_{\mu\nu}\right)  . \label{anomeq}%
\end{equation}

This famous result (which can be shown to be independent of the regularization
procedure) expresses the gist of the ``axial anomaly'': although the axial
current is conserved in classical chromodynamics, the corresponding symmetry
is broken at the quantum level, i.e. in QCD, due to the necessity of
regularization and renormalization\footnote{Readers who are familiar with the
original derivation of the anomaly in terms of explicit Feynman graphs will
have noted that we just calculated the celebrated triangle graph Fig.
\ref{triangle} in a somewhat implicit, functional fashion. The functional
approach allowed us to get the desired result without introducing the
machinery of perturbative quantum field theory (i.e. Feynman rules).}.

\subsection{Quark zero modes and index theorem}

The existence of the axial anomaly has crucial implications for fermions in
the background of instanton fields. In order to exhibit them, let us take a
closer look at the Dirac spectrum%
\begin{equation}
\mathcal{D}\psi_{n}\left(  x\right)  =\lambda_{n}\psi_{n}\left(  x\right)
\end{equation}
introduced in the last section. Since the (hermitean) Dirac operator
anticommutes with $\gamma_{5}$,
\begin{equation}
\left\{  \mathcal{D},\gamma_{5}\right\}  =0, \label{g5anticom}%
\end{equation}
each (normalized) eigenfunction $\psi_{n}$ with $\lambda_{n}\neq0$ implies the
existence of another eigenfunction
\begin{equation}
\psi_{-n}\left(  x\right)  \equiv\gamma_{5}\psi_{n}\left(  x\right)
\end{equation}
with eigenvalue%
\begin{equation}
\lambda_{-n}=-\lambda_{n},\text{ \ \ \ \ \ \ \ \ }(n>0).
\end{equation}

In other words, nonvanishing eigenvalues appear in pairs with opposite signs.
The eigenfunctions $\psi_{0,k}$, corresponding to the remaining eigenvalues
$\lambda_{k}=0$, are called zero modes. The zero modes can be made to
diagonalize $\gamma_{5}$ (due to (\ref{g5anticom})) by transforming to the
chiral basis%
\begin{equation}
\psi_{0\pm,k}=\frac{1}{2}\left(  1\pm\gamma_{5}\right)  \psi_{0,k}%
\end{equation}
so that%
\begin{equation}
\gamma_{5}\psi_{0\pm,k}=\pm\psi_{0\pm,k}.
\end{equation}
Since, finally, the QCD Dirac operator is flavor-independent, i.e.
\begin{equation}
\left[  \mathcal{D},\mathcal{T}_{a}\right]  =0,
\end{equation}
(where $\mathcal{T}_{a}$ are the generators of the flavor group) each zero
mode comes in $N_{f}$ copies (one of each flavor).

The zero modes $\psi_{0,\pm}$ are normalizable and inherit several
characteristic properties of the instanton, including the localization in
space and time and the coupling of spin and color. In the so-called singular
gauge, their explicit form \cite{tho76} is
\begin{equation}
\psi_{0,\pm}(x)=\frac{\rho}{\pi}\frac{1\pm\gamma_{5}}{(r^{2}+\rho^{2})^{3/2}%
}\,\frac{\rlap{/}  {r}}{r}\,u.
\end{equation}
(The spin-color coupling matrix $u$ is defined by $(\vec{\sigma}+\vec{\tau
})\,u=0$, and $r=x-x_{0}$.) For our subsequent discussion only qualitative
properties of the zero modes will matter, though.

The two main characteristics of the Dirac spectrum established above - paired
eigenvalues of opposite sign and chiral zero modes - have profound
consequences. One of those becomes explicit when one integrates the anomaly
equation (\ref{anomeq}) over spacetime and uses Eq. (\ref{specsum})
(multiplied by a factor $1/\left(  2N_{f}\right)  $) to write
\begin{align}
Q &  \equiv\frac{g^{2}}{16\pi^{2}}\int d^{4}xtr_{c}\left(  G_{\mu\nu}\tilde
{G}_{\mu\nu}\right)  \label{topcharge}\\
&  =-\sum_{n}e^{-\varepsilon\lambda_{n}^{2}}\int d^{4}x\bar{\psi}_{n}%
\gamma_{5}\psi_{n}.\label{qquark}%
\end{align}
Above we have defined the ``topological charge'' $Q$ whose meaning will be
clarified below. Since the eigenfunctions $\psi_{n}$ and $\gamma_{5}\psi_{n}$
of $\mathcal{D}$ have different eigenvalues ($\lambda_{n}$ and $-\lambda_{n}$)
and are thus orthogonal to each other, the integral in (\ref{qquark}) vanishes
for all $n$ with $\lambda_{n}\neq0,$ leaving us with an integral over the zero
modes only:%
\begin{align}
Q &  =-\sum_{k}\int d^{4}x\bar{\psi}_{0,k}\gamma_{5}\psi_{0,k}\\
&  =\sum_{m=1}^{n_{-}}\int d^{4}x\bar{\psi}_{0-,m}\psi_{0-,m}-\sum
_{m=1}^{n_{+}}\int d^{4}x\bar{\psi}_{0+,m}\psi_{0+,m}%
\end{align}
where $n_{\pm}$ is the number of right- (left-) handed zero modes per flavor
in the given background gluon field. Since the eigenfunctions of $\mathcal{D}$
(including the zero modes) are normalized, we finally obtain%
\begin{equation}
Q=n_{-}-n_{+}.\label{atsing}%
\end{equation}

The above formula, relating a property $Q\left[  G\right]  $ of the gauge
field to the number of unpaired quark zero modes of the Dirac operator
(\ref{diraceq}) in its background, is a special case of the celebrated
Atiyah--Singer index theorem \cite{ati68,nas91} for the Euclidean Dirac
operator. It implies that $Q$ must be integer and thus cannot change under
smooth variations of the background gluon field. Such a behavior is
characteristic for topological properties of the gluon field and prompts us to
have a closer look at those in the next section.

Before doing so, however, we should mention a few phenomenological
consequences of the instanton-induced zero modes in the quark spectrum. The
integrated anomaly equation (see Eq. (\ref{anomint}) below) reveals that the
$Q=1$ instanton is accompanied by the appearance of $2N_{f}$ units of axial
charge. This can happen either by creating $N_{f}$ left-handed quarks and
annihilating $N_{f}$ right-handed antiquarks in zero-mode states or, in the
alternative (crossed) time-ordering, by helicity flips of quarks propagating
in the zero-mode state. Such processes are described by non-local $2N_{f}%
$-quark interactions known as 't Hooft vertices \cite{tho76}. They generate
the dominant instanton effects in the light-quark sector and crucial
interactions between instantons. As a reflection of the anomaly, the 't Hooft
vertices manifestly break the axial $U\left(  1\right)  $ symmetry of the QCD
Lagrangian with $m_{q}=0$ and thereby resolve (at least qualitatively) the
so-called ``$U(1)$ problem'', i.e. they explain why the $\eta^{\prime}$ meson
has almost twice the mass of the $\eta$ meson and cannot be considered a
(quasi) Goldstone boson. More generally, the quark interactions mediated by 't
Hooft vertices have a characteristic channel dependence which makes them,
e.g., strongly attractive in spin-0 quark-antiquark channels but mute in
vector and axial-vector channels. These patterns are reflected in the
instanton contributions to hadronic amplitudes. We will come back to this
issue in the final lecture.

\section{Instanton topology}

So far, we have established two important implications of the axial anomaly:
first, in order for the anomaly to have physical significance there must be
gauge fields with non-vanishing topological charge $Q$, and second, such
fields imply the existence of unpaired zero modes in the spectrum of the
associated Dirac operator. Our next tasks will be to better understand the
topological charge, the gluon fields which carry it, and the impact of both on
the vacuum structure.

\subsection{Topological charge}

We start by noting that the integrand of the topological charge $Q$ (cf.
(\ref{topcharge})) can be written as a total derivative
\begin{equation}
\frac{g^{2}}{16\pi^{2}}tr_{c}\left(  G_{\mu\nu}\tilde{G}_{\mu\nu}\right)
=\partial_{\mu}K_{\mu} \label{topcur}%
\end{equation}
(we do not need the explicit expression for the ``Chern-Simons current''
$K_{\mu}$ here), which implies that only fields with nontrivial behavior at
the spacetime boundary $x^{2}\rightarrow\infty$ (typical for topologically
active fields) can carry a finite $Q$. Furthermore, the index theorem
(\ref{atsing}) implies that $Q$ can only take integer values and cannot
therefore change under continuous deformations of the gluon background field.
Thus, besides the homotopy classification of the pure gauges encountered in
Section \ref{vactop}, gluon fields carry another topological property
specified by their $Q$-values. As we have seen above, fields with nonvanishing
$Q$ ``activate'' the anomaly by generating quark zero-modes and are therefore,
via Eq. (\ref{anomeq}), associated with a non-conservation of the axial charge
$q_{5}\left(  \tau\right)  =\int d^{3}xJ_{5,4}\left(  x|G\right)  $:
\begin{align}
Q  &  =\frac{g^{2}}{16\pi^{2}}\int d^{4}xtr_{c}\left(  G_{\mu\nu}\tilde
{G}_{\mu\nu}\right)  =\frac{1}{2N_{f}}\int d^{4}x\partial_{\mu}J_{5,\mu
}\left(  x|G\right) \\
&  =\frac{1}{2N_{f}}\int d^{3}x\int d\tau\left[  \partial_{4}J_{5,4}\left(
x|G\right)  +\vec{\nabla}\vec{J}_{5}\left(  x|G\right)  \right]  =\left.
\frac{1}{2N_{f}}\int d^{3}xJ_{5,4}\left(  x|G\right)  \right|  _{\tau=-\infty
}^{\tau=\infty}\\
&  =\frac{1}{2N_{f}}\Delta q_{5}. \label{anomint}%
\end{align}
where%
\begin{equation}
\Delta q_{5}=q_{5}\left(  \tau=\infty\right)  -q_{5}\left(  \tau
=-\infty\right)
\end{equation}
is the amount of axial charge created by the gluon field with topological
charge $Q$.

We will now show that an integer value of $Q$ can be assigned to any gluon
field with finite (Euclidean) action (only those are relevant for the SCA),
and that this value has a transparent physical interpretation. For the action
(\ref{ymact}) to be finite, the field strength $G_{\mu\nu}$ has to be square
integrable, i.e. it must vanish towards the boundary $S^{3}(x^{2}=\infty)$ of
Euclidean spacetime, i.e. for $\left|  x\right|  \rightarrow\infty$, faster
than $1/x^{2}$. Now we know from Section \ref{vactop} that the Euclidean
action has its absolute minimum at $G_{\mu\nu}=0$, i.e. for pure gauges. As a
consequence, finite-action fields satisfy the boundary condition
\begin{equation}
\lim_{x^{2}\rightarrow\infty}G_{\mu}=\frac{-i}{g}U^{\left(  n\right)
}(x)\,\partial_{\mu}U^{\left(  n\right)  -1}(x) \label{bcinst}%
\end{equation}
where $U^{\left(  n\right)  }$ is an element of the gauge group ${\mathcal{G}%
}$ with winding number $n$. (Not surprisingly, instantons are of this type
(cf. (\ref{bc3}), (\ref{bc4})) as they interpolate between pure gauges of
different winding number for $\tau\rightarrow\pm\infty$.)

The essential consequence of Eq. (\ref{bcinst}) is that it implies a
time--independent topological classification of all finite--action gauge
fields. This classification is based on (but not equal to) the homotopy
analysis of the gauge group elements $U^{\left(  n\right)  }$ which we went
through in Section \ref{vactop}. Indeed, Eq. (\ref{bcinst}) implies that any
finite--action gauge field defines a map $S^{3}(x^{2}=\infty)\rightarrow
{\mathcal{G}}$, and such maps fall into disjoint homotopy classes according to
the third homotopy group\footnote{In mathematical terms, $\pi_{3}%
({\mathcal{G}})$ classifies principal ${\mathcal{G}}$ fibre bundles over
$S^{4}$ \cite{nas91}.} $\pi_{3}({\mathcal{G}})$ of ${\mathcal{G}}$. For
${\mathcal{G}}=SU(N)$ with $N\geq2$, which includes QCD, we have $\pi
_{3}(SU(N))=Z$ as in Section \ref{vactop} while, e.g., (non-compact) QED has
$\pi_{3}(U(1))=1$. As a consequence, all finite-action gluon fields in
Euclidean QCD fall into topologically distinct equivalence classes labeled by
the topological charge $Q$ (which is technically the ``Pontryagin index'' of
the gauge field)\footnote{At this point, it might be useful to emphasize the
differences between the topological classification discussed here and that of
the pure gauges in section \ref{vactop}. The winding number $n$ of the latter
classifies maps from compactified space $S_{space}^{3}$ (on which static gauge
transformations with unit-value at the boundary are defined) into the gauge
group, while $Q$ characterizes maps from the boundary $S^{3}$ of Euclidean
spacetime into the gauge group.}.

For instantons, the topological charge $Q$ has a particularly transparent
meaning: it is just the difference between the winding numbers $n$ and $m$ of
the pure gauges between which the instanton interpolates,
\begin{equation}
Q=m-n. \label{qdif}%
\end{equation}
Instead of proving this expression in general (which would not be difficult),
let us check it for the explicit instanton solution (\ref{yminst}) found
above. We first calculate the classical instanton action (from which we know
already that it is finite and that it is minimal in its $Q$-sector) by
plugging the field strength
\begin{equation}
G_{I,\mu\nu}\left(  x\right)  =\frac{-4\rho^{2}}{g}\frac{\eta_{a\mu\nu}t_{a}%
}{[\left(  x-x_{0}\right)  ^{2}+\rho^{2}]^{2}} \label{instfs}%
\end{equation}
(obtained by inserting the expression (\ref{yminst}) for the instanton into
the Yang-Mills field strength tensor (\ref{fstrength})) into the Yang-Mills
action (\ref{ymact}). The result is
\begin{equation}
S_{I}=\frac{1}{2}\int d^{4}xtr_{c}\left(  G_{I,\mu\nu}G_{I,\mu\nu}\right)
=\frac{8\pi^{2}}{g^{2}}.
\end{equation}
We note that the classical action is additive for multi-instantons and
multi-anti-instantons, $S_{Q}=\left|  Q\right|  \times S_{I}$ (but not for
combinations of both), and that it does not depend on the instanton's
collective coordinates $z_{\mu}$, $U$ and $\rho$. The latter is a direct
consequence of the translation, gauge, and scale invariance of classical
Yang-Mills theory. At the quantum level, the scale invariance gets broken by
the trace anomaly (yet another variety of anomaly) and the effective quantum
action becomes $\rho$-dependent. This generates a $\rho$-dependent weight
$n\left(  \rho\right)  $ in functional integrals which can be identified with
the instanton size density introduced at the end of Section \ref{instsoln}.

Having calculated the instanton action, we can easily check whether the
instanton has indeed $Q=1$. In fact, this becomes trivial once we have
established another essential instanton property, namely the self-duality of
its field strength. We now turn to this issue.

\subsection{Self duality}

Inspection of the instanton's field strength (\ref{instfs}) reveals that it
depends on the Lorentz indices only via the 't Hooft symbol $\eta_{a\mu\nu}$.
Moreover, the expression (\ref{thsymb}) for $\eta_{a\mu\nu}$ shows that it is
self-dual (recall the definition (\ref{dualfs})), i.e.
\begin{equation}
\tilde{\eta}_{a\mu\nu}=\frac{1}{2}\varepsilon_{\mu\nu\rho\sigma}\eta
_{a\rho\sigma}=\eta_{a\mu\nu},
\end{equation}
which directly carries over to the field strength of the instanton:%
\begin{equation}
G_{I,\mu\nu}=\tilde{G}_{I,\mu\nu}. \label{sdeq}%
\end{equation}
Note that self-duality is a gauge invariant property. It implies $E_{i}%
^{a}=B_{i}^{a}$, i.e. the equality of the chromoelectric and chromomagnetic
fields (\ref{ebfields}), and yields immediately
\begin{equation}
Q_{I}=\frac{g^{2}}{16\pi^{2}}\int d^{4}xtr_{c}\left(  G_{I,\mu\nu}\tilde
{G}_{I,\mu\nu}\right)  =\frac{g^{2}}{16\pi^{2}}\int d^{4}xtr_{c}\left(
G_{I,\mu\nu}G_{I,\mu\nu}\right)  =\frac{g^{2}}{8\pi^{2}}S_{I}=1.
\end{equation}

The anti-instanton is anti-selfdual, i.e. $G_{\bar{I},\mu\nu}=-\tilde
{G}_{\bar{I},\mu\nu}$, and thus has $Q_{\bar{I}}=-1$. (Anti-) Self-duality is
a mathematically very powerful property. Self-dual fields automatically
satisfy, for instance, the Yang-Mills equation (\ref{ymeq}). This is
immediately obvious when combining the self-duality equation (\ref{sdeq}) with
the Bianchi identity%
\begin{equation}
D_{\mu}\tilde{G}_{\mu\nu}=0 \label{bianchi}%
\end{equation}
which holds for any gauge field, due to the form (\ref{fstrength}) of the
field strength tensor. In practice, it is often advantageous to deal with Eq.
(\ref{sdeq}), instead of with the Yang-Mills equation itself, since it is of
first order. As a matter of fact, the original instanton solution was found in
\cite{bel75} by solving (\ref{sdeq}). Similarly, in the quantum mechanical
example of Section \ref{qminsoln} we derived the instanton solution by solving
the corresponding first-order equation (\ref{eneq}). Self-duality has even
more profound mathematical implications, e.g. in Donaldson theory. It can be
shown, incidentally, that all minima of the Euclidean Yang-Mills action with
$S_{E}<\infty$ are (anti-) self-dual, and that they correspond to (multi-)
instanton solutions \cite{shi94}.

The index theorem (\ref{atsing}) implies that there must be one unpaired,
left-handed zero-mode per flavor in a gluon field with $Q=1$. This result can
be sharpened for (anti-) selfdual fields with $G_{\mu\nu}^{\left(  \pm\right)
}=\pm\tilde{G}_{\mu\nu}^{\left(  \pm\right)  }$. In order to decouple left-
and right-handed modes we consider the eigenvalue equation of the iterated
Dirac operator (cf. Eq. (\ref{dopsq})),
\begin{equation}
\mathcal{D}^{2}\psi_{n}=\left[  -\left(  \partial_{\mu}+igG_{\mu}\right)
^{2}-\frac{g}{2}\sigma_{\mu\nu}G_{\mu\nu}\right]  \psi_{n}=\lambda_{n}^{2}%
\psi_{n}. \label{direveq}%
\end{equation}
For (anti-) selfdual fields the second term in the square bracket can be
rewritten with the help of the identity
\begin{equation}
\sigma_{\mu\nu}=-\gamma_{5}\tilde{\sigma}_{\mu\nu}%
\end{equation}
as
\begin{equation}
\sigma_{\mu\nu}G_{\mu\nu}^{\left(  \pm\right)  }=\mp\sigma_{\mu\nu}G_{\mu\nu
}^{\left(  \pm\right)  }\gamma_{5}=\sigma_{\mu\nu}G_{\mu\nu}^{\left(
\pm\right)  }\frac{1}{2}\left(  1\mp\gamma_{5}\right)  .
\end{equation}
Restricted to quark zero modes and to (anti-) selfdual gluon fields, Eq.
(\ref{direveq}) then becomes
\begin{equation}
\mathcal{D}^{2}\psi_{0}=\left[  -\left(  \partial_{\mu}+igG_{\mu}^{\left(
\pm\right)  }\right)  ^{2}-\frac{g}{2}\sigma_{\mu\nu}G_{\mu\nu}^{\left(
\pm\right)  }\frac{1}{2}\left(  1\mp\gamma_{5}\right)  \right]  \psi_{0}=0.
\end{equation}
For selfdual fields, this implies%
\begin{equation}
-\left(  \partial_{\mu}+igG_{\mu}^{\left(  +\right)  }\right)  ^{2}\psi
_{0,+}=0,
\end{equation}
and since $-\left(  \partial_{\mu}+igG_{\mu}^{\left(  +\right)  }\right)
^{2}$ has a positive spectrum we conclude that in a selfdual field $\psi
_{0,+}=0$ and thus $n_{+}=0.$ Then the index theorem (\ref{atsing}) tells us
that there exists exactly one left-handed (right-handed) zero mode per flavor
in the background of an instanton (anti-instanton), and none of the opposite chirality.

\section{The angle $\theta$\label{theta}}

In the following sections we are going to elaborate on the instanton-induced
vacuum structure and the $\theta$-angle which we have encountered in Section
\ref{vactop}. Besides the semiclassical approach and its description of vacuum
tunneling discussed above, there are several other and partially complementary
angles from which one can study the emergence of $\theta$ and the implied
superselection rule for non-communicating (i.e. physically independent) Fock
spaces. Below we will consider two such approaches in more detail. Both of
them lead to new insights into the workings of the QCD vacuum, and both of
them allow the $\theta$ structure to be derived as an exact, formal result,
i.e. without recourse to the SCA or, in fact, any approximation. This is
reassuring since the conditions under which the SCA works in QCD depend on the
physical situation and are often difficult to check in the strongly coupled
regime. Possibilities to learn about the consequences of the nontrivial
gauge-group topology\ without relying on the SCA are therefore welcome.

Up to now we have approached QCD in the path integral framework, which
provides the most natural setting for the SCA. In the following we are going
beyond the SCA, with the aim of augmenting our earlier insights into the theta
vacua
\begin{equation}
\left|  \theta\right\rangle =\mathcal{N}\sum_{n}e^{-i\theta n}\left|
n\right\rangle \label{thetvac2}%
\end{equation}
as a familiy of (up to a phase) gauge-invariant ground states. Moreover, we
will be interested in the general behavior of physical states under
topologically nontrivial gauge transformations. It is therefore suggestive to
chose a quantization procedure in which the Fock states play a more explicit
role. Canonical quantization in the Schr\"{o}dinger picture will be used for
this purpose in Section \ref{thanglesec}. Although instantons are
indispensable for elucidating the tunneling mechanism behind the $\theta$
structure, they will not appear explicitly in this section. In the subsequent
Section \ref{thangleclus} we will return to the path integral and look at the
$\theta$ vacuum from yet another perspective, namely that of the cluster
decompositon principle. On the way, we will learn how to properly treat the
topological charge $Q$ of the instantons and other topologically nontrivial
gauge fields in functional integrals.

\subsection{$\theta$ structure from Gau\ss ' law\label{thanglesec}}

We begin this section with a brief summary of the\ Hamiltonian formulation of
QCD in Minkowski space\footnote{A potential disadvantage of the Hamiltonian
formulation is that explicit Lorentz invariance is lost because one has to
single out a spacelike reference surface on which to impose the canonical
commutation relations.} and the canonical quantization of gauge theories.
Several alternative methods for this purpose have been developed\footnote{In
QED, it can be shown that all quantization schemes lead to the same results,
and one expects the same to hold for QCD.}, including e.g.
constraint-quantization \`{a} la Dirac and the BRST\ formalism. In our context
quantization in Weyl (or temporal) gauge,
\begin{equation}
G_{0}^{a}=0,\text{ \ \ \ \ \ \ \ }G_{0i}^{a}=\partial_{0}G_{i}^{a}%
,\label{gcond}%
\end{equation}
which we have already encountered in Section \ref{vactop}, turns out to be the
most convenient. To simplify the discussion, we restrict ourselves to pure
Yang-Mills theory (i.e. QCD without quarks) with the gauge-fixed Lagrangian%
\begin{equation}
\mathcal{L}=-\frac{1}{4}G_{\mu\nu}^{a}G^{a,\mu\nu}=\frac{1}{2}\left(
\partial_{0}G_{i}^{a}\right)  ^{2}-\frac{1}{4}G_{ij}^{a}G_{ij}^{a}%
\end{equation}
since the topological effects to be discussed are rooted in the gluon sector.

In terms of the color electric and magnetic fields
\begin{equation}
E_{i}^{a}=G_{i0}^{a},\,\text{\ \ \ \ \ \ \ \ }B_{i}^{a}=-\frac{1}%
{2}\varepsilon_{ijk}G_{jk}^{a}.\text{ \ \ \ \ \ \ }\left(  G_{ij}%
^{a}=-\varepsilon_{ijk}B_{k}^{a}\right)
\end{equation}
the gauge-fixed Lagrangian (in Minkowski space) is just the difference between
the ``kinetic'' and ``potential'' gluon energy densities,%
\begin{equation}
\mathcal{L}=\frac{1}{2}\left(  \vec{E}_{a}^{2}-\vec{B}_{a}^{2}\right)  .
\end{equation}
The canonical variables, in Weyl gauge, are the spacial components $G_{i}^{a}$
of the gauge field. In terms of those and the associated conjugate momenta%
\begin{equation}
\pi_{i}^{a}=\frac{\delta\mathcal{L}}{\delta\left(  \partial_{0}G_{i}%
^{a}\right)  }=G_{0i}^{a}=-E_{i}^{a}%
\end{equation}
the Hamiltonian reads
\begin{equation}
H=\int d^{3}x\left(  \pi_{i}^{a}\partial_{0}G_{i}^{a}-\mathcal{L}\right)
=\frac{1}{2}\int d^{3}x\left(  \vec{E}_{a}^{2}+\vec{B}_{a}^{2}\right)  .
\label{ymham}%
\end{equation}

Note that, due to the antisymmetry of the field strength, there is no momentum
conjugate to $G_{0}^{a}$, i.e. $\pi_{0}^{a}=\delta\mathcal{L}/\delta\left(
\partial_{0}G_{0}^{a}\right)  \equiv0$. The corresponding Yang-Mills (i.e.
Euler-Lagrange) equation, Gau\ss ' law%
\begin{equation}
D_{i}^{ab}G_{0i}^{b}=-\vec{D}^{ab}\vec{E}^{b}=0,
\end{equation}
is therefore just a constraint. In other words, $G_{0}$ does not propagate
since no kinetic term for it appears in the Lagrangian. This is the root of
several complications in the canonical quantization of gauge theories. One
finds, in particular, that Gau\ss ' law, as an equation at fixed time, does
not appear as one of Hamilton's equations. How should one then implement the
full Yang-Mills dynamics in the canonical formulation? Certainly not by simply
adding Gau\ss ' law as an operator equation: this would be inconsistent
because it does not commute with the canonical variables. Instead, it must be
imposed as a constraint on the Fock space, restricting the physical states to
those eigenstates of $H$ which satisfy
\begin{equation}
D_{i}^{ab}G_{0i}^{b}\left|  \psi\right\rangle =0.
\end{equation}
(This is analogous to how one eliminates the non-physical ghost states with
negative norm, e.g., in Gupta-Bleuler quantization of QED or in the covariant
quantization of strings.)

In order to get a more explicit representation of the Fock space, we will now
adopt the Schr\"{o}dinger picture in which the physical states become
functionals $\Psi\left[  \mathbf{G}\right]  $ of the canonical variables, i.e.
of the spacial components of the gauge potential. Accordingly, the
$\Psi\left[  \mathbf{G}\right]  $ are eigenfunctionals of the
Hamiltonian\footnote{The canonical variables are now the classical fields
$G_{i}^{a}\left(  \vec{x}\right)  $ and the corresponding momenta
\begin{equation}
\pi_{i}^{a}\left(  \vec{x}\right)  =-E_{i}^{a}\left(  \vec{x}\right)
=-i\frac{\delta}{\delta G_{i}^{a}\left(  \vec{x}\right)  },
\end{equation}
so that the usual canonical commutation relations are satisfied. All other
(time-independent) operators, including the Hamiltonian, can be constructed
from the $G_{i}^{a}$ and $\pi_{i}^{a}$.},%
\begin{equation}
H\Psi\left[  \mathbf{G}\right]  =E\Psi\left[  \mathbf{G}\right]  ,
\end{equation}
and satisfy the Gau\ss \ law constraint%
\begin{equation}
\left(  D_{i}^{ab}G_{0i}^{b}\right)  \Psi\left[  \mathbf{G}\right]
=0.\label{gaussonfunc}%
\end{equation}

Let us now establish the appearance of the $\theta$ angle as a consequence of
the nontrivial topology of the gauge group $G$ (i.e. $\pi_{3}\left(  G\right)
=Z\neq1$) in this framework \cite{jac80}. The unitary gauge transformations
$U$ with%
\begin{equation}
G_{\mu}\rightarrow\text{ }^{U}G_{\mu}\equiv U\left(  x\right)  \left[
\frac{1}{ig}\partial_{\mu}+G_{\mu}\right]  U^{-1}\left(  x\right)
\label{gtrf}%
\end{equation}
are restricted in Weyl gauge to those ``residual'' ones which leave the gauge
condition (\ref{gcond}) intact, i.e. to the time-independent (and in this
sense global) transformations%
\begin{equation}
U=e^{i\omega^{a}\left(  \vec{x}\right)  t^{a}}=U\left(  \omega^{a}\left(
\vec{x}\right)  \right)  ,
\end{equation}
where\ the parameters $\omega^{a}$ specify a given group element. The
gauge-fixed Hamiltonian (\ref{ymham}) is still invariant under those
transformations. This implies, after quantization, that the Hamilton operator
commutes with the operators $\mathcal{U}\left(  \omega\left(  \vec{x}\right)
\right)  $ which furnish a unitary representation of the group on the
$\Psi\left[  \mathbf{G}\right]  $:%
\begin{equation}
\left[  H,\mathcal{U}\left(  \omega\left(  \vec{x}\right)  \right)  \right]
=0.
\end{equation}

Thus we choose the physical states $\Psi\left[  \mathbf{G}\right]  $ to be
simultaneously eigenstates of $H$ and $\mathcal{U}$. At first, this might not
seem necessary since the operator $D_{i}^{ab}G_{0i}^{b}$ appearing in Gau\ss '
law is the generator of infinitesimal gauge transformations\footnote{Note that
for small $\omega$ the gauge transformation (\ref{gtrf}) reduces to
\begin{equation}
G_{\mu}\rightarrow\text{ }G_{\mu}+\frac{1}{ig}D_{\mu}\omega+O\left(
\omega^{2}\right)  .
\end{equation}
},
\begin{equation}
i\left[  \int d^{3}x^{\prime}\omega^{a}\left(  \vec{x}^{\prime}\right)
\frac{1}{ig}D_{i}^{ab}G_{0i}^{b}\left(  \vec{x}^{\prime}\right)  ,G_{j}\left(
\vec{x}\right)  \right]  =\frac{1}{ig}D_{i}^{ab}\omega^{b}\left(  \vec
{x}\right)  =\delta_{\omega}G_{j}\left(  \vec{x}\right)  .
\end{equation}
Therefore, Eq. (\ref{gaussonfunc}) implies that ``small'' gauge
transformations, defined to be those which can be built from (up to infinitely
many) infinitesimal ones and are thus continuously connected to the unit in
the gauge group, leave the physical states invariant. However, we have seen in
Section \ref{vactop} that not all time-independent gauge transformations in
QCD belong to this class. Those which do not, the so-called ``large'' (i.e.
topologically nontrivial) gauge transformations, could change the phase of the
physical states.

Let us study this issue in more detail. As we have discussed in Section
\ref{vactop}, there is in fact an enumerably infinite number of homotopy
classes of gauge transformations (i.e. of the map $S^{3}\rightarrow S^{3}$)
which are labeled by an integer $n$. Members of the classes with $n\neq0$
cannot be continuously connected to the unit element of the group. The classic
example of a gauge transformation with $n=1$ (in the group $SU\left(
2\right)  $ which is embedded in the color gauge group $SU\left(  3\right)  $)
is the ``hedgehog''%
\begin{equation}
U^{\left(  1\right)  }\left(  \vec{x}\right)  =\exp\left(  \frac{i\pi
x^{a}\tau^{a}}{\sqrt{x^{2}+c^{2}}}\right)
\end{equation}
where $c$ is a real constant. Note that the corresponding $\omega_{\left(
1\right)  }^{a}\left(  \vec{x}\right)  =2\pi x^{a}/\sqrt{x^{2}+c^{2}}$ do not
vanish at spacial infinity ($\left|  \vec{x}\right|  \rightarrow\infty$), that
they contain a singularity (here a branch cut), and that the limiting value
$U^{\left(  1\right)  }\left(  \vec{x}\right)  \overset{\left|  \vec
{x}\right|  \rightarrow\infty}{\longrightarrow}-1$ is angle-independent. This
is characteristic for topologically nontrivial gauge transformations since the
exponential can then neither be smoothly deformed to the unit element nor
expanded for all $\vec{x}$. A representative of the class $n$ can be
immediately obtained from $U^{\left(  1\right)  }$ by using the additivity of
the winding number under group multiplication\footnote{More generally, any
gauge transformation $U=\exp\left[  if\left(  \vec{x}^{2}\right)  \hat{x}%
^{a}\tau^{a}\right]  $ with $f\left(  0\right)  =0$ and $f\left(
\infty\right)  =n\pi$ has winding number $n$.},
\begin{equation}
U^{\left(  n\right)  }\left(  \vec{x}\right)  =\left[  U^{\left(  1\right)
}\left(  \vec{x}\right)  \right]  ^{n}=\exp\left(  \frac{in\pi x^{a}\tau^{a}%
}{\sqrt{x^{2}+c^{2}}}\right)  .
\end{equation}

Thus, although all physical states are annihilated by Gau\ss ' law,
$\mathcal{U}$ can have nontrivial eigenvalues for large gauge transformations
with $n\neq0$. (An explicit example is given in Appendix \ref{ginvlargesmall}%
.) Since the eigenvalues of unitary operators must be phases, we then have
\begin{equation}
\mathcal{U}^{\left(  n\right)  }\left(  \omega\left(  \vec{x}\right)  \right)
\Psi\left[  \mathbf{G}\right]  =\Psi\left[  ^{U^{\left(  n\right)  }\left(
\omega\left(  \vec{x}\right)  \right)  }\mathbf{G}\right]  =e^{i\beta
_{n}\left(  \omega\left(  \vec{x}\right)  \right)  }\Psi\left[  \mathbf{G}%
\right]  .\label{gtrfschroed}%
\end{equation}
Concatenating two gauge transformations, we furthermore obtain
\begin{equation}
\mathcal{U}^{\left(  n\right)  }\left(  \omega\right)  \mathcal{U}^{\left(
m\right)  }\left(  \chi\right)  \Psi\left[  \mathbf{G}\right]  =\Psi\left[
^{U^{\left(  n\right)  }U^{\left(  m\right)  }}\mathbf{G}\right]
=\mathcal{U}^{\left(  n+m\right)  }\left(  \xi\right)  \Psi\left[
\mathbf{G}\right]
\end{equation}
(where the group parameters $\omega$, $\chi$, and $\xi$ depend on $\vec{x}$),
and with (\ref{gtrfschroed}) this implies
\begin{equation}
\beta_{n}\left(  \omega\right)  +\beta_{m}\left(  \chi\right)  =\beta
_{n+m}\left(  \xi\right)  .\label{addlaw1}%
\end{equation}
However, the transformation behavior of $\Psi$ under gauge transformations is
in fact even simpler. The existence of the inverse in the gauge group allows
us to write the unit element as $1=U_{0}U_{0}^{-1}$ for any $U_{0}%
\in\mathcal{G}$ with $n=0$, so that%
\begin{equation}
\mathcal{U}^{\left(  n\right)  }\left(  \omega\right)  \Psi\left[
\mathbf{G}\right]  =\mathcal{U}^{\left(  0\right)  }\left(  \chi\right)
\mathcal{U}^{\left(  n\right)  }\left(  \omega^{\prime}\right)  \Psi\left[
\mathbf{G}\right]  =e^{i\beta_{n}\left(  \omega^{\prime}\right)  }\Psi\left[
\mathbf{G}\right]
\end{equation}
where $\mathcal{U}^{\left(  n\right)  }\left(  \omega^{\prime}\right)
=\left[  \mathcal{U}^{\left(  0\right)  }\left(  \chi\right)  \right]
^{-1}\mathcal{U}^{\left(  n\right)  }\left(  \omega\right)  $. Together with
(\ref{gtrfschroed}) we then have
\begin{equation}
\beta_{n}\left(  \omega\right)  =\beta_{n}\left(  \omega^{\prime}\right)
\end{equation}
and since $\omega^{\prime}$ can be chosen independently of $\omega$ by
selecting the appropriate $\chi$, we conclude that the phases $\beta$ cannot
depend on details of the gauge transformation (including its $\vec{x}%
$-dependence) but only on its winding number. As a consequence, (\ref{addlaw1}%
) simplifies to
\begin{equation}
\beta_{n}+\beta_{m}=\beta_{n+m}%
\end{equation}
which finally implies%
\begin{equation}
\beta_{n}=n\theta
\end{equation}
with an arbitrary, real and state-independent angle $\theta\in\left[
0,2\pi\right]  $. Altogether, we have therefore established%
\begin{equation}
\mathcal{U}^{\left(  n\right)  }\left(  \omega\right)  \Psi\left[
\mathbf{G}\right]  =e^{in\theta}\Psi\left[  \mathbf{G}\right]  \label{psitrf}%
\end{equation}
for any $\omega$ and for any physical state (not just the vacuum). The angle
$\theta$, induced by the topological properties of the\ gauge group (or, more
accurately, the residual, time-independent gauge transformations which
preserve the Weyl gauge), has now appeared in a more general fashion than in
our previous discussion based on the SCA.

In order to link the transformation behavior (\ref{psitrf}) to our
semiclassical tunneling picture for the $\theta$ vacuum (and thus to
instantons), we concentrate on the vacuum functional $\Psi_{0}\left[
\mathbf{G}\right]  $ which is characterized by the smallest energy eigenvalue
in the physical Fock space. Furthermore, we recall from Section \ref{vactop}
that, while the vacuum gauge fields $G_{\mu}^{\left(  m\right)  }$ before and
after a gauge transformation $U^{\left(  n\right)  }$ can be continuously
deformed into each other and are gauge-equivalent, the configurations
encountered at intermediate stages of the deformation are generally not. If
the initial and final fields correspond to the vacuum state, then at least
some of the intermediate states must have a larger energy expectation value
and thus form ``potential'' barriers. Those are penetrated by the
instanton-mediated vacuum tunneling processes discussed in Section
\ref{instsoln}. And indeed, the resulting semiclassical vacuum state
(\ref{thetvac2}) of Bloch type realizes the above transformation law
(\ref{psitrf}):
\begin{equation}
U^{\left(  n\right)  }\left|  \theta\right\rangle =\mathcal{N}\sum_{m=-\infty
}^{\infty}e^{-i\theta m}U^{\left(  n\right)  }\left|  m\right\rangle
=\mathcal{N}\sum_{m=-\infty}^{\infty}e^{-i\theta m}\left|  n+m\right\rangle
=\mathcal{N}\sum_{k=-\infty}^{\infty}e^{-i\theta\left(  k-n\right)  }\left|
k\right\rangle =e^{in\theta}\left|  \theta\right\rangle
\end{equation}
where $k=n+m$. The same holds for any physical state built on $\left|
\theta\right\rangle $, in accord with (\ref{psitrf}).

Let us note, incidentally, that the instanton-mediated tunneling picture of
the SCA provides not only insight into the mechanism which generates the
$\theta$-structure but, in suitable situations, also a convenient means for
explicit calculations. Again, there exists an analogy with the
quantum-mechanical periodic potential in, e.g., condensed-matter physics: the
Bloch-Floquet wave function is the exact solution of the Schr\"{o}dinger
equation in the periodic potential while the\ tight-binding approximation
provides an intuitive and effective tool for practical calculations.

To summarize, we have derived the essential aspects of the QCD $\theta
$-structure, which owes its existence to the nontrivial topology of the gauge
group $SU\left(  3\right)  $, as a consequence of Gau\ss' law. We have
established, without recourse to any approximation, that QCD has a free
angular parameter $\theta$ which does not show up in classical chromodynamics
or in the field equations and which gives rise to a superselection rule.
Physical consequences of this new parameter will be discussed in the next section.

\subsection{$\theta$ via cluster decomposition\label{thangleclus}}

The above discussion of the $\theta$ angle can be complemented by following a
different line of reasoning which starts from the cluster-decomposition
requirement for QCD amplitudes \cite{cal76}. As a side benefit, it gives new
insight into the role of instantons in the path integral and in physical
amplitudes. In particular, this approach provides an answer to the question of
how the different topological charge sectors should be treated in functional
integrals over the gluon fields. One might wonder, e.g., whether to restrict
the functional integration to specific $Q$-sectors like, for example, the
$Q=0$ sector in which perturbation theory takes place. Or do all sectors have
to be included, and if so, with which weights? Interestingly, these questions
can be answered by imposing the cluster decomposition principle\footnote{The
cluster decomposition principle is one of the basic requirements for any
physically sensible, local quantum field theory. It can be shown to hold under
rather general conditions on the form of the Hamiltonian if a unique vacuum
state exists \cite{wei95}.} which formalizes the requirement that
spacially\ distant measurements yield uncorrelated results.

At the Greens function level, the cluster decomposition principle requires
that connected, time-ordered expectation values of products of local operators
at distant positions factorize. Let us see what this implies for the vacuum
expectation value of an operator $\mathcal{O}\left[  q,\bar{q},G\right]  $
which is composed of QCD fields and which we assume to be strongly localized
in a spacetime volume $\Omega_{1}$. (In other words, $\mathcal{O}$ has support
only in a small volume inside $\Omega_{1}$.) In order to keep an open mind on
the question of which $Q$-sectors to include, we write
\begin{equation}
\left\langle 0\left|  \mathcal{O}\right|  0\right\rangle _{\Omega}%
=\frac{\sum_{Q=-\infty}^{\infty}w\left(  Q\right)  \int\mathcal{D}%
q\mathcal{D}\bar{q}\mathcal{D}G_{Q}\mathcal{O}\left[  q,\bar{q},G\right]
e^{-S\left[  q,\bar{q},G,\Omega\right]  }}{\sum_{Q=-\infty}^{\infty}w\left(
Q\right)  \int\mathcal{D}q\mathcal{D}\bar{q}\mathcal{D}G_{Q}e^{-S\left[
q,\bar{q},G,\Omega\right]  }}%
\end{equation}
where we have allowed for the most general case by summing the contributions
of all topological charge sectors with a yet to be determined weight function
$w\left(  Q\right)  $ (which may be zero for some $Q$). Now we split the total
Euclidean spacetime volume $\Omega$ as $\Omega=\Omega_{1}+\Omega_{2}$ and note
that the topological charge $Q$ of a gluon field in the total volume $\Omega$
may be approximately written as a sum $Q=Q_{1}+Q_{2}$ over its topological
charges $Q_{1}$ in $\Omega_{1}$ and $Q_{2}$ in $\Omega_{2}$. (Strictly
speaking, $Q_{1,2}$ do not have to be integer since the condition
(\ref{bcinst}) only holds at the boundary of $\Omega.$ However, due to the
localization of the topological charge density they approximately are.) The
action is additive, too, $S\left[  \Omega\right]  =S\left[  \Omega_{1}\right]
+S\left[  \Omega_{2}\right]  $, and the functional ``measure'' over any field
$\phi$ on $\Omega$ factorizes as%
\begin{equation}
\int\mathcal{D}\phi^{\left(  \Omega\right)  }=\prod_{x\in\Omega}\int
d\phi\left(  x\right)  =\prod_{x_{1}\in\Omega_{1}}\int d\phi\left(
x_{1}\right)  \prod_{x_{2}\in\Omega_{2}}\int d\phi\left(  x_{2}\right)
=\int\mathcal{D}\phi^{\left(  \Omega_{1}\right)  }\times\int\mathcal{D}%
\phi^{\left(  \Omega_{2}\right)  }.
\end{equation}
Thus we have%
\begin{align}
\sum_{Q=-\infty}^{\infty}w\left(  Q\right)  \int\mathcal{D}G_{Q}^{\left(
\Omega\right)  }  &  =\sum_{Q=-\infty}^{\infty}w\left(  Q\right)  \sum
_{Q_{1}=-\infty}^{\infty}\int\mathcal{D}G_{Q_{1}}^{\left(  \Omega_{1}\right)
}\sum_{Q_{2}=-\infty}^{\infty}\int\mathcal{D}G_{Q_{2}}^{\left(  \Omega
_{2}\right)  }\delta_{Q,Q_{1}+Q_{2}}\\
&  =\sum_{Q_{1},Q_{2}=-\infty}^{\infty}w\left(  Q_{1}+Q_{2}\right)
\int\mathcal{D}G_{Q_{1}}^{\left(  \Omega_{1}\right)  }\int\mathcal{D}G_{Q_{2}%
}^{\left(  \Omega_{2}\right)  }%
\end{align}
and can therefore rewrite the above matrix element as
\begin{equation}
\left\langle 0\left|  \mathcal{O}\right|  0\right\rangle _{\Omega}%
=\frac{\sum_{Q_{1},Q_{2}}w\left(  Q_{1}+Q_{2}\right)  \int\left[
\mathcal{D}q\mathcal{D}\bar{q}\mathcal{D}G_{Q_{1}}\right]  ^{\left(
\Omega_{1}\right)  }\mathcal{O}\left[  q,\bar{q},G_{Q_{1}}\right]
e^{-S\left[  \Omega_{1}\right]  }\times\int\left[  \mathcal{D}q\mathcal{D}%
\bar{q}\mathcal{D}G_{Q_{2}}\right]  ^{\left(  \Omega_{2}\right)  }e^{-S\left[
\Omega_{2}\right]  }}{\sum_{Q_{1},Q_{2}}w\left(  Q_{1}+Q_{2}\right)
\int\left[  \mathcal{D}q\mathcal{D}\bar{q}\mathcal{D}G_{Q_{1}}\right]
_{\Omega_{1}}e^{-S\left[  \Omega_{1}\right]  }\times\int\left[  \mathcal{D}%
q\mathcal{D}\bar{q}\mathcal{D}G_{Q_{2}}\right]  _{\Omega_{2}}e^{-S\left[
\Omega_{2}\right]  }}. \label{oexp0}%
\end{equation}

Now comes the crucial step: since the operator $\mathcal{O}$ is strongly
localized in $\Omega_{1}$, cluster decomposition requires that $\left\langle
\mathcal{O}\right\rangle $ must be independent of what is going on in the far
separated volume $\Omega_{2}$. This is the case only if
\begin{equation}
w\left(  Q_{1}+Q_{2}\right)  =w\left(  Q_{1}\right)  w\left(  Q_{2}\right)
\text{ \ \ \ \ \ \ }\Rightarrow\text{ \ \ \ \ \ \ }w\left(  Q\right)
=e^{iQ\theta}%
\end{equation}
where the free angle $\theta$ is real since $w\left(  Q\right)  $ has to
remain finite for all $Q\in\left[  -\infty,\infty\right]  $. As a consequence,
(\ref{oexp0}) reduces to
\begin{equation}
\left\langle 0\left|  \mathcal{O}\right|  0\right\rangle _{\Omega}%
=\frac{\sum_{Q}e^{iQ\theta}\int\mathcal{D}q\mathcal{D}\bar{q}\mathcal{D}%
G_{Q}\mathcal{O}\left[  q,\bar{q},G_{Q}\right]  e^{-S\left[  \Omega
_{1}\right]  }}{\sum_{Q}e^{iQ\theta}\int\mathcal{D}q\mathcal{D}\bar
{q}\mathcal{D}G_{Q}e^{-S\left[  \Omega_{1}\right]  }}, \label{oexp2}%
\end{equation}
which shows that we indeed have to integrate over the gluon fields of all
topological charge sectors, with a given, $\theta$-dependent weight.
Incidentally, the above $\theta$-dependence is exactly what one would expect
for matrix elements of gauge-invariant operators between the $\theta$-vacuum
states (\ref{thetvac2}):%
\begin{align}
\left\langle 0\left|  \mathcal{O}\right|  0\right\rangle  &  =\frac{\sum
_{m,n}e^{im\theta}e^{-in\theta}\left\langle m\left|  \mathcal{O}\right|
n\right\rangle }{\sum_{m,n}e^{im\theta}e^{-in\theta}\left\langle
m|n\right\rangle }\\
&  =\frac{\sum_{Q}e^{i\theta Q}\left[  \sum_{n}\left\langle n+Q\left|
\mathcal{O}\right|  n\right\rangle \right]  }{\sum_{Q}e^{i\theta Q}\left[
\sum_{n}\left\langle n+Q|n\right\rangle \right]  },
\end{align}
where $Q=m-n$ and the expressions in the square brackets collect the matrix
elements which connect pure-gauge sectors with winding number
difference\footnote{Recall from Eq. (\ref{qdif}) that $Q=n_{\tau
\rightarrow\infty}-n_{\tau\rightarrow-\infty}$ is the number of potential
barriers which are tunneled through.} $Q$, just as the functional integrals in
Eq. (\ref{oexp2}).

Let us now interrupt for a moment the discussion of the functional integral in
order to illustrate by an explicit example that a restriction to gluon fields
of fixed $Q$ is in conflict with cluster decomposition \cite{lue80}. To this
end, consider the two-point function of the pseudoscalar gluonic operator%
\begin{equation}
p\left(  x\right)  =G_{\mu\nu}^{a}\left(  x\right)  \tilde{G}_{\mu\nu}%
^{a}\left(  x\right)
\end{equation}
and restrict to the $Q=0$ sector containing an instanton and an anti-instanton
of fixed size $\rho$. According to the cluster decomposition principle, one
would expect%
\begin{equation}
\Pi_{I+\bar{I}}\left(  \left|  x\right|  \rightarrow\infty\right)  \equiv
\lim_{\left|  x\right|  \rightarrow\infty}\left\langle 0\left|  Tp\left(
x\right)  p\left(  0\right)  \right|  0\right\rangle _{I+\bar{I}}\overset
{?}{=}\left\langle p\right\rangle ^{2}\equiv\lim_{\left|  x\right|
\rightarrow\infty}\left\langle 0\left|  p\left(  x\right)  \right|
0\right\rangle _{I+\bar{I}}\left\langle 0\left|  p\left(  0\right)  \right|
0\right\rangle _{I+\bar{I}}. \label{clustertest}%
\end{equation}
Let us check this hypothetical equation by evaluating both sides
independently. For the left-hand side we find
\begin{align}
\Pi_{I+\bar{I}}\left(  \left|  x\right|  \rightarrow\infty\right)   &
=\lim_{\left|  x\right|  \rightarrow\infty}\frac{1}{2}\left[  \left\langle
0\left|  Tp\left(  x\right)  p\left(  0\right)  \right|  0\right\rangle
_{I}+\left\langle 0\left|  Tp\left(  x\right)  p\left(  0\right)  \right|
0\right\rangle _{\bar{I}}\right] \\
&  =\frac{1}{2}\lim_{\left|  x\right|  \rightarrow\infty}\left[  \left\langle
0\left|  p\left(  x\right)  \right|  0\right\rangle _{I}\left\langle 0\left|
p\left(  0\right)  \right|  0\right\rangle _{I}+\left\langle 0\left|  p\left(
x\right)  \right|  0\right\rangle _{\bar{I}}\left\langle 0\left|  p\left(
0\right)  \right|  0\right\rangle _{\bar{I}}\right]  =\lambda^{2}%
\end{align}
since $\tilde{G}_{I,\mu\nu}=G_{I,\mu\nu}$ and $\tilde{G}_{\bar{I},\mu\nu
}=-G_{\bar{I},\mu\nu}$ imply%
\begin{equation}
\left\langle 0\left|  p\left(  x\right)  \right|  0\right\rangle
_{I}=-\left\langle 0\left|  p\left(  x\right)  \right|  0\right\rangle
_{\bar{I}}\equiv\lambda.
\end{equation}
On the right-hand side of Eq. (\ref{clustertest}) we have instead
\begin{equation}
\left\langle p\right\rangle ^{2}=\lim_{\left|  x\right|  \rightarrow\infty
}\frac{1}{2}\left[  \left\langle 0\left|  P\left(  x\right)  \right|
0\right\rangle _{I}+\left\langle 0\left|  P\left(  0\right)  \right|
0\right\rangle _{\bar{I}}\right]  \frac{1}{2}\left[  \left\langle 0\left|
P\left(  x\right)  \right|  0\right\rangle _{I}+\left\langle 0\left|  P\left(
0\right)  \right|  0\right\rangle _{\bar{I}}\right]  =0.
\end{equation}
Obviously the one-instanton approximation violates cluster decomposition, and
gluon fields of higher $Q$-sectors, as included in the path integral
(\ref{oexp2}), are indispensable to restore it.

We now continue our above line of reasoning by recasting Eq. (\ref{oexp2})
into a more familiar form. To this end, we define%
\begin{equation}
\sum_{Q}\mathcal{D}G_{Q}\equiv\mathcal{D}G,\text{ \ \ \ \ \ \ }S_{QCD}%
^{^{\prime}}\equiv S_{QCD}-iQ\theta, \label{sprime}%
\end{equation}
which implies%
\begin{equation}
\left\langle 0\left|  \mathcal{O}\right|  0\right\rangle =\frac{\int
\mathcal{D}q\mathcal{D}\bar{q}\mathcal{D}G\mathcal{O}\left[  q,\bar
{q},G\right]  e^{-S_{QCD}^{^{\prime}}}}{\int\mathcal{D}q\mathcal{D}\bar
{q}\mathcal{D}Ge^{-S_{QCD}^{^{\prime}}}}.
\end{equation}

Analytically continuing back to Minkowski space (with $d^{4}x_{E}=id^{4}x_{M}%
$, $G_{ai4}=-iG_{ai0}$, $\varepsilon^{0123}=+1$) and recalling Eq.
(\ref{topcharge}), the generalized action (\ref{sprime}) becomes%
\begin{equation}
S_{QCD,M}^{^{\prime}}=S_{QCD,M}-\frac{\theta g^{2}}{16\pi^{2}}\int
d^{4}xtr_{c}\left(  G_{\mu\nu}\tilde{G}^{\mu\nu}\right)
\end{equation}
which amounts to adding the term%
\begin{equation}
\mathcal{L}_{\theta}=-\frac{\theta g^{2}}{16\pi^{2}}tr_{c}\left(  G_{\mu\nu
}\tilde{G}^{\mu\nu}\right)  \label{thterm}%
\end{equation}
to the QCD Lagrangian (in Minkowski space). This new, renormalizable
interaction had been discarded during the initial development of QCD since it
is a total derivative (cf. (\ref{topcur})). The latter implies that it plays
no role in perturbation theory (perturbative fields have $Q=0$) or more
generally for any globally trival gauge field, and in the field
equations\footnote{Hence the classical instanton solutions are unaffected by
the presence of $\mathcal{L}_{\theta}$, too.}. Only the later discovery of
instantons showed explicitly that this term can have physical consequences,
the most dramatic being strong CP\ violation. Indeed, (\ref{thterm}) breaks
the combined charge conjugation (C) and parity (P) symmetry of the QCD
Lagrangian or, equivalently, its time-reversal (T) invariance. The latter is
particularly obvious since (\ref{thterm}) is at most linear in the time derivative.

The physical realization of strong CP violation does not depend solely on
$\mathcal{L}_{\theta}$, however. This can be seen by invoking a
flavor-dependent chiral $U\left(  1\right)  $ redefinition
\begin{equation}
q_{f}\rightarrow e^{i\alpha_{f}\gamma_{5}}q_{f} \label{chredef}%
\end{equation}
of the quark field of flavor $f$ in the functional integral. Due to the axial
anomaly, such a transformation produces a nontrivial change in the measure of
the quark fields, which effects a shift in the value of $\theta$:%
\begin{equation}
\theta\rightarrow\theta+2\sum_{f}\alpha_{f}.
\end{equation}
(This can be shown by arguments almost identical to those of Section
\ref{axanomsec}.) Since the quark mass term
\begin{equation}
\mathcal{L}_{m}=-\frac{1}{2}\sum_{f}\left[  m_{f}\bar{q}_{f}\left(
1+\gamma_{5}\right)  q_{f}+m_{f}^{\ast}\bar{q}_{f}\left(  1-\gamma_{5}\right)
q_{f}\right]  ,
\end{equation}
written here in its most general, CP and T non-conserving form with complex
``mass'' parameters $m_{f}$ (note that for real $m_{f}$ the $\gamma_{5}$ part
vanishes and the standard (CP-even) mass term remains), explicitly breaks
chiral invariance, it also changes under (\ref{chredef}):%
\begin{equation}
m_{f}\rightarrow e^{2i\alpha_{f}}m_{f}%
\end{equation}
where we have used%
\begin{equation}
e^{2i\alpha\gamma_{5}}\left(  1\pm\gamma_{5}\right)  =e^{\pm2i\alpha}\left(
1\pm\gamma_{5}\right)  .
\end{equation}

Since a redefinition of the path integration variables is not allowed to
change physical properties, the latter can depend only on the invariant
combination%
\begin{equation}
e^{-i\theta}\prod_{f}m_{f}\equiv e^{-i\bar{\theta}}\prod_{f}\left|
m_{f}\right|  \label{invpar}%
\end{equation}
of $\theta$ and the mass parameters $m_{f}$, or equivalently on%
\begin{equation}
\bar{\theta}=\theta-\sum_{f}\arg\left(  m_{f}\right)  . \label{thetabar}%
\end{equation}
Eq. (\ref{invpar}) shows that a finite $\theta$ would have no observable
consequences (no CP violation, in particular) if at least one quark mass would
be zero. (Although this seems quite unlikely in QCD, it cannot be firmly ruled
out at present.)

From the experimental bounds on the (CP-violating) dipole moment of the
neutron we know that $\bar{\theta}$ has to be exceedingly small\footnote{or
very close to $\bar{\theta}=\pi$, which seems however to be ruled out by meson
phenomenology}, $\bar{\theta}<10^{-9}$. How can the two seemingly independent
terms in (\ref{thetabar}) cancel so (almost) perfectly? The unknown physical
mechanism behind this cancellation and the unexplained smallness of
$\bar{\theta}$ are referred to as the ``strong CP problem''. Several
theoretical ideas for its solution \cite{pec91}, most prominently the axion
models, have been proposed. Unified theories of Nature embedding QCD (e.g.
string theory?) must even predict the value of $\bar{\theta}$ from the very
first principles. This is a highly nontrivial challenge.

\chapter{Outlook: Instantons and hadron physics}

We are coming to the end of these lectures and still have barely scratched the
surface of instanton physics. This was to be expected, given the extent and
variety of this field. However, in a workshop on hadronic physics it would
hardly be appropriate to close the discussion without having given at least a
glimpse of what impact instantons have on hadrons.

We have argued above that instantons play an essential role in shaping the QCD
vacuum. And, together with other intense, strongly correlated vacuum fields,
they render this ground state truly complex. As a case in point, the
``elementary excitations'' of such ground states are typically not the
canonical degrees of freedom in which we formulate the microscopic dynamics
(in QCD the quarks and gluons) but rather bound states or ``collective''
degrees of freedom (in QCD the hadrons) which can be considered as
disturbances of the vacuum ``medium''.

Condensed-matter physics supplies many interesting examples of such composite
elementary excitations. One of them is the propagation of phonons in a
crystal: the properties of phonon spectra and wave functions are intimately
linked to the structure of the underlying crystal ground state (e.g. to its
geometry, its distance scales, the interactions between the specific ions,
etc.). As a consequence, a thorough understanding of excitation (phonon)
properties (beyond simple mean-field approximations a la Debye) requires
detailed knowledge of the ground state. Practically all quantum systems with
many degrees of freedom share this requirement, and QCD is very likely no
exception. Thus, knowing an important ingredient of the QCD vacuum wave
functional, the instantons, it is natural to ask what this knowledge implies
for hadron structure.

This question is difficult to answer since interacting instanton ensembles,
strongly coupled to other vacuum fields over large distances, do not lend
themselves easily to a systematic and model-independent treatment (this is
common to just about any infrared-sensitive problem in QCD). As a consequence,
unequivocal and quantitative evidence for instanton effects in hadrons turned
out to be difficult to establish and the role of instantons in hadronic
physics has remained elusive until long after their discovery.

Nonetheless, several complementary approaches have inbetween significantly
improved our theoretical understanding of this role. One of the first lines of
attack was to include instanton-induced interactions into hadron models (like
MIT-bag and quark-soliton models) \cite{hor278}. Instanton vacuum models
\cite{sch98,dia84} start at a more fundamental level and approch the physics
of the instanton ensemble by approximating the field content of the vacuum
solely as a superposition of instantons and anti-instantons. This approach,
which neglects other (including perturbative) vacuum fields, has been
developed for almost two decades and can describe an impressive amount of
hadron phenomenology (from static properties and correlation functions to
parton distributions \cite{sch98,dia00}).

As we have already noted, QCD lattice simulations recently began to complement
such vacuum model studies by isolating instantons in equilibrated lattice
configurations and by studying their size distribution and their impact on
hadron correlators \cite{lat}. While results obtained from different,
currently developed lattice techniques have not yet reached quantitative
agreement, they do confirm the overall importance of instantons and some bulk
properties of their distribution in the vacuum. However, despite some initial
attemps it will still take considerable time and effort before these
numerically intensive simulations can establish reliable links to hadron
structure. Moreover, lattice ``measurements'' usually do not give insight into
the physical mechanisms which generate their data. For such purposes it is
useful to resort to more transparent, analytical approaches.

Over the last years, I have been involved in the development of such an
approach, which has the capacity to relate the instanton component of the
vacuum rather directly to hadron properties \cite{for00}. This largely
model-independent method has led to several qualitative and quantitative
insights into the role which instantons play in hadronic physics. Some of
those I will now briefly discuss.

The central idea is to calculate hadronic correlations functions%
\begin{equation}
\Pi_{1,\dots,n}(x_{1},\dots,x_{n})=\langle0|TJ_{1}(x_{1})J_{2}(x_{2})\cdots
J_{n}(x_{n})|0\rangle, \label{corr1}%
\end{equation}
i.e. vacuum expectation values of hadronic currents $J(x)$ (these are
composite QCD operators with hadron quantum numbers, e.g. $J_{M}\left(
x\right)  =\bar{q}(x)\Gamma q(x)$ for mesons) at short, spacelike distances by
means of a generalized operator product expansion which sytematically
implements instanton contributions (IOPE). The pivotal merit of this expansion
is that it factorizes (at short distances $\left|  x_{i}\right|  \ll
\Lambda_{QCD}^{-1}$) the contributions of all field modes to (\ref{corr1})
into soft ones (with momenta below a given ``operator renormalization scale''
$\mu,$ $k_{i}<\mu$) and hard ones ($k_{i}\geq\mu$). The soft contributions,
from instantons as well as from other soft vacuum fields, are summarily
accounted for by vacuum expectation values of colorless operators (the
``condensates''), while the hard contributions, originating from perturbative
fluctuations and from small (or ``direct'') instantons, are calculated
explicitly. Input are the phenomenologically known values of a few condensates
and of the two bulk properties (\ref{nbar}) and (\ref{rhobar}) of the
instanton distribution. On this basis, the IOPE provides a model-independent,
controlled approximation to the correlation functions at distances $\left|
x\right|  \lesssim0.2-0.3$ fm. In particular, it achieves a unified QCD
treatment of instanton contributions in conjunction with contributions from
long-wavelength vacuum fields and perturbative fluctuations.

Hadron properties are obtained from the IOPE by matching it to a dual,
hadronic description of the correlators. The latter is based on a
parametrization of the corresponding spectral functions in terms of hadron
properties and local quark-hadron duality (which is essentially a consequence
of asymptotic freedom). The specifics of the matching between both
descriptions rely on techniques developed for QCD sum rules (like e.g. the use
of the Borel transform) \cite{shi79}. While the application range of the IOPE
approach is smaller than that of instanton vacuum model and lattice
calculations, it has the advantages of being transparent, largely
model-independent and fully analytical.

Over the last years several hadronic channels have been studied in this
framework, including those of pions \cite{for95,for300}, baryons
\cite{for93,for97,aw99,for400} and glueballs \cite{for200}. In addition to
quantitative predictions for hadron properties, these investigations have led
to various qualitative insights into how instantons manifest themselves in
hadron structure:

\begin{enumerate}
\item In several hadron channels the direct instanton contributions were found
to be of substantial size and impact. This adds to the evidence from other
sources for the importance of instantons in hadron structure. More
specifically, the results show that nonperturbative effects can strongly
affect hadron structure already at surprisingly small distances $\left|
x\right|  <0.2$ fm, and that most of these effects can be attributed to
(semi-hard) instantons.

\item Instanton effects are strongly hadron-channel selective and favor
especially spin-0 meson and glueball channels. Moreover, various invariant
amplitudes of the same correlator can receive qualitatively and quantitatively
different instanton contributions. In such situations the neglect of hard
instantons (which is standard practice in conventional QCD sum rules) leads to
reduced stability or failure of those sum rules which are more strongly
affected by instantons. Implementing the missing direct-instanton
contributions led, in particular, to the resolution of long-standing stability
problems in the chirally-odd nucleon sum-rule \cite{for93}, one of the
magnetic-moment sum rules \cite{aw99}, the pseudoscalar pion sum rule
\cite{for300} and the (lowest moment)\ scalar glueball sum rule \cite{for200}.

\item Stable and predictive IOPE-based sum rules exist even for correlators
which do not permit a conventional QCD sum rule analysis. In particular, the
first sum rule for the pion form-factor based on pseudoscalar currents could
be established in the IOPE\ approach \cite{for95}. Its prediction for the
form-factor agrees well with experiment in the full range of accessible
momentum transfers.

\item In the pseudoscalar pion sum rule, the direct-instanton contributions
induce new, chiral-symmetry breaking operators which play a crucial role in
generating the exceptionally small (pseudo-Goldstone) pion mass \cite{for300}.

\item Instanton effects enhance the magnetic susceptibility of the quark
condensate. The resulting values are in line with predictions of other,
independent approaches. Moreover, direct instantons have a strong impact on
one of the magnetic-moment sum rules of the nucleon and considerably improves
their overall stability and consistency \cite{aw99}.

\item An interesting new mechanism for instanton-enhanced isospin-breaking in
hadrons was found in \cite{for97}. Although instantons, being gluon fields,
are ``flavor blind'', they can strongly amplify isospin violation effects
which originate from other, soft vacuum fields (most prominently from the
difference of the up- and down-quark condensate) and which manifest
themselves, e.g., in the proton neutron mass difference. Even sophisticated
quark models which include instanton--induced quark interactions miss such effects.

\item The scalar glueball can be (over-) bound by the instanton contributions
alone \cite{for200}. In fact, this channel provides the first example for a
sum rule which can be stabilized solely by the contributions from instantons.
This result lends support to the findings of instanton vacuum models
\cite{sch95} which neglect the remaining contributions. Incidentally, a mainly
instanton-bound $0^{++}$ glueball fits naturally to the particularly small
glueball radius $r_{G}$ found on the lattice \cite{def92}, which is of the
order of the instanton size:%
\begin{equation}
r_{G}\sim\bar{\rho}.
\end{equation}

The IOPE also provides the first set of $0^{++}$ glueball sum rules which are
overall consistent with the low-energy theorem which governs the zero-momentum
limit of the corresponding correlator \cite{for200}.

\item Direct instantons generate by far the dominant contributions to the
IOPE\ of the scalar glueball correlator. This makes it possible to establish
approximate scaling relations between the bulk features of the instanton
distribution and the mass $m_{G}$ and decay constant $f_{G}$ of the scalar
glueball \cite{for200}:
\begin{align}
m_{G}  &  \sim\bar{\rho}^{-1},\\
f_{G}^{2}  &  \sim\bar{n}\bar{\rho}^{2}.
\end{align}
These relations are the first of their kind and provide a particularly direct
link between instanton and hadron properties.
\end{enumerate}

The above findings, together with those from other sources, are beginning to
assemble into a comprehensive picture of how instantons manifest themselves in
hadron structure and interactions. This picture will certainly become richer
and more detailed in the future, incorporating physics ranging from nuclear
and quark matter to hard processes, and it will very likely also teach us more
about how to deal with nonperturbative QCD in general.\medskip

\bigskip

\textbf{Acknowledgement}

It is a pleasure to thank the organizer, Prof. Yogiro Hama, for this pleasant
and informative workshop and Profs. Krein and Chiapparini for the invitation
to lecture on instantons.

\appendix         

\chapter{Gauge invariance: large vs. small\label{ginvlargesmall}}

In this appendix we show that Schr\"{o}dinger wave functionals which satisfy
Gau\ss' law may still not be invariant under ``large'' gauge transformations,
i.e. those which are not continuously connected to the unit element of the
gauge group. To this end, consider the functional
\begin{equation}
W\left[  \mathbf{G}\right]  =\int d^{3}xK^{0}=-\frac{1}{8\pi^{2}}\int
d^{3}x\epsilon_{ijk}tr\left[  G_{i}\left(  \partial_{j}G_{k}+\frac{2}{3}%
G_{j}G_{k}\right)  \right]
\end{equation}
($K^{\mu}$ is the ``topological current'' (here in Minkowski space) whose
divergence we have encountered in Eq. (\ref{topcur}), the integrand is known
as the ``Chern-Simons 3-form''). One easily calculates that
\begin{equation}
\frac{\delta W\left[  \mathbf{G}\right]  }{\delta G_{i}^{a}\left(  \vec
{x}\right)  }=-\frac{1}{4\pi^{2}}\epsilon_{ijk}tr\left(  t_{a}\left[
\partial_{j}G_{k}+G_{j}G_{k}\right]  \right)  =\frac{1}{8\pi^{2}}\tilde
{G}_{i0}^{a}%
\end{equation}
so that
\begin{equation}
\left(  -\vec{D}^{ab}\vec{E}^{b}\right)  W\left[  \mathbf{G}\right]
=-iD_{i}^{ab}\frac{\delta W\left[  \mathbf{G}\right]  }{\delta G_{i}%
^{a}\left(  \vec{x}\right)  }=-\frac{i}{8\pi^{2}}D_{i}^{ab}\tilde{G}_{i0}%
^{a}=0
\end{equation}
vanishes identically as a consequence of the Bianchi identity (\ref{bianchi}).
This implies that $W\left[  \mathbf{G}\right]  $ varies under gauge
transformations $\delta G_{i}^{a}=D_{i}^{ab}\omega^{b}\left(  \vec{x}\right)
$ with arbitrary gauge functions $\omega$ as%
\begin{align}
\delta_{\omega}W\left[  \mathbf{G}\right]   &  =\int d^{3}x\frac{\delta
W\left[  \mathbf{G}\right]  }{\delta G_{i}^{a}\left(  \vec{x}\right)  }%
D_{i}^{ab}\omega^{b}\left(  \vec{x}\right)  =\int d^{3}x\left(  -D_{i}%
^{ab}\frac{\delta W\left[  \mathbf{G}\right]  }{\delta G_{i}^{a}\left(
\vec{x}\right)  }\right)  \omega^{b}\left(  \vec{x}\right)  +\int_{\Sigma
}d^{2}\Omega_{i}\frac{\delta W\left[  \mathbf{G}\right]  }{\delta G_{i}%
^{a}\left(  \vec{x}\right)  }\omega^{b}\left(  \vec{x}\right) \\
&  =\int_{\Sigma}d^{2}\Omega_{i}\frac{\delta W\left[  \mathbf{G}\right]
}{\delta G_{i}^{a}\left(  \vec{x}\right)  }\omega^{b}\left(  \vec{x}\right)
\end{align}
where we have used Gau\ss' law. Since the surface $\Sigma$ lies at spacial
infinity we see that ``small'' gauge transformations with $\omega^{a}\left(
\vec{x}\right)  \overset{\vec{x}\rightarrow\infty}{\longrightarrow}0$ leave
$W$ invariant. The parameters of large gauge transformations, in contrast, do
not vanish at spacial infinity and can therefore be topologically ``active''.
Under such topologically nontrivial gauge transformations the above surface
term stays finite,%
\begin{equation}
\delta_{U}W\left[  \mathbf{G}\right]  =W\left[  ^{U}\mathbf{G}\right]
-W\left[  \mathbf{G}\right]  =n\left[  U\right]  ,
\end{equation}
where
\begin{equation}
n\left[  U\right]  =\frac{1}{24\pi^{2}}\int d^{3}x\epsilon_{ijk}tr\left[
\left(  U^{-1}\partial_{i}U\right)  \left(  U^{-1}\partial_{j}U\right)
\left(  U^{-1}\partial_{k}U\right)  \right]
\end{equation}
provides an explicit expression for the winding number of the gauge
transformation $U$.

\end{document}